\documentclass[%
 reprint,
 superscriptaddress,
preprintnumbers,
nofootinbib,
 amsmath,amssymb,
 aps,
]{revtex4-1}

\usepackage{tikz}
\usepackage{graphicx}
\usepackage{dcolumn}
\usepackage{bm}
\usepackage{hyperref}
\usepackage{color}

\usepackage{multirow, array}

\hypersetup{
    colorlinks = true
}
\usepackage{aas_macros}

\renewcommand{\vec}{\bm}

\def\fun#1#2{\lower3.6pt\vbox{\baselineskip0pt\lineskip.9pt
        \ialign{$\mathsurround=0pt#1\hfill##\hfil$\crcr#2\crcr\sim\crcr}}}
\newcommand{\lexp}{\mathop{\langle}}    
\newcommand{\rexp}{\mathop{\rangle}}    

\newcommand{\beq}{\begin{equation}}
\newcommand{\eeq}{\end{equation}}
\newcommand{\beqa}{\begin{eqnarray}}
\newcommand{\eeqa}{\end{eqnarray}}
\newcommand{\be}{\begin{equation}}
\newcommand{\ee}{\end{equation}}
\newcommand{\bea}{\begin{eqnarray}}
\newcommand{\eea}{\end{eqnarray}}
\newcommand{\nn}{\nonumber}

\newcommand{\vpt}{\mbox{\sf VPT}}

\DeclareMathOperator{\arccot}{arc\,cot}
\DeclareMathOperator{\arccoth}{arc\,coth}

\begin{document}

\preprint{TUM-HEP 1423/22}

\title{Perturbation theory with dispersion and higher cumulants:\\ framework and linear theory}

\author{Mathias Garny}
\email{mathias.garny@tum.de}
\affiliation{Physik Department T31, Technische Universit\"at M\"unchen, James-Franck-Stra\ss{}e 1, D-85748 Garching, Germany
}%

\author{Dominik Laxhuber}
\email{dominik.laxhuber@tum.de}
\affiliation{Physik Department T31, Technische Universit\"at M\"unchen, James-Franck-Stra\ss{}e 1, D-85748 Garching, Germany
}%

\author{Rom\'an Scoccimarro}
\email{rs123@nyu.edu}
\affiliation{
 Center for Cosmology and Particle Physics, Department of Physics, New York University, NY 10003, New York, USA
}%

\date{October 14, 2022}

\begin{abstract} 
The standard perturbation theory (SPT) approach to gravitational clustering is based on a fluid approximation of the underlying Vlasov-Poisson dynamics,
taking only the zeroth and first cumulant of the phase-space distribution function into account (density and velocity fields). This assumption breaks down when dark matter particle orbits cross and leads to well-known problems, e.g. an anomalously large backreaction of small-scale modes onto larger scales that compromises predictivity. We  extend SPT  by incorporating second and higher cumulants  generated by orbit crossing. For collisionless matter, their equations of motion are completely fixed by the Vlasov-Poisson system, and thus we refer to this approach as \emph{Vlasov Perturbation Theory} (\vpt). 
Even cumulants develop a background value, and they enter the hierarchy of coupled equations for the fluctuations. The background values are in turn sourced by power spectra of the fluctuations. The latter can be brought into a form that is formally analogous to SPT, but with an extended set of variables and linear as well as non-linear terms, that we derive explicitly.  In this paper, we focus on linear solutions, which are far richer than in SPT, showing that modes that cross the dispersion scale set by the second cumulant are highly suppressed. We derive stability conditions on the background values of even cumulants from the requirement that exponential instabilities be absent. We also compute the expected magnitude of averaged higher cumulants for various halo models and show that they satisfy the stability conditions. 
Finally, we derive self-consistent solutions of perturbations and background values for a scaling universe and study the convergence of the cumulant expansion. 
The \vpt~framework provides a conceptually straightforward and deterministic extension of SPT that accounts for the decoupling of small-scale modes.
\end{abstract}

\maketitle

\tableofcontents

\section{Introduction}
\label{sec:introduction}

Some aspects of gravitational clustering in cosmology are still poorly understood, despite decades of work and progress. Two shortcomings of standard perturbation theory (hereafter SPT, see~\cite{BerColGaz02} for a review) in particular, motivate the present work. 

Simulations have established over two decades ago that for initial conditions with blue spectra non-linear growth is {\em suppressed} compared to linear~\cite{ColBouHer96}. This is in contrast with red or cold dark matter (CDM)-like initial spectra which show the familiar enhancement at small scales. Unfortunately, SPT quickly breaks down before it can actually provide any useful understanding of this remarkable property of gravitational clustering, giving UV divergences for spectral indices $n_s\geq -1$~\cite{MakSasSut92,ScoFri9612}.  In addition,  the bluer the spectra the worse these UV divergences become,  when in fact small-scale clustering is actually {\em most suppressed}  compared to linear expectations.

A better understanding of the non-linear regime must also address how non-linear modes decouple from large-scale quasilinear modes. A quick look at simulations makes  clear that small-scale regions form fairly stable objects that decouple from the expansion of the universe: dark matter halos. What is the backreaction of these halos on large-scale structure? In SPT there is none, as the theory is expanded about free linear modes that know nothing about halo formation, even for the shortest wavelengths. In halo models of gravitational clustering~\cite{CooShe02} prescriptions are built to marry SPT at large scales with halos at small scales, but these provide little  insight into the physics by which such decoupling takes place. Measurements of the response function in simulations~\cite{NisBerTar1611,NisBerTar1712} also highlight that the non-linear power depends on linear modes weaker than SPT dictates, presumably related to decoupling. 

Recent work on large-scale structure has focussed on effective field theory (EFT)~\cite{CarHerSen1206,BauNicSen1207}, which adds to SPT counter-terms consistent with the symmetries whose amplitudes are determined from fitting simulation measurements of clustering statistics. These counter-terms are derived from a derivative expansion of the stress tensor, valid at large scales. 
When allowing for the most general form compatible with symmetries, the counter-terms parameterize the way how small-scale physics can modify SPT predictions in the large-scale limit, while deliberately being ignorant about the origin of these modifications.
By construction, an EFT approach therefore can neither explain nor take advantage of the decoupling mentioned above. Since the counter-terms also correct the leading sensitivities to the highly non-linear regime that SPT has as a result of loop integrations over free linear modes, there is little physical insight to be extracted from their amplitude, and they need to be treated as free parameters in practice. While this approach is certainly possible and useful in particular applications, it is nevertheless tempting to try to take advantage of the decoupling of UV modes in order to obtain a predictive framework of perturbation theory that systematically improves over SPT without the need to introduce a large set of free parameters.

The shortcomings of SPT can be traced back to the key assumption that the equations of motion for CDM correspond to a pressure-less perfect fluid (PPF) at all times. This ignores the physics of orbit crossing (or shell crossing in the spherical dynamics language), which generates at once all higher cumulants of the phase-space distribution function (DF) beyond the density and velocity fields, giving rise to the Vlasov hierarchy~\cite{PueSco0908}. As we shall discuss in this paper, when this physics is incorporated one can expand the equations of motion about a new linear theory that knows about small-scale orbit crossing through the {\em expectation values} of the cumulants of the DF. This  simultaneously incorporates the decoupling of large from small-scale modes and explains the trends of clustering in the non-linear regime with spectral index~\cite{cumPT2}, showing that the motivations cited above are two sides of the same coin. 

The purpose of the present paper is to introduce the main ideas behind our approach based on the Vlasov hierarchy, and in particular explore in detail its linearized solutions which are far richer than that in SPT. We also highlight the connections between certain physical quantities (expectation values of cumulants of the DF) that can be estimated from dark matter halos and use these results to gain some intuition about the size of the effects beyond SPT and the stability of the linear solutions. 
Furthermore, we develop the formalism for a systematic perturbative expansion when taking second and higher cumulants into account, and provide explicit results for the non-linear terms in the corresponding extended set of equations of motion.  We will commonly refer to the perturbative expansion incorporating second or higher cumulants as \emph{Vlasov Perturbation Theory} (\vpt). Solutions taking non-linear corrections into account are studied in detail in a companion work~\cite{cumPT2} (hereafter paper II).

Our main aim in these papers is to investigate to what extent perturbative techniques for gravitational clustering can be improved by taking advantage of the underlying collisionless dynamics. This builds on previous results in the literature that explored some aspects of the Vlasov hierarchy truncated at the second cumulant, in particular corrections to large-scale modes using a low-$k$ expansion~\cite{McD1104} and the more systematic approach in~\cite{ErsFlo1906,Erschfeld:2021kem}. An active field of related research has been to study  the growth of velocity dispersion and other shell-crossing aspects from the point of view of  Lagrangian perturbation theory~\cite{BucDom9807,AdlBuc9903,TatSudMae0209,MorTat0112,Col1501,Avi1603,CusTanDur1703,TarCol1710,RamFri1710,McDVla1801,SagTarCol1812,HalNisTar2012,RamHah2102}. Along similar lines,~\cite{Val1009,ValNis1103,ValNis1108,ValNisTar1304,SelVla1506} investigate how to match Lagrangian perturbation theory to dark matter halos at small scales to incorporate some aspects of shell-crossing. Complementary insights to Vlasov collisionless dynamics have resulted from the development of numerical codes to follow the phase-space distribution function~\cite{AbeHahKae1211,HahAbeKae1309,ColTou1407,HahAng1601,SouCol1609,Stucker:2019txm,Col2103,SagTarCol2208,AngHah2212}, as well as using the Schr\"odinger equation to model
collisionless dynamics~\cite{WidKai9310,Schive:2014dra,Kopp:2017hbb,GarKon1801,Uhl1810,Mocz:2018ium,Li:2018kyk,UhlRamGos1904,Garny:2019noq,May:2021wwp}.

This work is organized as follows. In Sec.\,\ref{sec:halo} we present a computation of average values of cumulants of the DF for two familiar static halo models, in order to gain some intuition on the expected magnitude of the cumulant expectation values.
After this prelude, we start to develop the extended framework of cosmological perturbation theory for large-scale structure referred to as \vpt~in Sec.\,\ref{sec:eqs}, where we review the cumulant generating function and the underlying Vlasov dynamics. Next, in Sec.\,\ref{sec:sigma}, we derive the extension of SPT taking the second cumulant into account, being the velocity dispersion tensor. We discuss the decomposition into expectation value and perturbations and derive their general non-linear equations of motion. The extension of this framework to higher cumulants is discussed in Sec.\,\ref{sec:pi}. In the remaining part of this work, we focus on the linear approximation. We start in Sec.\,\ref{sec:linear}, discussing analytical solutions in the second cumulant approximation, and changes when taking the third and fourth cumulant into account. We then proceed to derive coupled equations up to arbitrary cumulant order in linear approximation in Sec.\,\ref{sec:hierarchy}. They involve expectation values for all even cumulants. We analytically derive conditions on their size from the requirement of stability, and compare them to the values for halos derived in the beginning. Finally, we apply the formalism to a scaling universe in Sec.\,\ref{sec:powerlaw}, where certain simplifications occur. This allows us to derive self-consistent solutions for the cumulant expectation values (up to the eighth cumulant), and discuss the convergence with respect to the truncation order of the cumulant expansion. We conclude in Sec.\,\ref{sec:conclusions}. The appendices contain further details and a collection of results that are too lengthy to present in the main text.

\section{Halo model for dispersion and higher cumulants}
\label{sec:halo}

We are interested in understanding what one can expect about the order of magnitude of the cumulants of the phase-space distribution function $f(\bm{r},\bm{p},t)$. For collisionless, non-interacting dark matter particles, the phase-space density is conserved along the particle trajectories, $0=df/d\tau$, which yields the Vlasov (or collisionless Boltzmann) equation. As mentioned above, while SPT neglects dispersion (second cumulant) and higher cumulants of $f$, these are all generated at once by orbit crossing~\cite{PueSco0908} at small scales and therefore are non-trivial in dark matter halos. To gain some insight into the properties expected from cumulants in halos we consider two models, with different (approximately orthogonal) approximations. Both of these models are solutions to the Vlasov equation
\beq
{\partial f \over \partial t} +\bm{p}\cdot {\partial f \over \partial \bm{r}} - \nabla\Phi\cdot {\partial f \over \partial  \bm{p}} = \bm{p}\cdot {\partial f \over \partial \bm{r}} - \nabla\Phi\cdot {\partial f \over \partial  \bm{p}} =0 \,,
\label{SteadyVlasov}
\eeq
in the steady-state limit $f(\bm{r},\bm{p},t)=f(\bm{r},\bm{p})$. 
This is so because the distribution function is a function of phase-space coordinates only through integrals of motion, such as energy and angular momentum. The gravitational potential obeys the Poisson equation, 
\beq
\nabla^2 \Phi = 4\pi G\, \rho= 4\pi G\, \int f(\bm{r},\bm{p})\, d^3p\,,
\label{PoissonHalos}
\eeq
with $\rho$ the density profile of the halo. Note in this section we consider isolated halos, so it is most convenient to work with the full density field and physical coordinates and momenta. In the next section, where we discuss the perturbative approach to the time-dependent Vlasov equation in structure formation we switch to comoving coordinates and momenta, and work with dimensionless density perturbations.

The first halo model we consider is an axisymmetric halo which can be non-spherical, with a flat rotation curve. It has a simple analytic form for the distribution function~\cite{Eva9301,Eva9403} that depends on energy and the $z$-component of the angular momentum and in which the expectation value for the cumulants can be calculated analytically as a function of halo shape. It has been used to model the dark matter halo of the Milky Way to infer deviations from spherical symmetry from microlensing observations~\cite{FriSco9408,AlcAllAxe9508,CalGarDav1809}. Apart from its analytic interest, this model gives us some insight about the dependence of cumulants on deviations of spherical symmetry. 

The second halo model is the Navarro-Frenk-White (NFW) profile~\cite{NavFreWhi97}, which is a reasonable fit to well-relaxed spherically averaged CDM halos in cosmological N-body simulations. Under the assumption of spherical symmetry, the distribution function depends on phase-space coordinates through energy and the square of the angular momentum; in particular we consider the case of constant anisotropy where the angular momentum dependence is a power-law.  While we loose the halo-shape information, this approach has the advantage that one can integrate over the halo mass function calibrated from simulations to obtain realistic estimates for the expectation values of cumulants in a given cosmology.

\subsection{Evans Halos}
\label{subsec:Evans}

Let us start from the phase-space density $f$ for an axisymmetric halo, due to Evans~\cite{Eva9301},  which is a function of phase-space coordinates through the energy $E=p^2/2+\Phi$ and angular momentum $L_z=p_\phi$, where $\Phi$ is assumed to be logarithmic (see Eq.~\ref{PhiLog} below) and use units where the particle mass is unity. It reads,

\beq
f(\bm{r},\bm{p})=(A\, L_z^2 + B) \exp(-2E/\sigma^2) + C \exp(-E/\sigma^2)\,,
\label{fEvans}
\eeq
where $\sigma$ is a characteristic (constant) velocity dispersion scale, and $A,B,C$ are constants specified below in a different parametrization. For the most part, we are interested in normalized cumulants, so the overall value of $\sigma$ will drop out from the quantities we are interesting in. We can rewrite Eq.~(\ref{fEvans}) in a more convenient form for our purposes as follows,
\beqa
f = \rho \ &\Big[& \Big(w_a\, {p_\phi^2\over {\sigma^2/2}} + w_b\Big)  \, f^{(1/2)}_{\rm G}(\bm{p}) +
 w_c\,  f^{(1)}_{\rm G}(\bm{p}) \Big]\,,
\label{fEvans2}
\eeqa
where $\rho$ is the density profile, $f^{(n)}_{\rm G}$ denotes a Gaussian probability distribution with variance $n\sigma^2$ in each dimension. Note that each of the three momentum-dependent terms in Eq.~(\ref{fEvans2}) integrates to unity over all phase-space. The weights $w_i$ are functions of the location inside the halo (characterized by cylindrical coordinates $R, \phi, z$) and add up to unity, i.e. $w_a+w_b+w_c=1$. They are given by
 \beq
 w_a \equiv {a\over a+b+c}, \ \ \ \ \  w_b \equiv {b\over a+b+c}, \ \ \ \ \  w_c \equiv {c\over a+b+c}\,,
 \label{weights}
 \eeq
with
\beq
a \equiv {2(1-q^2) R^2 \over \xi^4}, \ \ \ \ \ b \equiv {2 R_c^2 \over \xi^4}, \ \ \ \ \ c \equiv {(2q^2-1) \over \xi^2}\,,
\label{abc}
\eeq
and
\beq
\xi^2 \equiv R^2 + R_c^2 + z^2/q^2\,,
\label{xisq}
\eeq
where $R_c$ is a core radius and $q$ is a shape parameter that describes oblate ($q<1$), spherical ($q=1$) and prolate ($q>1$) halos. High-resolution simulations show that dark matter halos are generically triaxial, but on average closer to prolate, although backreaction from baryons makes them more spherical overall and closer to oblate~\cite{ChuPilVog1903,PraForGra1912}. Imposing positivity of the density everywhere and of the distribution function itself, leads to the constraint $1/\sqrt{2} \leq q \leq 1.08$~\cite{Eva9301}.

The density profile is given by $\rho = A\, (a+b+c)$, with $A\equiv \sigma^2/(2\pi G q^2)$, and thus,
\beq
\rho = A\ \Big[ \frac{R^2+(2q^2+1) R_c^2+(2q^2-1) z^2/q^2}{\xi^4} \Big]\,,
\label{EvansProfile}
\eeq
which satisfies the Poisson equation for the logarithmic Newtonian potential
\beq
\Phi = \sigma^2\, \ln(\xi^2) + {\rm const}\,.
\label{PhiLog}
\eeq
From Eqs.~(\ref{EvansProfile},~\ref{PhiLog}) we see that the Evans density profile has an oblateness/prolateness that depends on location in the halo, it is rather the equipotentials that are oblate/prolate independent of location inside the halo.

\begin{figure}[t]
  \begin{center}
  \includegraphics[width=\columnwidth]{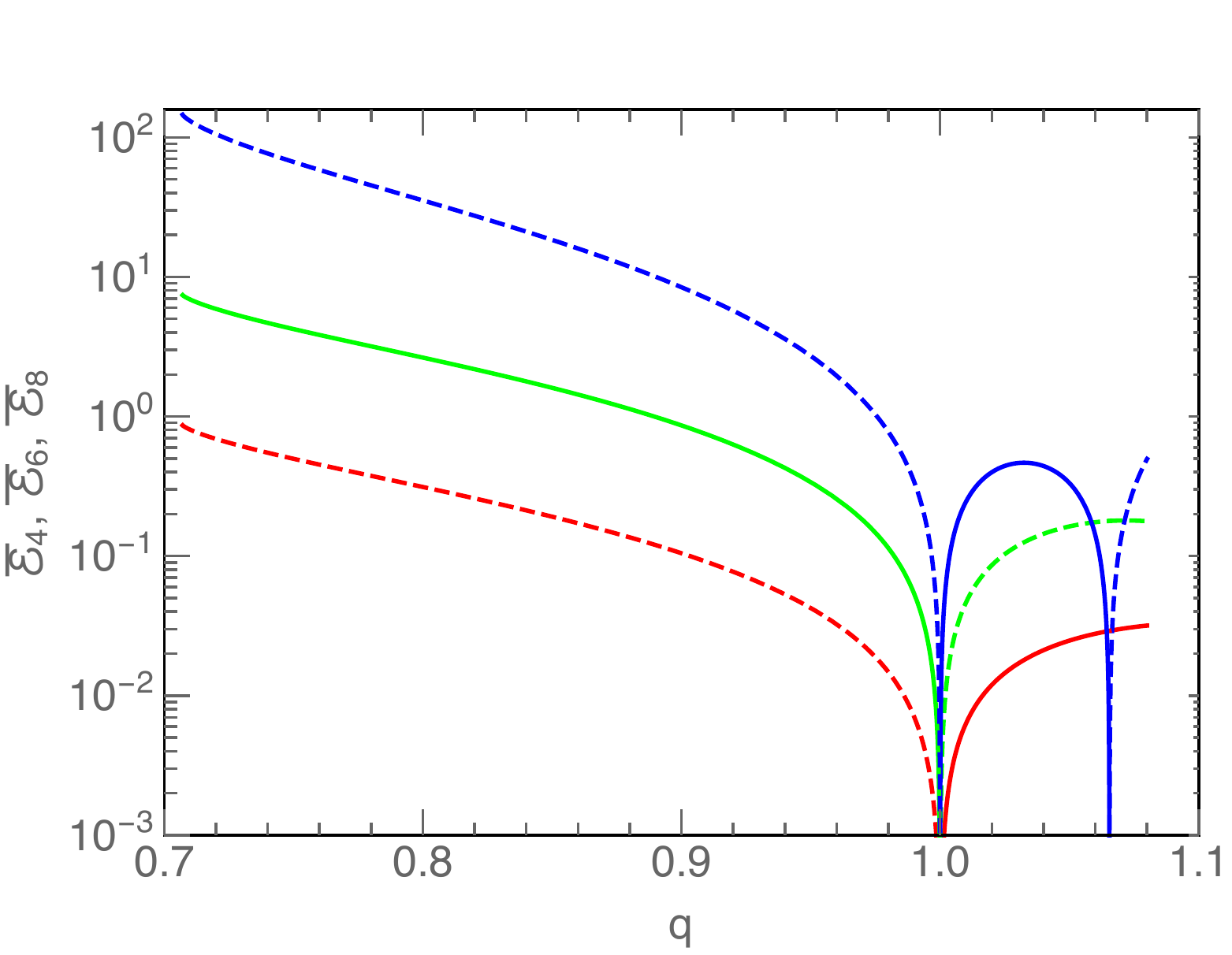}
  \end{center}
  \caption{\label{fig:E4E6E8Evans} Normalized cumulants $\bar{\cal E}_4, \bar{\cal E}_6, \bar{\cal E}_8$ (bottom to top) for Evans infinite halos as a function of halo-shape parameter $q$. Dashed lines denote negative values. When halos are spherical ($q=1$), the distribution function becomes Gaussian and the density profile isothermal.}
\end{figure}

Note that $f$ given by  Eq.~(\ref{fEvans}) is even in $L_z$, one can always add to this $f$ a contribution odd in $L_z$ (which will correspond to adding net rotation to the halo). This will not change the density profile, but will induce odd moments of $f$, which for Eq.~(\ref{fEvans}) are all zero. To compute the cumulants, it is convenient to calculate directly the cumulant generating function (CGF)
\beqa
{\cal C}(\bm{r},\bm{l}) &\equiv& \ln \Big[  \int d^3p  \ e^{\bm{l}\cdot\bm{p}} \ f(\bm{r},\bm{p}) \Big] \nonumber \\ &= & 
\ln\rho + {\cal C}_i l_i + {\cal C}_{ij} \, {l_i l_j \over 2!}+ {\cal C}_{ijk}\, {l_i l_j l_k \over 3!}+ \ldots \,,
\label{CGF}
\eeqa
where in the second expression we introduced the Taylor expansion that identifies the cumulant themselves, i.e. ${\cal C}_{i\ldots j}(\bm{r})\equiv \nabla_{l_i}\dots \nabla_{l_j} {\cal C}|_{\bm{l}=0}$. Given the expression of $f(\bm{r},\bm{p})$ in terms of Gaussian distributions, Eq.~(\ref{fEvans2}), the CGF can be obtained right away,

\beqa
{\cal C}(\bm{r},\bm{l}) &=& \ln \Big\{  \Big[w_a \Big(1+ \frac{(\bm{l}\cdot\hat{\phi})^2}{2} \Big) + w_b \Big] \ e^{\bm{l}^2/4} +  w_c \, e^{\bm{l}^2/2}\Big\}
 \nonumber \\& & + \ln \rho \,,
\label{CGFEvans}
\eeqa
where 
$\hat{\phi}$ is the unit vector in the $\phi$ direction and we have scaled out the $\sigma$ dependence, i.e. $\bm{l}\sigma \to \bm{l}$; this corresponds to calculating dimensionless cumulants normalized by the corresponding power of the constant $\sigma$ in this model. Since the CGF is quadratic in its argument, only even cumulants are non-trivial in this model. For example, the normalized velocity dispersion tensor $\hat{\sigma}_{ij}\equiv \sigma_{ij}/\sigma^2={\cal C}_{ij}$ gives, 
\beq
\hat{\sigma}_{ij} = \Big({1\over 2}w_a+{1\over 2}w_b+w_c\Big) \delta_{ij} + w_a \, \hat{\phi}_i \hat{\phi}_j\,,
\label{sigmahat}
\eeq
which in turn yields a trace of $\hat{\sigma}_{ii}=5w_a/2+3w_b/2+3w_c$.  As mentioned earlier, we are mostly interested  here in the {\em expectation values} of the cumulants, which should be dominated by halo contributions where shell crossing is most severe, and where perturbation theory is least reliable. By symmetry only even cumulants have non-zero expectation values, and they are spatially homogeneous. In halo models they correspond to averaging over the halo. We introduce the  expectation values of even cumulants ${\cal E}_{2n}$,
\beq
{\cal E}_{2n} \equiv {\lexp {\cal C}_{i_1i_1 \dots i_n i_n} \rexp \over {2n+1} }\,.
\label{E2n}
\eeq
Indeed, this definition gives for the expectation value of Eq.~(\ref{CGF}) the usual definition of cumulants from the Taylor series of the CGF
\beq
\lexp {\cal C} \rexp = \lexp \ln\rho  \rexp+ \sum_{n=1}^\infty {\ell^{2n} \over (2n)!} \ {\cal E}_{2n}\,,
\label{expCGF}
\eeq
where $\ell^2 = \bm{l}\cdot \bm{l}$. The lowest cumulant expectation value corresponds to the average velocity dispersion,
\beq
3\, {\cal E}_2 =  \frac{5}{2} \lexp w_a\rexp + \frac{3}{2} \lexp w_b\rexp +3 \lexp w_c \rexp \,,
\label{E2Evans}
\eeq
where $\lexp\rexp$ denotes averaging the weights over the halo. The case of the fourth cumulant is instructive in two respects. It reads,
\beqa
5\, {\cal E}_4 &=& \Big\langle \frac{35}{4}  w_a+  \frac{15}{4} w_b+ 
 15  w_c \Big\rangle  - \Bigg[ \frac{47}{4} \lexp w_a^2\rexp   
+ \frac{15}{4} \lexp w_b^2\rexp  \nonumber \\ && + 15 \lexp w_c^2\rexp + \frac{25}{2} \lexp w_a w_b\rexp
 + 15 \lexp w_b w_c\rexp + 15 \lexp w_c w_a\rexp \Bigg] \,. \nonumber \\ &&
\label{E4Evans}
\eeqa
The first thing to note is that terms being subtracted in square brackets correspond to the usual subtraction in going from the fourth moment (first three terms in Eq.~\ref{E4Evans}) to the fourth cumulant. But since momentum is a vector field, this does not correspond to the usual kurtosis of a scalar, instead we have ${\cal C}_{iijj}={\cal M}_{iijj}-2 {\cal M}_{ij} {\cal M}_{ij}- {\cal M}_{ii} {\cal M}_{jj}$, where the ${\cal M}_{i\ldots j}=\nabla_{l_i} \ldots \nabla_{l_j} {\cal M}|_{\bm{l}=0}$ are the moments with ${\cal M}=\exp ({\cal C})$ the moment generating function.  The second observation is that the subtraction requires to average over the halo quadratic combinations of the weights $w_i$. Let us briefly discuss this average. We can perform an average of the weights over a sphere of radius $r$ as,
\beq
\bar{w}_i(r) \equiv \frac{3}{r^3} \int_0^r dz \int_0^{\sqrt{r^2-z^2}} w_i(R,z) \, R\, dR\,,
\label{SphAvg}
\eeq
which can be done analytically. We also need, for the cumulants, averages of non-linear combinations of weights. These however do not appear to have analytic expressions, except in the case of an infinite halo, which is what we shall consider from now on. In this case, it is easy to check from simple scaling that this corresponds to setting $w_b=0$, i.e. the core radius $R_c$ (fixed as $r\to\infty$) drops out, and the halo-shape parameter $q$ remains the only quantity in the problem. Since the weights add up to unity, for infinite halos everything can be written in terms of $w_a$, which simplifies the expressions. Equations~(\ref{E2Evans}-\ref{E4Evans}) become then 
\beq
{\cal E}_2 = 1- {1\over 6} \lexp w_a \rexp, \ \ \ \ \ {\cal E}_4 = - {1\over 4} \lexp w_a \rexp -{7\over 20} \lexp w_a^2 \rexp \,,
\label{E2E4Inf}
\eeq
whereas the full expressions for the sixth and eighth cumulants, presented in App.~\ref{app:halo}, become
\beqa
\label{E6Inf}
{\cal E}_6 &=& \frac{15}{8} \lexp w_a \rexp + \frac{27}{8} \lexp w_a^2 \rexp - \frac{27}{28} \lexp w_a^3 \rexp \\ 
{\cal E}_8 &=& -\frac{175}{16} \lexp w_a \rexp - \frac{791}{16} \lexp w_a^2 \rexp + \frac{15}{4} \lexp w_a^3 \rexp - \frac{107}{8} \lexp w_a^4 \rexp \,. \nonumber \\ & & 
\label{E8Inf}
\eeqa
The halo averages $\lexp w_a^n \rexp$ are functions of the halo-shape parameter $q$, resulting in reasonably simple expressions for ${\cal E}_{2n}(q)$ as given in App.~\ref{app:halo}. We can introduce the standard normalized cumulants ($n>1$)
\beq
\bar{\cal E}_{2n} \equiv {{\cal E}_{2n} \over {\cal E}_2^n}\,,
\label{E2nbar}
\eeq
which correspond to the (dimensionless) kurtosis ($n=2$), etc, and characterize how non-Gaussian the distribution function is. Figure~\ref{fig:E4E6E8Evans} shows these as a function of $q$ for the whole range allowed by positivity constraints. We see that from the point of view of the kurtosis, non-Gaussianity is always weak ($|\bar{\cal E}_4| <1$) independent of halo shape  but the same is not true for higher cumulants in the case of oblate halos. Spherical halos are Gaussian as the distribution function becomes Maxwellian in this case ($w_a=0$). We discuss the implication of these results for the stability of the linear perturbative solutions in Sec.~\ref{subsec:stability}.

\subsection{NFW Halos}
\label{subsec:NFW}

We now assume halos with a spherical density profile, the distribution function therefore depends on energy and the square the angular momentum, $f=f(E,L^2)$, or  $f(\bm{r},\bm{p})=f(r,p,\bm{p}\cdot\bm{r})$. We are particularly interested in the simple case of constant anisotropy, where the distribution function takes a simple form
\beq
f(E,L^2)=L^{-2\beta} f_E(E)\,,
\label{fEL}
\eeq
where $-\infty\leq\beta\leq 1$ is a constant that characterizes deviations from isotropy ($\beta=0$). In particular, $\beta$ measures the anisotropy of the velocity dispersion tensor,
\beq
\beta = 1- \frac{\sigma_{\theta\theta}+\sigma_{\phi\phi}}{2 \sigma_{rr}}=1- \frac{\sigma_{\theta\theta}}{\sigma_{rr}}  \,,
\label{betadef}
\eeq
where the last equality uses spherical symmetry. Models with $\beta>0$ are said to be radially biased, while those with $\beta<0$ are tangentially biased~\cite{BinTre08}. If all orbits are circular, then $\sigma_{rr}=0$ and $\beta=-\infty$. If all orbits are radial $\sigma_{\theta\theta}=\sigma_{\phi\phi}=0$ and $\beta=1$.  

Our results in this section hold for general $\beta={\rm const}$ and {\em any} spherical density profile. However, when computing the expectation value of cumulants, we will take for simplicity the case of $\beta=0,1/2$ (as this brackets the radial dependence of $\beta$ in cosmological simulations~\cite{HanMoo0603,WojlokMam0808,SpaHan1210}) and an NFW density profile. More sophisticated models of the distribution function with radial dependent $\beta$ have been developed (e.g.~\cite{Osi7901,Mer8506,WojlokMam0808}); in our case, however, we are interested in expectation values of cumulants which are dominated by the outer parts of the halo where $\beta \approx 1/2$. 
The NFW density profile for a halo with mass $m$ is given by~\cite{NavFreWhi97}, 
\beq
\rho(r) = m\, u(r) =  {c^3f(c) \over 4\pi \, r_{\rm vir}^3} \ \frac{m}{x(1+x)^2}\,,
\label{NFWrho}
\eeq
where $x\equiv cr/r_{\rm vir}$, and $f^{-1}(c) \equiv \ln(1+c)-c/(1+c)$ guarantees that $\int d^3r\, u(r)=1$, where the integral is  over the volume enclosed by virial radius of the halo $r_{\rm vir}$. The parameter $c$ denotes the concentration of the halo, which determines the scale radius $a_s\equiv r_{\rm vir}/c$ where the slope of the density profile is $-2$. For an infinite halo, the corresponding potential that obeys the Poisson equation is,
\beq
\Phi(r) = - \frac{Gm}{r_{\rm vir}}\  c f(c)\ {\ln(1+x) \over x}\,. 
\label{PhiInf}
\eeq

\begin{figure*}[t!]
  \begin{center}
        \includegraphics[width=\columnwidth]{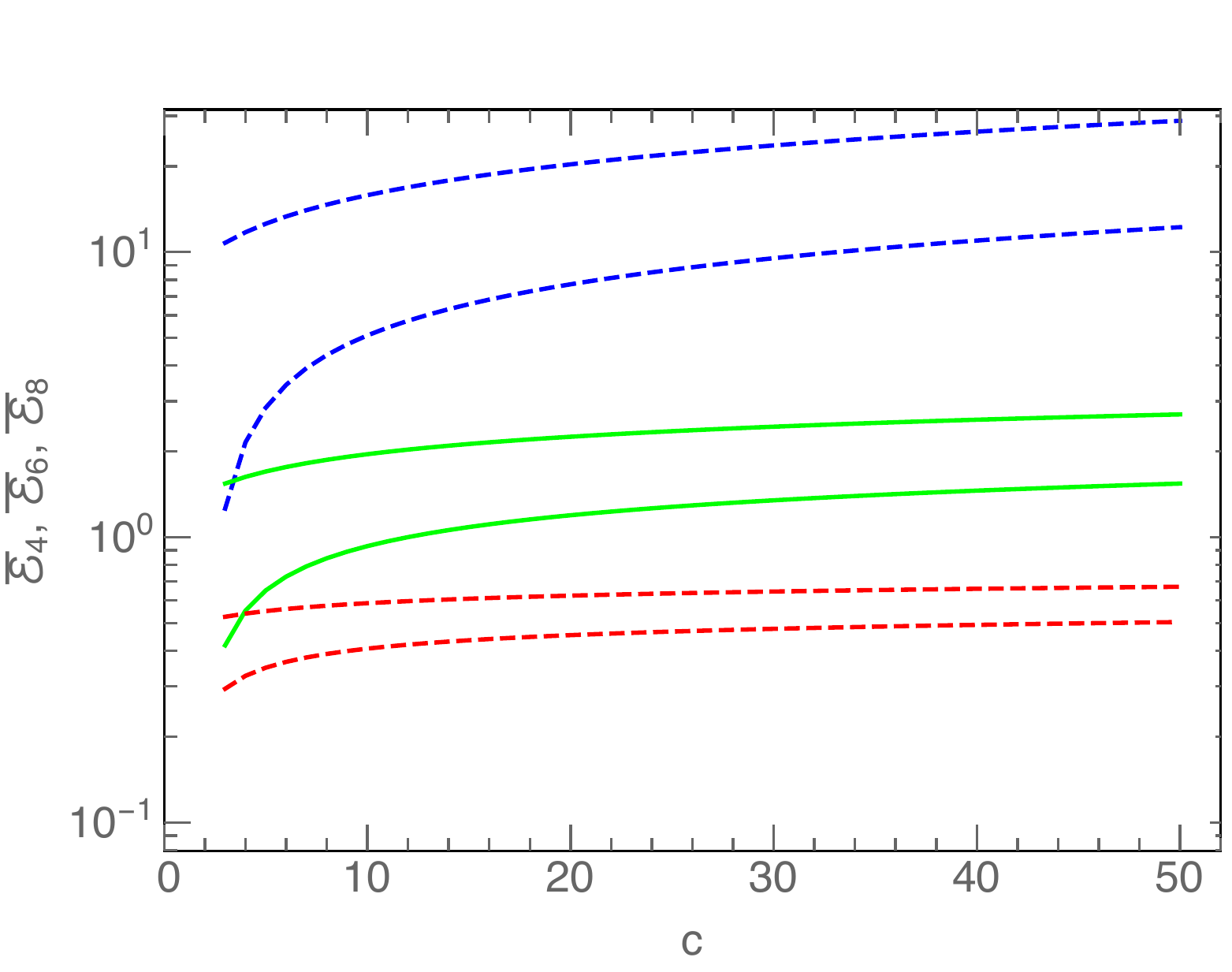}
        \includegraphics[width=\columnwidth]{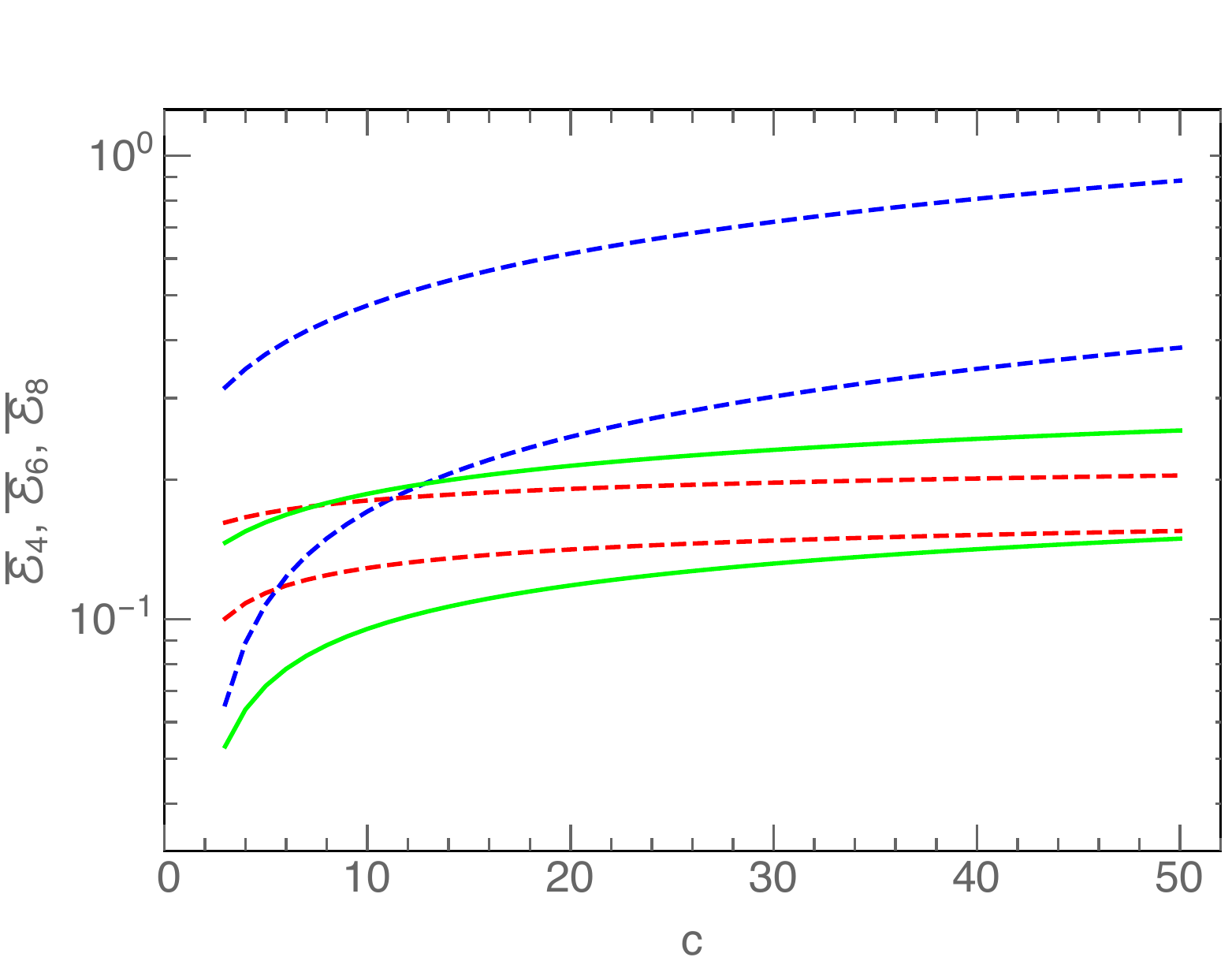}
  \end{center}
  \caption{\label{fig:E4E6E8NFW} Normalized cumulants $\bar{\cal E}_4, \bar{\cal E}_6, \bar{\cal E}_8$ (bottom to top) for NFW halos as a function of halo concentration parameter $c$. Dashed lines denote negative values. For each case we show results for two values of the anisotropy parameter, $\beta=0$ (lower value) and $\beta=1/2$ (higher value). The left panel corresponds to  integrating out to the virial radius, whereas the right panel corresponds to when we associate with the halo a region of 1.5 times the virial radius.}
\end{figure*}

The  steady Vlasov equation, Eq.~(\ref{SteadyVlasov}), for a distribution function of the form given by Eq.~(\ref{fEL}) can be written in the simple form:
\beq
p^2 \Big[ {\partial f\over \partial r} + {2\beta\over r} f \Big] = {d\Phi\over dr} \Big[ {\partial f\over \partial \ln p}+2\beta  f \Big]\,.
\label{VlasovSphAniso}
\eeq
Multiplying by $p^{2(n-1)}$ and integrating over all momenta, we obtain after some algebra a simple recursion relation for the even moments $m_{2n}$ we are interested in,
\beq
{d\over dr}\Big[\rho r^{2\beta} m_{2n}  \Big] = (2n-1)\Big({2\beta\over 2n+1}-1\Big){d\Phi\over dr} \, \rho r^{2\beta}m_{2(n-1)} \,,
\label{recrel}
\eeq
where, in correspondence to Eq.~(\ref{E2n}), we defined
\beq
m_{2n} \equiv { {\cal M}_{i_1i_1 \dots i_n i_n} \over {2n+1} }\,,
\label{m2nDef}
\eeq
for $n\geq 1$, with $m_0 \equiv 1$. Equation~(\ref{recrel}) can be integrated to give
\beqa
m_{2n}&=& {(2n-1)\over \rho r^{2\beta}}\Big(1-{2\beta\over 2n+1}\Big) \int_r^\infty {d\Phi\over dr}\rho r^{2\beta} m_{2(n-1)}  dr \,,\nonumber \\ & & 
\label{recrel2}
\eeqa
where we imposed the boundary condition that $m_{2n}\to 0$ at infinity. As it turns out, it is more realistic to do this rather than truncating the halo at the virial radius, as simulations show that non-truncated halo predictions for the velocity dispersion are more accurate than truncated ones~~\cite{ColLac9607,SheHuiDia01}.

Using Eq.~(\ref{recrel2}) we can construct all moments of the distribution function starting from any density-potential pair and any value of $\beta$. Integrating over the halo, one can then obtain their expectation values and thus the expectation values of the cumulants ${\cal E}_{2n}$. For $n=1$, this gives us the velocity dispersion, which we shall use in paper~II to compute the dispersion scale from halos by integrating this result over the halo mass function (see~Sec.~V.B. in~\cite{cumPT2}). In this paper, we are mostly interested in the (ensemble averaged) non-Gaussianity of $f$, which plays an important role in determining the stability of the linear solutions we shall discuss below. For this reason, we concentrate on the normalized cumulants.

Figure~\ref{fig:E4E6E8NFW} shows the results of such calculation for the normalized cumulants $\bar{\cal E}_{2n}$ as a function of the halo concentration $c$ for a broad range of values expected from low to high-mass halos in cosmological simulations ($c=3-50$). The left panel shows $\bar{\cal E}_{2n}$ for the case where the average is done up to the virial radius, with the upper (lower) limit for each cumulant corresponding to $\beta=1/2$ ($\beta=0$). We see that, similarly to Evans halos, the kurtosis is weak ($|\bar{\cal E}_4| <1$) but that higher normalized cumulants can be larger than unity in absolute value. Indeed, Eq.~(\ref{fEL}) suggests and 
App.~\ref{app:halo} shows explicitly that the shape of the distribution function significantly differs from a Maxwellian and thus in general there is no parameter that controls non-Gaussianity, unlike the case of Evans halos where the shape parameter $q$ plays that role. 

On the other hand, it is worth exploring to what extent this result is robust to reasonable changes. One obvious issue is that  velocity dispersion and higher cumulants do not sharply become zero outside halos, therefore there are more regions that contribute to cumulants than those captured by our calculation so far. This motivates extending the region of integration beyond the virial radius when calculating expectation values. In addition, it has been long known from simulations that at least for the second cumulant, predictions from NFW halos remain reasonable far outside the virial radius~\cite{ColLac9607}. Thus, the right panel in Fig.~\ref{fig:E4E6E8NFW} shows the normalized cumulants in the case we associate with a given halo a region of 1.5 times the virial radius. Compared to the left panel, we see a suppression of the $\bar{\cal E}_{2n}$, which remain below unity in absolute value for all concentrations. This suppression is due to the increase in contributions to each cumulant (roughly proportional to the volume), which means that {\em normalized cumulants} will get suppressed increasingly with $n$ due to normalization by increasing powers of ${\cal E}_2$. Note that we still normalize the volume average by the halo volume (up to the virial radius) as we are associating to the halo a larger region where dispersion and higher cumulants are non-zero rather simply redefining the halo size. We shall see in Sec.~\ref{subsec:stability} that in the space of the $\bar{\cal E}_{2n}$'s 
 such modifications for NFW halos give rise to a sequence that mimics varying concentration as normalized cumulants are driven to zero (see Fig.~\ref{fig:cum4vs6} below).

\section{Perturbation theory with higher cumulants}
\label{sec:eqs}

We now proceed with a systematic development of the perturbative approach to the Vlasov equation. In contrast to the previous section we switch to an expanding Friedmann-Lema\^itre-Robertson-Walker background and comoving coordinates from now on. For scales smaller than the Hubble radius, relevant for non-linear large-scale structure formation, the non-relativistic limit suffices, which reads
\be\label{eq:Vlasov}
  0=\frac{\partial f}{\partial\tau}+\frac{p_i}{a}\frac{\partial f}{\partial x_i}-a(\nabla_i\Phi)\frac{\partial f}{\partial p_i}\,,
\ee
with conformal time $\tau$, scale-factor $a$, comoving momentum $\bm{p}$ (per unit particle mass), comoving coordinates $\bm{x}$, and gravitational potential fluctuation $\Phi$ obeying the Poisson equation
\be
  \nabla^2\Phi=\frac{3}{2}{\cal H}^2\Omega_m\delta \,,
\ee
where ${\cal H}$ is the conformal Hubble rate, $\Omega_m$ the time-dependent matter density parameter, and $\delta$ the density contrast given by the zeroth moment of the distribution function,
\be
  1+\delta = \int d^3p\, f(\tau,\bm{x},\bm{p}) \,.
\ee
Taking the zeroth and first moment of the Vlasov equation yields the coupled continuity and Euler equations
\bea\label{eq:fluid}
  \partial_\tau\delta+\nabla_i[(1+\delta) v_i] &=& 0 \,, \\
  \partial_\tau v_i +{\cal H}v_i+v_j\nabla_j v_i + \nabla_i\Phi 
  &=& -\nabla_j \sigma_{ij} -\sigma_{ij}\nabla_j\ln(1+\delta) \,, \nn
\eea
for the density contrast and peculiar velocity field
\be
  v_i = \frac{1}{1+\delta}\int d^3p\, \frac{p_i}{a} \, f(\tau,\bm{x},\bm{p}) \,.
\ee
The widely used framework of standard perturbation theory (SPT) is based on a perturbative solution of these equations obtained when neglecting the right-hand side of the Euler equation,
that contains the velocity dispersion tensor
\be
  \sigma_{ij} = \frac{1}{1+\delta}\int d^3p\, \frac{p_i}{a}\frac{p_j}{a} \, f(\tau,\bm{x},\bm{p}) - v_iv_j \,.
\ee
However, even for initially (almost) vanishing velocity dispersion, as appropriate for cold dark matter, it is well-known that velocity dispersion is generated in the process of
non-linear structure formation via orbit crossing~\cite{PueSco0908}.

In this work, we develop the extension of SPT that includes velocity dispersion and higher cumulants of the distribution function.
The equation of motion for $\sigma_{ij}$ can be obtained by taking the second moment of the Vlasov equation, and reads
\bea\label{eq:eomsigmaij}
  \partial_\tau \sigma_{ij} &+& 2{\cal H}\sigma_{ij}+v_k\nabla_k\sigma_{ij}+\sigma_{jk}\nabla_k v_i+\sigma_{ik}\nabla_k v_j   \nn\\
 &=& - \nabla_k{\cal C}_{ijk}  - {\cal C}_{ijk}\nabla_k\ln(1+\delta) \,.
\eea
It depends on the third cumulant ${\cal C}_{ijl}$, that in turn depends on the fourth cumulant, and so on, leading to an infinite set
of coupled equations, reminiscent of the BBGKY hierarchy in kinetic theory.
A solution can only be obtained by a suitable truncation, and we explore the impact of higher cumulants in our approach.

The density, velocity, dispersion and all higher cumulants can be obtained from the generating function for cumulants of the distribution function,
\be\label{eq:genfunc}
 e^{ {\cal C}(\tau,\bm{x},\bm{l}) } = \int d^3p\, e^{ \bm{l}\cdot\bm{p}/a }\, f(\tau,\bm{x},\bm{p})\,.
\ee
This is the analog of Eq.~(\ref{CGF}) for the more appropriate choice of coordinates to discuss time-dependent structure formation. 
As discussed earlier in the halo case, the cumulants of the distribution function are obtained by taking derivatives with respect to the auxiliary vector $\bm{l}$,
\be
  {\cal C}_{ijk\cdots}(\tau,\bm{x}) = \nabla_{l_i}\nabla_{l_j}\nabla_{l_k}\cdots  {\cal C}|_{\bm{l}=0} \,,
\ee
in particular
\be
  {\cal C}|_{\bm{l}=0}=\ln(1+\delta),\
  {\cal C}_i = v_i,\
  {\cal C}_{ij} = \sigma_{ij}\,.
\ee
The Vlasov equation yields an equation for the generating function given by~\cite{PueSco0908}
\be\label{eq:genfunceom}
 \partial_\tau{\cal C}+{\cal H}(\bm{l}\cdot\nabla_l){\cal C}+(\nabla{\cal C})\cdot(\nabla_l{\cal C})+(\nabla\cdot\nabla_l){\cal C}=-\bm{l}\cdot\nabla\Phi\,,
\ee
from which the hierarchy of equations for the cumulants can be obtained by taking derivatives with respect to $\bm{l}$ and setting $\bm{l}=0$,
yielding the continuity and Euler equations Eq.~\eqref{eq:fluid} as well as Eq.~\eqref{eq:eomsigmaij} for the zeroth, first and second derivative, respectively.

The  pressureless perfect fluid approximation, on which SPT is based, corresponds to an ansatz for ${\cal C}$ containing only constant and linear terms in $\bm{l}$ in the Taylor expansion of the cumulant generating function, see Eq.~(\ref{CGF}),
\be
  {\cal C}_{\text{SPT}} = \ln(1+\delta) + \bm{l}\cdot\bm{v}\,.
\ee
Remarkably, this ansatz is preserved under time-evolution by Eq.~(\ref{eq:genfunceom}), which does not generate higher powers of $\bm{l}$, 
i.e. velocity dispersion and all higher cumulants remain exactly zero at all times for which the solution exists. 
However, a careful analysis of the Vlasov equation solution shows that this ansatz breaks down once orbit crossing occurs, and this solution formally ceases to exist due to a singularity in the density contrast. In reality, this singularity is regulated by an arbitrarily small initial velocity dispersion, which then from Eq.~(\ref{eq:genfunceom}) generates all cumulants~\cite{PueSco0908}. After orbit crossing, the superposition of orbits at any point leads to the generation of a sizable velocity dispersion, as well as higher cumulants which in turn generates vorticity, the curl of the peculiar velocity field. This is one key observable that gives us a unique window into orbit crossing~\cite{PicBer9903}.

In the next section we focus on the inclusion of velocity dispersion, and then extend the formalism to include also higher cumulants in Sections~\ref{sec:pi} and\,\ref{sec:hierarchy}.

\section{Second cumulant}
\label{sec:sigma}

\subsection{Background value and perturbations of the velocity dispersion tensor}

It is convenient to work with the normalized quantities
\bea\label{eq:uepspi}
  u_i &=& \frac{v_i}{-{\cal H}f}\,,\nn\\
  \epsilon_{ij} &=& \frac{\sigma_{ij}}{({\cal H}f)^2}\,,\\
  \pi_{ijk} &=& \frac{{\cal C}_{ijk}}{-({\cal H}f)^3}\,,\nn
\eea
in terms of which the equations for the zeroth, first and second cumulants read
\bea\label{eq:fluidnorm}
  \label{eq:continuitynorm}
  \delta' &=& \theta  + \nabla_i[\delta \, u_i] \,,\\
  \label{eq:Eulernorm}
  u_i'+\left(\frac32\frac{\Omega_m}{f^2}-1\right)u_i &=& \nabla_i\hat\Phi + u_j\nabla_j u_i \nn\\
  &+& \nabla_j \epsilon_{ij} +\epsilon_{ij}\nabla_j\ln(1+\delta) \,, \\
  \label{eq:dispersionnorm}
  \epsilon_{ij}'+2\left(\frac32\frac{\Omega_m}{f^2}-1\right)\epsilon_{ij} &=& u_l\nabla_l\epsilon_{ij}+\epsilon_{jl}\nabla_l u_i+\epsilon_{il}\nabla_l u_j\nn\\
  && {} + \nabla_l \pi_{ijl} + \pi_{ijl}\nabla_l\ln(1+\delta)\,, \nn\\
\eea
where $'=d/d\eta$, $\eta=\ln D$, $f=d\ln D /d\ln a $, $D$ is the usual linear growth factor and 
$\hat\Phi = \Phi/({\cal H}f)^2$ the rescaled gravitational potential satisfying $\nabla^2\hat\Phi=\frac32\frac{\Omega_m}{f^2}\delta$.

Mass conservation as well as statistical isotropy guarantee that the average values of the density contrast and peculiar velocity fields are exactly zero. 
However, in general, higher cumulants are expected to possess a non-zero average value, that can depend only on time (not space) due to statistical homogeneity, in analogy to e.g. the square of the density contrast or the density of a population of biased tracers.
Furthermore, the expectation value has to be compatible with isotropy, corresponding to rotationally invariant objects. The first example occurs at the level of the second cumulant,
\be\label{eq:epsilon}
  \langle \epsilon_{ij}(\eta,\bm{x})\rangle = \epsilon(\eta)\, \delta_{ij}\,,
\ee
with time-dependent, homogeneous expectation value $\epsilon(\eta)$ proportional to the $3\times 3$ unit matrix $\delta_{ij}$, which corresponds to the background value of the velocity dispersion.
The equation of motion for $\epsilon(\eta)$ can be obtained by taking the trace as well as the statistical ensemble average
of Eq.~\eqref{eq:dispersionnorm}, giving
\be\label{eq:epseom}
  \epsilon' +2\left(\frac32\frac{\Omega_m}{f^2}-1\right)\epsilon = Q(\eta)\,,
\ee
with source term
\be\label{eq:Qdef}
  Q(\eta) \equiv \frac{1}{3}\langle u_l\nabla_l \epsilon_{ii}\rangle + \frac{2}{3}\langle \epsilon_{il}\nabla_l u_i \rangle 
  + \frac13\langle \pi_{iil}\nabla_l\ln(1+\delta)\rangle\,,
\ee
using $\langle \pi_{ijk}\rangle=0$ due to isotropy (see also~\cite{ErsFlo1906,Erschfeld:2021kem} for the case without third cumulant). Therefore, velocity dispersion is sourced by the cross power spectrum of
peculiar velocity and the perturbations of $\epsilon_{ij}$, as well as a cross spectrum between the third cumulant and the logarithm of the density field perturbations. It is important to note that what enters into $Q$ are these various spectra \emph{integrated over all momenta} (see Eq.~\ref{eq:Q2nd} below). This means that there is no sense in which $\epsilon$ can be taken as a small quantity in general, on equal footing with density or velocity fluctuations. This should also be clear from the halo discussion in the previous section. 

Our strategy for solving this system perturbatively is then as follows: we first split all quantities in a background value and perturbations, in particular
\be\label{eq:epsdecomp}
  \epsilon_{ij}(\eta,\bm{x}) = \epsilon(\eta)\,\delta_{ij} + \delta\epsilon_{ij}(\eta,\bm{x})\,,
\ee
for the velocity dispersion tensor. Then we define a zeroth order (linear) solution to these equations by keeping all terms linear in \emph{perturbations}, while
formally treating the background quantities (viz. $\epsilon(\eta)$) as quantities similar to the Hubble rate or $\Omega_m(\eta)$. That is, we are explicitly agnostic about the size of velocity dispersion effects in our treatment. This allows us, as we shall see, to  obtain the expected decoupling of UV modes due to halo formation at small scales discussed in the introduction. 

Expanding around these solutions and including non-linear terms in the perturbation variables,
one obtains a perturbative solution to the coupled equations of motion for the fluctuations of the cumulants. Finally, these solutions may be used to evaluate the source term (viz. $Q(\eta)$). A self-consistent solution within perturbation theory
then requires that $\epsilon(\eta)$ is chosen such that it satisfies Eq.~\eqref{eq:epseom}. We come back to this final step in Sec.\,\ref{sec:powerlaw}. Alternatively, $\epsilon(\eta)$ may also be given as an external input, for example from simulation measurements or theoretical input, such as the halo model discussed in Sec.\,\ref{sec:halo} (see also paper II~\cite{cumPT2}).

\subsection{SVT decomposition and equations of motion}

We now proceed to derive equations for $\delta\epsilon_{ij}$ as well as $\delta$ and $u_i$, while treating $\epsilon(\eta)$ as given.
We start with the velocity divergence $\theta=\nabla_iu_i$. Taking the divergence of Eq.~\eqref{eq:Eulernorm}, and inserting
the decomposition~\eqref{eq:epsdecomp} in the second line yields
\be\label{eq:euler2}
  \theta'+\left(\frac32\frac{\Omega_m}{f^2}-1\right)\theta -  \frac32\frac{\Omega_m}{f^2}\delta
   = \nabla_i(u_j\nabla_j u_i) +  q_\theta\,,
\ee
with an extra term compared to SPT given by
\be\label{eq:qtheta}
  q_\theta = \epsilon(\eta) \nabla^2 A+ \nabla_i\nabla_j \delta\epsilon_{ij} +\nabla_i(\delta\epsilon_{ij}\nabla_j A) \,,
\ee
where we introduced a short-hand notation for the log-density field,
\be
  A \equiv \ln(1+\delta)\,.
\ee
For given $\epsilon(\eta)$, the first two terms on the right-hand side of Eq.~\eqref{eq:qtheta} can be viewed as contributing to the zeroth order (``linear'') solution in presence of velocity dispersion, while
the last term yields an additional non-linear term. Further non-linearities are generated when Taylor expanding $A$, and we systematically take those terms into account by a method discussed below.

Due to the presence of velocity dispersion terms in Eq.~\eqref{eq:Eulernorm}, also a non-zero rotational component of the peculiar velocity field, i.e. vorticity
\be
  w_i = \varepsilon_{ijk}\nabla_ju_k=(\bm{\nabla}\times \bm{u})_i\,,
\ee
is generated~\cite{PueSco0908}, unlike the case of SPT where the Euler version of Eq.~\eqref{eq:Eulernorm} preserves vanishing vorticity. Indeed, taking the curl of Eq.~\eqref{eq:Eulernorm} yields an evolution equation for the vorticity,
\be\label{eq:euler2w}
  w_i'+\left(\frac32\frac{\Omega_m}{f^2}-1\right)w_i  
   = (\bm{\nabla} \times (u_j\nabla_j\bm{ u}))_i +  (q_w)_i\,,
\ee
with
\be\label{eq:qw}
  (q_w)_i = \varepsilon_{ijk}\nabla_j (\nabla_l \delta\epsilon_{kl} +\delta\epsilon_{kl}\nabla_l A )\,,
\ee
where $\varepsilon_{ijk}$ is the Levi-Civita symbol.
We can formally write the decomposition of the peculiar velocity into scalar (divergence) and vector (vorticity) contributions as (denoted as usual by $S$ and $V$)
\be\label{eq:uSV}
  u_i = u_i^S + u_i^V = \frac{\nabla_i}{\nabla^2}\theta-\frac{\varepsilon_{ijk}\nabla_j }{\nabla^2}w_k\,,
\ee
where $\nabla_iw_i=0$. Note that operationally the inverse Laplacian is easily written in Fourier space as $-1/k^2$ acting on fields at wavenumber $\bm{k}$.  
Analogously, we decompose the perturbations of the velocity dispersion tensor into scalar, vector and tensor modes (denoted as $S$, $V$ and $T$),
\bea\label{eq:epsSVT}
  \delta\epsilon_{ij} &=& \delta\epsilon_{ij}^S + \delta\epsilon_{ij}^V + \delta\epsilon_{ij}^T\,,
\eea
with
\bea
  \delta\epsilon_{ij}^S &=& \delta_{ij}\,\delta\epsilon+\frac{\nabla_i\nabla_j}{\nabla^2}g \,,\\
  \delta\epsilon_{ij}^V &=& -\frac{\varepsilon_{ilk}\nabla_l \nabla_j}{\nabla^2}\nu_k 
    -\frac{\varepsilon_{jlk}\nabla_l \nabla_i}{\nabla^2}\nu_k \,,\\
 \delta\epsilon_{ij}^T &=& t_{ij}\equiv P_{ij,ls}^T\delta\epsilon_{ls}\,,
\eea
with $\nabla_i\nu_i=0$, $t_{ii}=0$, $\nabla_i t_{ij}=\nabla_j t_{ij}=0$, and the tensor projection operator
\bea\label{eq:Ptensor}
  P^T_{ij,ls} &=& \frac12\left(\delta_{is}-\frac{\nabla_i\nabla_s}{\nabla^2}\right) \left(\delta_{jl}-\frac{\nabla_j\nabla_l}{\nabla^2}\right)\nn\\
  &&{} +\frac12\left(\delta_{js}-\frac{\nabla_j\nabla_s}{\nabla^2}\right) \left(\delta_{il}-\frac{\nabla_i\nabla_l}{\nabla^2}\right)\nn\\
  &&{} -\frac12\left(\delta_{ij}-\frac{\nabla_i\nabla_j}{\nabla^2}\right) \left(\delta_{ls}-\frac{\nabla_l\nabla_s}{\nabla^2}\right)\,.
\eea
The perturbations of the velocity dispersion tensor are therefore fully characterized by
\be
\delta\epsilon,\,g,\,\nu_i,\,t_{ij}\,.
\ee
They describe two scalar modes encoded by $\delta\epsilon$ and $g$ (2~dof), a divergence-free vector $\nu_i$ (2 dof), and a traceless-transverse tensor $t_{ij}$ (2 dof), comprising all six degrees of freedom of the symmetric velocity dispersion tensor. 
We choose the notation $\delta\epsilon$ to discriminate the scalar perturbation mode of the velocity dispersion tensor that is
proportional to the unit matrix $\delta_{ij}$ from the homogeneous background value $\epsilon(\eta)$.
Since the expectation value of $\epsilon_{ij}$ is proportional to the unit matrix, the other contributions, and in particular $g$, are guaranteed to
vanish in the ensemble average.

For the remainder of this section, we neglect $\pi_{ijk}$ and higher cumulants, coming back to them in Sec.\,\ref{sec:pi} and\,\ref{sec:hierarchy}.
Inserting the decomposition of Eq.~\eqref{eq:epsSVT} into the source term Eq.~\eqref{eq:Qdef} for $\epsilon(\eta)$ then yields (see also~\cite{ErsFlo1906})
\be\label{eq:Q2nd}
  Q(\eta) = \frac{1}{3}\int d^3k \left( P_{\theta \tilde g}(k,\eta) + 2 P_{w_i\nu_i}(k,\eta)\,\right)\,,
\ee
where $\tilde g \equiv g - \delta \epsilon$, and $P_{w_i\nu_i}(k,\eta)$ is the cross power spectrum of
vorticity and the vector perturbation, summed over $i=1,2,3$.
To obtain equations of motion for $g$ and $\delta\epsilon$ we subtract Eq.~\eqref{eq:epseom} (multiplied by the unit matrix) from Eq.~\eqref{eq:dispersionnorm}
and contract it with $\delta_{ij}$ or $\nabla_i\nabla_j/\nabla^2$, respectively. 
Taking suitable linear combinations of the resulting two equations yields
\bea\label{eq:delepsS}
  \delta\epsilon'+2\left(\frac32\frac{\Omega_m}{f^2}-1\right)\delta\epsilon  &=& 
  q_\epsilon\,, \nn\\
  g'+2\left(\frac32\frac{\Omega_m}{f^2}-1\right)g  -2\epsilon\theta &=& 
  q_g\,,
\eea
where
\bea
  q_\epsilon &=& \frac12 u_l\nabla_l(3\delta\epsilon+g) +\delta\epsilon_{il}\nabla_l u_i 
  -\frac12\frac{\nabla_i\nabla_j}{\nabla^2}(u_l\nabla_l\delta\epsilon_{ij})\nn\\
  && {} - \frac{\nabla_i\nabla_j}{\nabla^2}(\delta\epsilon_{il}\nabla_l u_j)-Q(\eta)\,,\nn\\
  q_g &=& - \frac12 u_l\nabla_l(3\delta\epsilon+g) -\delta\epsilon_{il}\nabla_l u_i
   + \frac32\frac{\nabla_i\nabla_j}{\nabla^2}(u_l\nabla_l\delta\epsilon_{ij})\nn\\
   && {} + 3\frac{\nabla_i\nabla_j}{\nabla^2}(\delta\epsilon_{il}\nabla_l u_j)\,.
\eea
The term $2\epsilon\theta$ in the equation for $g$ implies that this mode is generated
in presence of a background dispersion $\epsilon(\eta)$, even when neglecting non-linear
terms in perturbations, as opposed to $\delta\epsilon$ that is a decaying mode in linear theory (when $q_\epsilon=0$ and assuming $\Omega_m/f^2>2/3$). In turn, the $g$ mode leads to a non-zero source term
$Q(\eta)$ through Eq.~\eqref{eq:Q2nd}, indicating that a self-consistent solution can exist even within the linear 
approximation. Beyond the linear level, the terms in $q_\epsilon$ and $q_g$ give further
contributions, as well as the vorticity and vector modes. 
Note that the source term $Q(\eta)$ contributes to $q_\epsilon$. It arises from subtracting the
background to obtain an equation for the perturbations $\delta\epsilon_{ij}=\epsilon_{ij}-\epsilon(\eta)\,\delta_{ij}$.
Therefore, this term ensures that
$\delta\epsilon$ maintains a vanishing average value (enforcing $\langle q_\epsilon\rangle=0$), as appropriate for a perturbation variable.
Technically, it removes so-called tadpole contributions, see Appendix~A in paper II~\cite{cumPT2} for more details on this.

To obtain an equation for the vector mode $\nu_i$, we use 
\be
  \nu_i = \varepsilon_{ijk}\frac{\nabla_j\nabla_l}{\nabla^2}\delta\epsilon_{kl} \,,
\ee
and contract Eq.~\eqref{eq:dispersionnorm} with $\varepsilon_{ijk}\nabla_j\nabla_l/\nabla^2$ (after renaming $ij\to kl$ in Eq.~\ref{eq:dispersionnorm}).
Contracting instead with Eq.~\eqref{eq:Ptensor} in addition yields an equation for the tensor mode $t_{ij}$,
\bea\label{eq:nu}
  \nu_i'+2\left(\frac32\frac{\Omega_m}{f^2}-1\right)\nu_i -\epsilon\, w_i 
  &=& \varepsilon_{ijk}\frac{\nabla_j\nabla_l}{\nabla^2}q_{kl}\,,\nn\\
  {t_{ij}}'+2\left(\frac32\frac{\Omega_m}{f^2}-1\right)t_{ij} 
  &=& P_{ij,kl}^T\ q_{kl}\,,
\eea 
where non-linear terms in perturbations are contained in
\be
  q_{kl}\equiv u_n\nabla_n\delta\epsilon_{kl}+\delta\epsilon_{ln}\nabla_n u_k+\delta\epsilon_{kn}\nabla_n u_l\,.
\ee
Note that the term $\epsilon w_i$ involving the vorticity field leads to a mixing of the vector and vorticity modes when
solving the equations of motion.

In total, up to the second cumulant, we therefore obtain the following perturbation modes:
\be
  \begin{array}{lll@\qquad l@\qquad l}
   \delta &\theta &\delta\epsilon,g & \text{scalar} & 4\times 1\ \text{dof}\,,\nn\\
          & w_i   & \nu_i & \text{vector} & 2\times 2\ \text{dof}\,,\nn\\
          &       & t_{ij} & \text{tensor} & 1\times 2\ \text{dof}\,.
  \end{array}
\ee

For practical reasons, as discussed in Sec.~\ref{logdelta} below, we also include an extra scalar representing the log-density field, but of course this is not an independent degree of freedom. 
In Fourier space the equations of motion can be written in a way that resembles SPT.
In particular, inserting the velocity decomposition in Eq.~\eqref{eq:uSV} into the continuity equation Eq.~\eqref{eq:continuitynorm} yields
\be\label{eq:eom2}
  \delta_k'-\theta_k =  \int_{pq} \left\{ \alpha_{pq}\theta_p\delta_q  + \frac{(\bm{p}\times \bm{q})\cdot \bm{w}_p}{p^2}\delta_q\right\} \, ,
\ee
where $\alpha_{pq}=(\bm{p}+\bm{q})\cdot\bm{p}/p^2$ is the standard  expression for the only non-linearity in the continuity equation in SPT. 
The second term on the right-hand side of Eq.~\eqref{eq:eom2} describes the backreaction of vorticity
on the density contrast, i.e. at the non-linear level the scalar $\delta$ is coupled to the  vector mode of the velocity. 
In Eq.~\eqref{eq:eom2} subscripts of the perturbation variables denote the corresponding Fourier wavevector, and
we use the shorthand notation
\be
  \int_{pq}=\int\, d^3p\, d^3q\, \delta^{(3)}(\bm{k}-\bm{p}-\bm{q})\,.
\ee

The  equation of motion Eq.~\eqref{eq:euler2} for the velocity divergence can be rewritten analogously, using Eq.~\eqref{eq:uSV} as well as the decomposition Eq.~\eqref{eq:epsSVT} of the velocity dispersion tensor $\delta\epsilon_{ij}$,
\bea\label{eq:euler3}
 \lefteqn{ \theta_k'+\left(\frac32\frac{\Omega_m}{f^2}-1\right)\theta_k -  \frac32 \frac{\Omega_m}{f^2}\delta_k + k^2\left(\delta\epsilon_k+  g_k +  \epsilon A_k\right)  }\nn\\
  &=& {}  \int_{pq}  \Bigg\{ \beta_{pq}\theta_p\theta_q 
   {} + \left(1+\frac{2p\cdot q}{q^2}\right) \frac{(\bm{p}\times \bm{q})\cdot \bm{w}_p}{p^2} \theta_q \nn\\
  && \qquad\qquad {} - \frac{(\bm{p}\times \bm{q})\cdot \bm{w}_p}{p^2} \frac{(\bm{p}\times \bm{q})\cdot \bm{w}_q}{q^2}\nn\\
  && {} -(\bm{p}+\bm{q})\cdot \bm{p} A_p \delta\epsilon_q 
   {} -(\bm{p}+\bm{q})\cdot \bm{q}\ \frac{\bm{q}\cdot \bm{p}}{q^2} A_p  g_q \nn\\
  && {} + \left(1+\frac{2p\cdot q}{q^2}\right) A_p (\bm{p}\times \bm{q})\cdot \bm{\nu}_q 
   - A_p p_i p_j t_{q,ij}\Bigg\}\, , \nn\\
\eea
where $\beta_{pq}=(\bm{p}+\bm{q})^2\bm{p}\cdot\bm{q}/(2p^2q^2)$ is the standard  expression for the only non-linearity in the Euler equation in SPT. Note that 
already at the linear level, the terms proportional to $k^2$ in the first line describe a ``Jeans-like'' 
suppression arising from a non-zero velocity dispersion at the
perturbation ($\delta\epsilon_k+  g_k$) as well as background ($\epsilon A_k$) level. In particular, the last term is formally analogous to a pressure or sound speed  contribution, which can be seen if one would expand $A_k=[\ln(1+\delta)]_k$ linearly. Nevertheless, the interpretation in terms of sound speed would be conceptually misleading, since we do not consider any microscopic interactions apart from gravity. The Jeans suppression in a fluid arises because pressure due to collisions resists gravitational collapse at small scales; in our case we have instead a collisionless system. The suppression is the damping of small-scale fluctuations because particles can cross each other without interacting (i.e. ``shell-cross"), thus the physical situation is in sharp contrast with that of a fluid despite the net effect being similar.  

Another noteworthy feature of this damping is  that it depends on the perturbation modes of the velocity dispersion tensor, and cannot be associated with the isotropic part of the  dispersion tensor alone as the anisotropic part $g_k$ contributes as well at the same order in perturbation theory. That background and perturbation modes of the dispersion tensor contribute to the same order  arises because the {\em stress tensor} contribution (whose divergence enters in momentum conservation) is given by $(1+\delta) \epsilon_{ij}$. An additional point of contact with fluids that is worth mentioning here is that the term from the anisotropic part $g_k$ looks superficially similar to viscosity in the Navier-Stokes equation, given that in linear theory $g \sim \epsilon\, \theta$ (see Eq.~\ref{eq:delepsS} and in particular Sec.~\ref{sec:linear} for a more detailed discussion). Again, this identification is misleading as there is no dissipation in the Vlasov equation. This is also made explicit by the form of the energy conservation equation, which contains no viscosity type contributions (see Sec.~\ref{sec:linear}). 

Given all these subtleties it is worth asking whether damping at small scales is always guaranteed. This is important because this damping will describe precisely what we termed earlier as the ``decoupling" of high-$k$ modes induced by small-scale orbit crossing, and has a direct impact on the convergence of \vpt~when considering loop corrections  (see paper II~\cite{cumPT2}).  In fact, the details of the high-$k$ linear response depend on the expectation value of the full distribution function and as we shall see when going beyond the second cumulant, it is in principle possible to have  instabilities, i.e. small-scale enhancement rather than suppression, which stands in sharp contrast with the response in normal fluids that are always stable at scales below the Jeans length. See Sec.~\ref{subsec:stability} for the discussion of stability conditions along these lines.

Finally, we note that the velocity divergence evolution in Eq.~\eqref{eq:euler3} is affected by additional non-linear terms as compared to the SPT contribution
$\beta_{pq}$. In particular, the second and third line describe vorticity backreaction on the divergence field, while the fourth and
fifth line contain the non-linear terms involving the scalar as well as vector and tensor perturbations of $\delta\epsilon_{ij}$, respectively. This makes clear that, as expected, at the non-linear level the scalar, vector and tensor modes are coupled to each other. 

The equations of motion for the remaining fields $\delta\epsilon, g, w_i,\nu_i$ and $t_{ij}$ can be written in a similar way in Fourier space. Since they become rather lengthy they are presented in App.~\ref{app:eom}.

\subsection{Treatment of $A=\ln(1+\delta)$}
\label{logdelta}

The  equation of motion for the velocity divergence and the vorticity involves the log-density field $A=\ln(1+\delta)$. In a perturbative solution, expanding the
logarithm would generate an infinite series of non-linear terms, that are inconvenient to treat. Another strategy could be to use $A$ instead of $\delta$ as a perturbation variable~\cite{SzaKai0301}. 
However, in that case one would have to express the density contrast as $\delta=e^A-1$ within the Poisson term on the left-hand side of Eq.~\eqref{eq:euler3}, which again entails an infinite tower
of non-linear terms when solving perturbatively in powers of $A$. In addition, this choice would be inconvenient for computing the matter density power spectrum $P_{\delta\delta}$, although dealing with $A$ as an observable rather than $\delta$ has some interesting statistical advantages (e.g. see~\cite{NeySzaSza0906,Wang:2011fj,CarSza1310,Rubira:2020inb}). 

Therefore, we follow a different approach here. We keep $\delta$ as independent variable, such that the Poisson term in Eq.~\eqref{eq:euler3} can be easily evaluated, and allowing
us to compute $P_{\delta\delta}$ straightforwardly. In addition, we complement the set of variables by $A$, and solve the equation of motion for $A$ along with all other modes.
It can be obtained by dividing Eq.~\eqref{eq:fluidnorm} by $1+\delta$, and is similar in form to Eq.~\eqref{eq:eom2}, except for the coefficient of the first non-linear term,
\be\label{eq:eomA}
  A_k'-\theta_k =  \int_{pq} \left\{ \frac{\vec q\cdot\vec p}{p^2}\theta_pA_q  + \frac{(\bm{p}\times \bm{q})\cdot \bm{w}_p}{p^2}A_q\right\} \,.
\ee
In a perturbative solution this generates all contributions to $A$ evaluated at a given order in perturbation theory. The latter can in turn be used to compute the contributions on
the right-hand side of Eq.~\eqref{eq:euler3} that involve $A$. In practice, this means we have to solve for five instead of four scalar modes, being $\delta,\theta,\delta\epsilon,g,A$, each of them with an equation of motion that involves at most quadratic terms in the full set of perturbation variables. 

In contrast to $\delta$, $A$ possesses a non-zero average value, 
\be\label{eq:A}
  {\cal A}\equiv\langle A\rangle\,.
\ee 
Its equation can be derived by taking the ensemble average of Eq.~\eqref{eq:fluidnorm} divided by $1+\delta$,
\be\label{eq:QA}
  {\cal A}' = Q_A(\eta),\quad Q_A(\eta)\equiv -\int d^3 k P_{\theta A}(k,\eta)\,.
\ee
Since only spatial derivatives of $A$ enter in Eq.~\eqref{eq:euler3} (this can also be seen from Eq.~\ref{eq:Eulernorm}), the homogeneous part ${\cal A}$
drops out in the Euler equation, and is not needed. 
In practice, this means we can use Eq.~\eqref{eq:eomA} and ignore the difference between $\delta A=A-{\cal A}$ and $A$ as long as we use the log-density field only as an input for solving
the equations of motion of the other perturbation variables. For a more detailed argument, we refer to Appendix~A in paper II~\cite{cumPT2}.

\subsection{Equations of motion in matrix form}

It is convenient to write the equations of motion in the familiar matrix form, by defining a vector of perturbation variables
\be
  \psi \equiv (\delta,\theta,g,\delta\epsilon,A,w_i,\nu_i,t_{ij})\,.
\ee
The equations of motion can then be brought into the standard form~\cite{Sco01}
\be\label{eq:eom}
  \psi_{k,a}'(\eta)+\Omega_{ab}(k,\eta)\,\psi_{k,b}(\eta) = \int_{pq} \gamma_{abc}(\vec p,\vec q)\psi_{p,b}(\eta)\psi_{q,c}(\eta)\,,
\ee
where the subscript labels the wavevector as well as the component of $\psi$.
Here the index $a$ is understood to run over all types of perturbations as well as their components, in case of vector and tensor modes.
Summation over repeated indices is implied. Non-linear terms are described by the coupling functions $\gamma_{abc}(\vec p,\vec q)$, that we refer to as \emph{vertices}. 

The linear evolution in presence of a background dispersion $\epsilon(\eta)$ is governed by the scale- and time-dependent matrix $\Omega_{ab}(k,\eta)$.
It has a block-diagonal form when grouping the perturbation vector $\psi=(\psi^S,\psi^V,\psi^T)$ into subsets of scalar, vector and tensor modes, respectively,
\be
  \Omega = \left(\begin{array}{ccc}
  \Omega^S &&\\
  & \Omega^V &\\
  && \Omega^T
  \end{array}\right)\,,
\ee
with vanishing off-diagonal entries implied by rotational symmetry. Using the approximation $\Omega_m/f^2\to 1$ (see paper II~\cite{cumPT2} for the general form), the scalar part is given by
\be\label{eq:OmegaS}
  \Omega^S = \left(\begin{array}{ccccc}
  & -1 \\
  -3/2 & 1/2 & k^2 & k^2 & k^2\epsilon \\
  & -2\epsilon & 1 \\
  & & & 1\\
  & -1 \\
  \end{array}\right)\,,
\ee
for $\psi^S=(\delta,\theta,g,\delta\epsilon,A)$. The upper left two-by-two submatrix corresponds to the familiar SPT case in the limit of vanishing background dispersion $\epsilon(\eta)$.
The second row corresponds to the Euler equation. Its third and fourth column capture the impact of scalar perturbation modes
$g$ and $\delta\epsilon$ of the velocity dispersion at linear level. Their equation of motion is contained in the third and fourth row. 
The fifth column contains the suppression term related to background dispersion as discussed above.
The linear part of the equations for $\delta$ (first row) and $A$ (last row) are identical, with a difference arising only at non-linear level due to differences in their vertices.
For the vector and tensor parts, one obtains
\be
  \Omega^V = \left(\begin{array}{cc}
  1/2 & k^2 \\
  -\epsilon & 1\\
  \end{array}\right)\,,
  \quad
  \Omega^T = 1\,,
\ee
where the $2\times 2$ vector matrix describes a mixing of vorticity $w_i$ and the vector mode $\nu_i$ of the velocity dispersion tensor.
They are understood to apply separately to each $i=x,y,z$ component of the doublet $(w_i,\nu_i)$, and for each $ij$ component of $t_{ij}$, respectively.
Therefore, components with different spatial indices evolve separately from each other at the linear level.

The non-linear vertices $\gamma_{abc}$ couple scalar, vector and tensor modes among themselves but also to each other, respecting rotational symmetry at the non-linear level.
From Eq.~\eqref{eq:eom2} one obtains for example
\be
  \gamma_{\delta\theta\delta}(\vec p,\vec q)=\frac12\alpha_{pq},\quad \gamma_{\delta w_i \delta}(\vec p,\vec q)=\frac12 \frac{(\bm{p}\times \bm{q})_i}{p^2}\,,
\ee
with the first one being the usual SPT expression, and the second a vorticity backreaction contribution to the density contrast.
Another example derived from the equation of motion for vorticity, Eq.~\eqref{eq:w} is
\be
  \gamma_{w_i A (\delta\epsilon)}=\frac12 (\bm{p}\times \bm{q})_i,\quad   \gamma_{w_i A g}=-\frac12 \frac{\bm{p}\cdot\bm{q}}{q^2} (\bm{p}\times \bm{q})_i\,.
\ee
These vertices capture the generation of vorticity from two scalar perturbations, related to the log-density as well as velocity dispersion perturbations. 
For a discussion of the generation of vorticity within this framework we refer to paper II~\cite{cumPT2}.

All other non-zero $\gamma_{abc}$ can be read off from the Fourier space equations of motion given above and in App.~\ref{app:eom} in a similar way.
We are free to assume that they are symmetrized,
\be
  \gamma_{abc}(\vec p,\vec q)=\gamma_{acb}(\vec q,\vec p)\,,
\ee
leading to factors of $1/2$ for $b\not= c$. Note that this property holds for all perturbation types, including vorticity, when interchanging both the wavenumber as well as the last two indices.
The full set of vertices is collected in App.~\ref{app:vert}.

The structure of the equation of motion Eq.~\eqref{eq:eom} suggests that a perturbative solution analogous to SPT is possible. Such a solution can indeed be obtained following a
strategy that is a generalization of the well-known recursion relations for non-linear kernels known from SPT~\cite{GorGriRey86,Bernardeau:2002}. However, apart from the fact that a separate kernel for
each perturbation mode is required, and a large number of vertices contributes, the recursion relations take the form of differential instead of algebraic equations due to the $\eta$-dependence of $\Omega_{ab}(k,\eta)$. 
An algorithm to deal with a time- and scale-dependent $\Omega_{ab}(k,\eta)$ matrix has been developed in~\cite{Blas:2015tla,Garny:2020ilv}, and we present non-linear solutions using an extension of this algorithm in paper II~\cite{cumPT2}.
We emphasize again that our hybrid treatment of including both $\delta$ and $A$ as variables allows us to capture \emph{all} non-linear terms by
contributions that are quadratic in $\psi$. This is an important requirement for an efficient algorithm to determine solutions at higher order in perturbation theory.

\section{Higher cumulants}
\label{sec:pi}

In this section we discuss how to incorporate cumulants of the distribution function above the velocity dispersion tensor.
While these are generically suppressed in a hydrodynamic context, where the distribution function is close to local thermal equilibrium,
non-linear processes related to shell-crossing generate a highly non-trivial distribution function, at least at small scales. One example of this situation was highlighted by halo models in Sec.~\ref{sec:halo} (see also App.~\ref{app:halo}). 
Nevertheless, within the domain of validity of perturbative methods, i.e. on sufficiently large scales, the total impact of higher cumulants on observables is expected to become more and more suppressed.
Therefore, it is important to quantify the impact of higher cumulants on the framework presented so far.

\subsection{Split of cumulant generating function into background values and perturbation modes}

For the discussion of higher cumulants it is convenient to use the generating functional, Eq.~\eqref{eq:genfunc}.
Here we define a rescaled version,
\be
  \bar{\cal C}(\eta,\bm{x},\bm{L})\equiv {\cal C}(\tau,\bm{x},\bm{l}))\,, \ \ \ \ \ \bm{l}={\bm{L}\over (-f{\cal H})}\,,
  \label{Ltol}
\ee
where $\eta=\ln(D)$. Setting
\be
  \bar{\cal C}_{i_1,\dots,i_n}\equiv \nabla_{L_{i_1}}\cdots\nabla_{L_{i_n}}\bar{\cal C}\big|_{\bm{L}=0}\,,
\ee
we directly obtain the rescaled peculiar velocity, velocity dispersion and higher cumulants,
\be
 u_i=\bar{\cal C}_i,\quad \epsilon_{ij}=\bar{\cal C}_{ij},\quad
 \pi_{ijk}=\bar{\cal C}_{ijk},\quad
 \Lambda_{ijkl}=\bar{\cal C}_{ijkl}\,,
\ee
in agreement with Eq.~\eqref{eq:uepspi}. In addition, we introduced also the fourth cumulant $\Lambda_{ijkl}$.
Assuming statistical isotropy the third cumulant has vanishing ensemble average, while the fourth cumulant can have an expectation value $\omega(\eta)$,
\bea\label{eq:omega}
  \langle \pi_{ijk} \rangle &=& 0\,, \nn\\
  \langle \Lambda_{ijkl}  \rangle &=& \left(\delta_{ij}\delta_{kl}+2\,{\rm cyc.}\right)\frac{\omega(\eta)}{5}\,,
\eea
with the normalization chosen such that $\langle \Lambda_{ijkk}  \rangle=\omega(\eta)\delta_{ij}$.
The equation for the expectation value $\omega(\eta)$ as well as the perturbations including higher cumulants can be derived conveniently
from the equation of motion Eq.~\eqref{eq:genfunceom} of the generating function.
Taking the time-dependent rescaling into account, it reads
\bea\label{eq:Cbar}
 \partial_\eta\bar{\cal C}+\left(\frac32\frac{\Omega_m}{f^2}-1\right)(\bm{L}\cdot\nabla_L)\bar{\cal C} && \nn\\
 -(\nabla\bar{\cal C})\cdot(\nabla_L\bar{\cal C})-(\nabla\cdot\nabla_L)\bar{\cal C} &=& \bm{L}\cdot\nabla\hat\Phi\,.
\eea
It is convenient to consider the ensemble average of the generating function itself.
Assuming statistical homogeneity and isotropy it is independent of $\bm{x}$ and can depend only on $L^2\equiv\bm{L}^2$,
\be
  {\cal E}(\eta,L^2)\equiv \langle \bar{\cal C}(\eta,\bm{x},\bm{L})\rangle\,.
\ee
Taking the ensemble average of Eq.~\eqref{eq:Cbar} and averaging over the direction of $\bm{L}$ yields its equation of motion
\be\label{eq:E}
  \left[\partial_\eta + \left(\frac32\frac{\Omega_m}{f^2}-1\right)\frac{\partial}{\partial\ln L}\right]{\cal E}=Q_{\cal E}\equiv\int\frac{d\Omega_L}{4\pi}\langle \nabla\bar{\cal C}\cdot\nabla_L\bar{\cal C}\rangle\,.
\ee
The equations for the expectation values of the individual cumulants can be obtained by Taylor expanding in $\bm{L}$, using
\bea\label{eq:Cbarexpansion}
  \bar{\cal C} &=& A+L_iu_i+\frac12 L_iL_j \epsilon_{ij}+\frac16 L_iL_jL_k\pi_{ijk}\nn\\
  &&  {} +\frac1{24} L_iL_jL_kL_l\Lambda_{ijkl}+\dots\,,
\eea
with the ellipsis standing for fifth and higher cumulants.
Taking the ensemble average and using the definitions of the expectations values in Eqs.~(\ref{eq:A},~\ref{eq:omega},~\ref{eq:epsilon}) of the zeroth, second and fourth cumulant, respectively, yields
\be\label{eq:Eexpansion}
  {\cal E}={\cal A}(\eta) + \frac12 \epsilon(\eta) L^2 + \frac{1}{24}\frac{3\omega(\eta)}{5}L^4+ {\cal O}(L^6)\,.
\ee
In Sec.\,\ref{sec:hierarchy} we discuss an extension of this expansion to higher cumulant orders, given by a Taylor expansion in even powers of $L$,
\be
  {\cal E} = \sum_{n}{\cal E}_{2n}(\eta)\frac{L^{2n}}{(2n)!}\,,
  \label{Ecalcosmo}
\ee
where\footnote{Note that while Eq.~\eqref{Ecalcosmo} is identical to Eq.~\eqref{expCGF} given Eq.~\eqref{Ltol}, the ${\cal E}_{2n}$'s have different normalization than in Sec.\,\ref{sec:halo} by powers of $(f {\cal H})^{2n}$. But the normalized cumulants $\bar{\cal E}_{2n}$ (see Eq.~\ref{E2nbar}) are of course the same.}
\bea
  {\cal E}_{0} &=& {\cal A}\,,\nn\\
  {\cal E}_{2} &=& \epsilon\,,\nn\\
  {\cal E}_{4} &=& \frac35\omega\,,
\eea
and ${\cal E}_6,\dots$ denote background values of the sixth and higher cumulants.
For the moment we restrict ourselves to the first three terms.
Inserting Eq.~\eqref{eq:Eexpansion} into Eq.~\eqref{eq:E} yields
\bea\label{eq:eomcumexpectation}
  \partial_\eta{\cal A} &=& Q_A \equiv Q_{\cal E}\big|_{L^0}\,,\nn\\
  \left[\partial_\eta + 2\left(\frac32\frac{\Omega_m}{f^2}-1\right)\right]\epsilon &=& Q \equiv 2Q_{\cal E}\big|_{L^2}\,,\nn\\
  \left[\partial_\eta + 4\left(\frac32\frac{\Omega_m}{f^2}-1\right)\right]\omega &=& Q_\omega \equiv 24\frac53 Q_{\cal E}\big|_{L^4}\,,\nn\\
\eea
where the right-hand side denotes the source term $Q_{\cal E}$ evaluated at a given order in powers of $L^2$.
The latter can be obtained by inserting Eq.~\eqref{eq:Cbarexpansion} into Eq.~\eqref{eq:Cbar}, evaluating the $\nabla_L$ derivative, and
performing the angular average using
\bea
  \int\frac{d\Omega_L}{4\pi} L_iL_j &=& \frac13 L^2\delta_{ij}\,,\nn\\
  \int\frac{d\Omega_L}{4\pi} L_iL_jL_kL_l &=& \frac1{15} L^4\left(\delta_{ij}\delta_{kl}+2\,{\rm cyc.}\right)\,.
\eea
After this integration $Q_{\cal E}$ depends only on even powers of $L$.
For ${\cal A}$ and $\epsilon$ we recover from Eq.~\eqref{eq:eomcumexpectation} the equations of motion Eqs.~(\ref{eq:QA},~\ref{eq:Qdef}) derived previously.
Going to order $L^4$ we find for the source term of the expectation value $\omega$ of the fourth cumulant,
\bea\label{eq:Qomega}
  Q_\omega &=& \frac{1}{3}\Big\{ \langle (\nabla_iA) \bar {\cal C}_{ijjkk}\rangle +4\langle (\nabla_iu_j)\Lambda_{ijkk}\rangle \nn\\
  && {} + 2\langle (\nabla_i\epsilon_{jj})\pi_{ikk}\rangle  + 4\langle (\nabla_i\epsilon_{jk})\pi_{ijk}\rangle \nn\\
  && {}  + 4\langle (\nabla_i\pi_{jkk})\epsilon_{ij}\rangle +  \langle (\nabla_i\Lambda_{jjkk})u_i\rangle
   \Big\}\,.
\eea
To obtain an equation for the perturbations around the expectation value we define
\be\label{eq:delC}
  \delta\bar{\cal C}(\eta,\bm{x},\bm{L}) \equiv \bar {\cal C}(\eta,\bm{x},\bm{L}) - {\cal E}(\eta,L^2)\,,
\ee
and using  Eqs.~(\ref{eq:Cbar},~\ref{eq:E}) we find the equation of motion
\bea\label{eq:dCbar}
   && \Bigg[\partial_\eta+\left(\frac32\frac{\Omega_m}{f^2}-1\right)(\bm{L}\cdot\nabla_L)-2\frac{\partial{\cal E}}{\partial L^2}\bm{L}\cdot\nabla\nn\\
   && -(\nabla\cdot\nabla_L)\Bigg]\delta\bar{\cal C} 
    = (\nabla\delta\bar{\cal C})\cdot(\nabla_L\delta\bar{\cal C})+\bm{L}\cdot\nabla\hat\Phi -Q_{\cal E}  \,.\nn\\
\eea
The term involving $\partial{\cal E}/\partial L^2=\epsilon/2+L^2\omega/20+{\cal O}(L^4)$ generates terms that can be viewed as a generalization of the ``Jeans-like'' term discussed above.
The equation can be Taylor expanded in $L_i$ to obtain equations of motion for the perturbation modes of the cumulants.
When taking up to the second cumulant into account, we find results consistent with those from Sec.\,\ref{sec:sigma}.

\subsection{Third and fourth cumulant}

In the following we work out the equations of motion when neglecting fifth and higher cumulants, and taking the complete set of \emph{scalar} perturbations of the third and fourth cumulant
into account. We refer to Sec.\,\ref{sec:hierarchy} for fifth and higher cumulants. We use the decomposition
\bea\label{eq:piLamdecomposition}
  \pi_{ijk}^S &=& -\left(\delta_{ij}\frac{\nabla_k}{\nabla^2}+2\,{\rm cyc.}\right)\frac{\chi}{5}-\frac{\nabla_i\nabla_j\nabla_k}{\nabla^4}(\pi-\chi)\,,\nn\\
  \delta\Lambda_{ijkl}^S &=& \left(\delta_{ij}\delta_{kl}+2\,{\rm cyc.}\right)\frac{\psi}{5}\nn\\
  && {} +\left(\delta_{ij}\frac{\nabla_k\nabla_l}{\nabla^2}+5\,{\rm cyc.}\right)\frac{\kappa-\xi-2\psi}{2}\nn\\
  && {} + \frac{\nabla_i\nabla_j\nabla_k\nabla_l}{\nabla^4}(7\psi+5\xi-4\kappa)\,,
\eea
with scalar modes $\pi,\chi,\kappa,\xi,\psi$ defined such that
$\pi=-\nabla_i\pi_{ijj}$, $\kappa=\Lambda_{iijj}$ and $\xi=\nabla_i\nabla_j/\nabla^2\Lambda_{ijkk}$.
Here $\delta\Lambda_{ijkl}=\Lambda_{ijkl}-\langle\Lambda_{ijkl}\rangle$ and the superscript indicates that we take the scalar contribution into account.
Inserting this decomposition in Eq.~\eqref{eq:Qomega} and using Eq.~\eqref{eq:eomcumexpectation} yields
\bea\label{eq:eomomegaepsilon}
  \lefteqn{ \left[\partial_\eta + 2\left(\frac32\frac{\Omega_m}{f^2}-1\right)\right]\epsilon }\nn\\
  &=& \frac13 \Big\{  \langle A\pi\rangle+\langle \theta(g-\delta\epsilon)\rangle +2 \langle w_i\nu_i\rangle \Big\} \,,\nn\\
  \lefteqn{ \left[\partial_\eta + 4\left(\frac32\frac{\Omega_m}{f^2}-1\right)\right]\omega  }\nn\\
  &=& \frac13\Big\{ \langle \theta(4\xi-\kappa)\rangle + 2 \langle (g+3\delta\epsilon)\pi\rangle -\frac8{5} \langle g\chi\rangle\Big\} \,,
\eea
where e.g. $\langle A\pi\rangle=\int d^3k\, P_{A\pi}(k,\eta)$. 
The only change in the equation for $\epsilon$ compared to Eq.~\eqref{eq:Q2nd} is
the additional $\langle A\pi\rangle$ term. We note that, for $\epsilon$, this is the complete source term, with no further terms arising even when relaxing the restriction to third and fourth cumulant scalar modes.
The equation for $\omega$ contains cross power spectra of the first and fourth as well as second and third cumulant perturbations, respectively.
In the present approximation contributions to the source term for $\omega$ from vector and tensor modes as well as cross spectra of the zeroth and fifth cumulant are neglected.

For the complete set of scalar perturbations, we find the following equations in Fourier space in the ``linear'' approximation when considering $\epsilon$ as well as $\omega$ as given,
\bea\label{eq:cum4lin}
  \partial_\eta \delta_k &=& \theta_k\,, \nn\\
  \left[\partial_\eta+\left(\frac32\frac{\Omega_m}{f^2}-1\right)\right]\theta_k &=& \frac32\frac{\Omega_m}{f^2}\delta_k -\epsilon k^2 A_k\nn\\
  && {} -k^2(g_k+\delta\epsilon_k)\,,\nn\\
  \left[\partial_\eta+2\left(\frac32\frac{\Omega_m}{f^2}-1\right)\right]g_k &=& 2\epsilon\theta_k-\pi_k+\frac35\chi_k\,,\nn\\
  \left[\partial_\eta+2\left(\frac32\frac{\Omega_m}{f^2}-1\right)\right]\delta\epsilon_k &=& -\frac15\chi_k\,,\nn\\
  \left[\partial_\eta+3\left(\frac32\frac{\Omega_m}{f^2}-1\right)\right]\pi_k &=& \omega k^2A_k+\epsilon k^2(3g_k+5\delta\epsilon_k)\nn\\
  && {} +k^2\xi_k\,,\nn\\
  \left[\partial_\eta+3\left(\frac32\frac{\Omega_m}{f^2}-1\right)\right]\chi_k &=& \omega k^2A_k+\epsilon k^2(5\delta\epsilon_k)\nn\\
  && {} + \frac12 k^2(5\kappa_k-5\xi_k-8\psi_k)\,,\nn\\
  \left[\partial_\eta+4\left(\frac32\frac{\Omega_m}{f^2}-1\right)\right]\kappa_k &=& 4\omega \theta_k -4\epsilon\pi_k\,,\nn\\
  \left[\partial_\eta+4\left(\frac32\frac{\Omega_m}{f^2}-1\right)\right]\xi_k &=& \frac{16}{5}\omega \theta_k -4\epsilon\pi_k+\frac45\epsilon\chi_k\,,\nn\\
  \left[\partial_\eta+4\left(\frac32\frac{\Omega_m}{f^2}-1\right)\right]\psi_k &=& 0\,,
\eea
as well as $\partial_\eta A_k=\theta_k$ such that $A_k=\delta_k$ in the linear approximation.

The full non-linear set of equations takes the form of Eq.~\eqref{eq:eom}, with an extended perturbation vector $\psi$.
In the approximation adopted here only its scalar part changes,
\be
  \psi^S=(\delta,\theta,g,\delta\epsilon,A,\pi,\chi,\kappa,\xi,\psi)\,.
\ee
The extended scalar evolution matrix is given in App.~\ref{app:higher}.
In addition, the set of vertices increases.
All additional vertices involving at least one of the third cumulant perturbations and only scalar modes are collected in App.~\ref{app:higher}.
For vertices involving $\pi$ or $\chi$ and vorticity or vector modes, we refer to future work.

After discussing the linear approximation in the next section, we extend the cumulant expansion to beyond the fourth order in Sec.\,\ref{sec:hierarchy}.

\section{Linear approximation}
\label{sec:linear}

As an illustrative example, we study the linear solution of the perturbation equations when taking the expectation values $\epsilon(\eta)$
and $\omega(\eta)$ of the second and fourth cumulant (and eventually also higher cumulants) as given, and neglecting non-linear couplings between perturbation modes.
We refer to paper II~\cite{cumPT2} for the non-linear case. Following the previous discussion, we expect a suppression of the density contrast for
wavenumbers $k\gtrsim \epsilon^{-1/2}$ or $k\gtrsim \omega^{-1/4}$ even at linear level, arising both from the direct impact of the
background dispersion $\epsilon(\eta)$ in the Euler equation, as well as the indirect impact via the perturbations $g$ and $\delta\epsilon$
of the velocity dispersion that are in turn generated in presence of a non-zero $\epsilon(\eta)$, and coupled to the higher cumulant modes.

\subsection{Second cumulant approximation}\label{sec:linanalyt}

Let us start by analyzing the approximation where the third and higher cumulants are neglected, see Sec.\,\ref{sec:sigma}.
In this case only the background dispersion $\epsilon(\eta)$ is relevant.
In linear approximation, scalar, vector and tensor modes evolve independently. We therefore focus on the scalar perturbations, that possess growing modes.
Inspecting their linear evolution equations described by Eq.~\eqref{eq:OmegaS}, one finds that the perturbation mode $\delta\epsilon$ of the velocity dispersion tensor
can be solved independently when disregarding non-linear terms and higher cumulants, and decays as $e^{-\eta}=1/D$. It can therefore be neglected in this particular approximation.
Furthermore, there is no difference between $A$ and $\delta$ at linear order. The system of equations therefore reduces to the three variables $\delta,\theta,g$ and
takes the form
\bea\label{eq:lin2nd}
  \delta_k' &=& \theta_k\,, \nn\\
  \theta_k' &=& -\frac12 \theta_k +\frac32\delta_k - k^2 g_k - \epsilon k^2\delta_k\,, \nn\\
  g_k' &=& -g_k+2\epsilon\theta_k\,.
\eea
The third equation has the formal solution
\be\label{eq:gksol}
  g_k(\eta) = \int^\eta d\eta'\, e^{\eta'-\eta}\,2\epsilon(\eta')\theta_k(\eta')\,.
\ee
Inserting this solution into the Euler equation in Eq.~\eqref{eq:lin2nd} yields a correction term relative to SPT that is
proportional to $k^2$ and non-local in time. These features provide a particular example for a modification that is
consistent with the most general structure allowed by symmetries~\cite{Abolhasani:2015mra}. Therefore, as expected, adding
second and also higher cumulants to the perturbative expansion and using the underlying Vlasov-Poisson dynamics yields a consistent, and a priori deterministic ``UV completion'' of SPT.

Note that naively replacing the non-local relation between $g_k$ and $\theta_k$ by a local ansatz of the
form $g_k \mapsto c_\text{vis}^2\theta_k$ would yield a dissipative viscosity term in the Euler equation. However, we stress that
the actual non-local relation Eq.~\eqref{eq:gksol} is derived from the collisionless dynamics of the Vlasov equation, being non-dissipative.
Indeed, it is possible to check that when using the non-local relation Eq.~\eqref{eq:gksol} (or equivalently the underlying equations Eq.~\ref{eq:lin2nd})
the dynamics does indeed obey the energy evolution equation for the sum of kinetic and potential energy, known as the cosmic energy equation~\cite{Pee80}, at linear
order in perturbation theory (see App.\,\ref{app:energy}). In contrast, a naive local replacement of the form mentioned above
would lead to a violation of the cosmic energy equation. Therefore, the non-local relation Eq.~\eqref{eq:gksol} cannot be naively interpreted in terms of a
fluid-like dissipative viscosity. Indeed, when extending the analysis to higher cumulants (see Sec.\,\ref{sec:pi}), additional terms in the last line of Eq.~\eqref{eq:lin2nd}
appear, that would modify Eq.~\eqref{eq:gksol}, but are still consistent with non-dissipative energy evolution (see App.\,\ref{app:energy}). This indicates that the collisionless dynamics is significantly more complex than fluid-like dissipative behavior, and this is indeed what we find further below.

Nevertheless, we observe that in the limit $k^2\epsilon\ll 1$ and at linear level we can approximate $\theta_k(\eta')\to \delta_k(\eta')\propto e^{\eta'}$
such that $g_k(\eta)\to 2\delta_k(\eta)\int^\eta d\eta'\, e^{2(\eta'-\eta)}\,\epsilon(\eta')$,
which yields an effective Jeans-like suppression scale
\be\label{eq:kJ}
  \frac{1}{k_\text{J-like}^2} = \epsilon(\eta) + 2\int^\eta d\eta'\, e^{2(\eta'-\eta)}\epsilon(\eta')\,.
\ee
Importantly, the second and higher cumulants affect also the non-linear evolution and go far beyond adding a Jeans-like term even at linear level, as stressed also above.
In particular, we are interested in the solution over the entire range of wavenumbers, including also the regime where $k^2\epsilon$ is not small and the
simplification leading to Eq.~\eqref{eq:kJ} cannot be used.

It is therefore most effective to directly solve the coupled system Eq.~\eqref{eq:lin2nd} including the $g_k$-mode explicitly.
The full linear solution of these equations studied here provides the starting point for a perturbative solution of the non-linear equations  (see paper II~\cite{cumPT2}).
Solving Eq.~\eqref{eq:lin2nd} in general requires some knowledge of the background dispersion.
For illustration we assume a power-law dependence
\be
  \epsilon(\eta) = \epsilon_0 e^{\alpha\eta}\,,
\ee
with some power-law index $\alpha$ and value $\epsilon_0$ today.
Apart from simplicity, this choice is relevant for the limit of a
scaling universe described by a power-law input spectrum $P_0(k) \sim k^{n_s}$, with exponent being given by $\alpha=4/(n_s+3)$ in that case (see Sec.\,\ref{sec:powerlaw}).
Furthermore, it can be viewed as an approximate description also within $\Lambda$CDM cosmology for a limited redshift interval.

\begin{figure}[t]
  \begin{center}
  \includegraphics[width=\columnwidth]{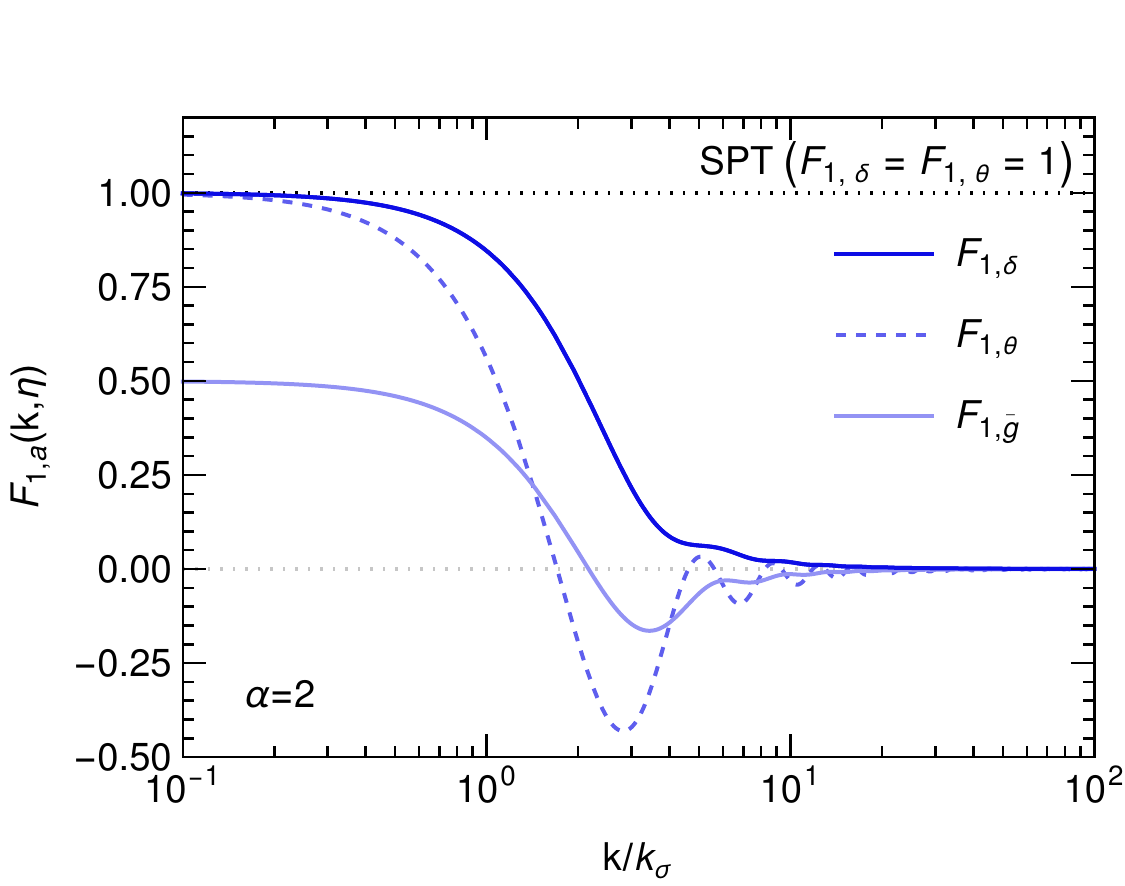}
  \end{center}
  \caption{\label{fig:T}
  Linear kernels $F_{1,a}(k,\eta)$ that describe the suppression relative to SPT for $a=\delta,\theta$ when taking the second cumulant into account. In addition, the rescaled scalar perturbation mode $\bar g_k=g_k/\epsilon$ of the dispersion tensor possesses a growing mode, shown for $a=\bar g$, with $F_{1,\bar g}\to 2/(2+\alpha)$ for $k\to 0$. Here $\alpha=2$ and $k_\sigma=1/\sqrt{\epsilon(\eta)}$ such that the time-dependence is scaled out.}
\end{figure}

It is convenient to use the rescaled variable
\be
  \bar g_k(\eta) \equiv g_k(\eta)/\epsilon(\eta)\,.
\ee
Setting $\bar\psi=(\delta_k,\theta_k,\bar g_k)$, Eq.~\eqref{eq:lin2nd} can be written in the form
\be\label{eq:psibareom}
  \bar\psi'+(\Omega_0+\epsilon k^2\Omega_1)\bar\psi=0\,,
\ee
where
\be
  \Omega_0 = \left( \begin{array}{ccc} 0&-1&0\\ -\frac32 & \frac12 & 0 \\ 0 & -2& 1+\alpha\end{array}\right)\,, \quad
  \Omega_1 = \left( \begin{array}{ccc} 0&0&0\\ 1 & 0 & 1 \\ 0 & 0& 0\end{array}\right)\,.
\ee
In the limit $k^2\epsilon\ll 1$ the growing mode solution reads
\be\label{eq:psi0}
  \bar\psi\to \bar\psi^{(0)}\equiv (1,1,2/(2+\alpha)) \, e^\eta\delta_{k0}\,,
\ee
in accordance with the previous discussion and Eq.~\eqref{eq:gksol}. Here $\delta_{k0}$ stands for the conventional linear density field.
We can obtain a general solution by an iteration in powers of $\Omega_1$, writing $\bar\psi=\sum_j\bar\psi^{(j)}$, with
\be\label{eq:recursion}
  \bar\psi^{(j)}(\eta) = \int^\eta d\eta' g_0(\eta-\eta')(-k^2\epsilon(\eta')\Omega_1)\bar\psi^{(j-1)}(\eta')\,.
\ee
Here $g_0$ is the Green function in the limit $\Omega_1\to 0$,
and contains the conventional SPT linear propagator in the upper left $2\times 2$ block,
\bea
 \lefteqn{ g_0(\eta-\eta') = \frac15\left( \begin{array}{ccc} 3&2&0\\ 3&2&0\\ \frac{6}{2+\alpha} & \frac{4}{2+\alpha} & 0\end{array}\right)e^{\eta-\eta'} }\nn\\
  && {} + \frac15\left( \begin{array}{ccc} 2&-2&0\\ -3&3&0\\ -\frac{12}{2\alpha-1} & \frac{12}{2\alpha-1} & 0\end{array}\right)e^{-3(\eta-\eta')/2}\nn\\
  && {} + \left( \begin{array}{ccc} 0&0&0\\ 0&0&0\\ \frac{6}{(2\alpha-1)(\alpha+2)} & \frac{-4(1+\alpha)}{(2\alpha-1)(\alpha+2)} & 1\end{array}\right)e^{-(1+\alpha)(\eta-\eta')}\,.\nn\\
\eea
Furthermore, compared to SPT, an additional decaying mode appears.
Using the growing mode initial condition Eq.~\eqref{eq:psi0}, one finds 
\be
  \bar\psi^{(j)}(\eta)=(\epsilon(\eta)k^2)^j\,e^\eta\delta_{k0}\, (c_\delta^{(j)},c_\theta^{(j)},c_{\bar g}^{(j)})\,,
\ee
with numerical coefficients that can be found recursively using Eq.~\eqref{eq:recursion},
\be
  c_\delta^{(j)} = -\frac{2(4+3\alpha j-2\alpha )}{\alpha j(5+2\alpha j)(2+\alpha j)}c_\delta^{(j-1)}\,,
\ee
and  $c_{\bar g}^{(j)}=\frac{2(1+\alpha j)}{2+\alpha j+\alpha }c_\delta^{(j)}$, $c_\theta^{(j)}=(1+\alpha j)c_\delta^{(j)}$.
The recursive solution can be found in explicit form,
\bea
   c_\delta^{(j)} &=& \left(-\frac{3}{\alpha^2}\right)^j\frac{1}{j!}\frac{\Gamma\left(j+\frac{4+\alpha}{3\alpha}\right)}{\Gamma\left(\frac{4+\alpha}{3\alpha}\right)}\frac{\Gamma\left(\frac{5}{2\alpha}+1\right)}{\Gamma\left(j+\frac{5}{2\alpha}+1\right)}\nn\\
   && {} \times \frac{\Gamma\left(\frac{2}{\alpha}+1\right)}{\Gamma\left(j+\frac{2}{\alpha}+1\right)}\,.
\eea
The sum can be expressed in terms of a generalized hypergeometric function ${}_1 F_2$. This gives a closed-form result for the evolution of the density contrast,
\be\label{eq:lincum2F1delta}
  \delta_k(\eta) = F_{1,\delta}(k,\eta)\,e^\eta\delta_{k0}\,,
\ee
with linear kernel
\be
  F_{1,\delta}(k,\eta) = \; {}_1 F_2\left(\frac{4+\alpha}{3\alpha};1+\frac{2}{\alpha},1+\frac{5}{2\alpha};\frac{-3k^2\epsilon(\eta)}{\alpha^2}\right)\,.
\ee
The solutions for $\theta_k$ and $\bar g_k$ can be written in a similar form,
\bea\label{eq:TthetaTg}
  \theta_k(\eta) &=& F_{1,\theta}(k,\eta)\,e^\eta\delta_{k0}\,,\nn\\
  \bar g_k(\eta) &=& F_{1,\bar g}(k,\eta)\,e^\eta\delta_{k0}\,,
\eea
with linear kernels given in App.\,\ref{app:lin}. 
They are shown for $\alpha=2$ in Fig.~\ref{fig:T}. The time-dependence can be scaled out by normalizing the wavenumber to the scale
\be
  k_\sigma \equiv \frac{1}{\sqrt{\epsilon(\eta)}}\,,
\ee
that characterizes the wave-number above which velocity dispersion becomes important.

In the limit $k^2\epsilon\to 0$ one has 
\bea\label{eq:Tsmall}
  \left(\begin{array}{c} F_{1,\delta} \\ F_{1,\theta} \\ F_{1,\bar g}\end{array} \right) &\to& 
  \left(\begin{array}{c} 1 \\ 1 \\ \frac{2}{2+\alpha}\end{array} \right)
  - \left(\begin{array}{c} 1 \\ 1+\alpha \\ 1\end{array} \right)\frac{2(4+\alpha)k^2\epsilon(\eta)}{\alpha(2+\alpha)(5+2\alpha)}\nn\\
  && {} +{\cal O}(k^4\epsilon^2)\,,
\eea
recovering the growing-mode SPT solution for $\delta$ and $\theta$ in the limit $\epsilon\to 0$, as well as a negative first-order correction in $\epsilon$ that describes the onset of ``Jeans-like'' suppression (assuming $\alpha>0$). Equation~\eqref{eq:Tsmall} agrees with the low-$k$ approach presented in~\cite{McD1104}. Note that the suppression given by the linear kernel is larger for the velocity divergence than for the density, in agreement with previous results in the literature~\cite{PueSco0908,McD1104}.  In the opposite limit $k^2\epsilon\gg 1$, we find that the linear kernels decay with a power-law behavior. As we shall see in Sec.~\ref{sec:hierarchy}, when including higher cumulants also an enhancement instead of a suppression can occur in general, a feature expected for a collisionless system~\cite{BinTre08}, and in contrast to fluids. Nevertheless, within the second cumulant approximation discussed here this behavior does not occur. The asymptotic expansion for large $k^2\epsilon\gg 1$ has the form
\be\label{eq:Tlarge}
  F_{1,a}(k,\eta) \to D_a s_k^{-d_a} + E_a s_k^{-e_a} \cos\left(\frac{2s_k}{\alpha}+\varphi_a\right)\,,
\ee
where $s_k^2\equiv 3k^2\epsilon(\eta)$. All coefficients, phases and exponents are given in Table~\ref{tab:Tasymptotic} in App.\,\ref{app:lin}.
One can rewrite it in the form
\be\label{eq:Tlarge2}
  \left(\begin{array}{c} F_{1,\delta} \\ F_{1,\theta} \\ F_{1,\bar g}\end{array} \right) \to 
  \left(\begin{array}{c} 1 \\ -\frac{1+\alpha}{3} \\ -1\end{array} \right)\frac{D_\delta}{s_k^{d_\delta}}
  + 2\,\text{Re}\left(\begin{array}{c} 1 \\ is_k \\ 2\end{array} \right)\frac{ E_\delta e^{i(\frac{2s_k}{\alpha}+\varphi_\delta)}}{s_k^{e_\delta}}\,.
\ee
At large $k\propto s_k$ the non-oscillating parts dominate for $F_{1,\delta}$ and $F_{1,\bar g}$ since
\be
  d_\delta = \frac{16+4\alpha}{6\alpha} < e_\delta = \frac{19+7\alpha}{6\alpha}\,,
\ee
for all $\alpha>0$. For $F_{1,\theta}$ this is only the case for $0<\alpha<1$, while the oscillating term dominates for $\alpha>1$ due to the additional factor $is_k$.

It is also possible to find the most general solution of Eq.~\eqref{eq:lin2nd}, including all eigenmodes. We find (see App.\,\ref{app:lin} for details)
\bea\label{eq:lincum2gensol}
 \delta_k &=&  A \, e^\eta\, {}_1 F_2\left(\frac{4+\alpha}{3\alpha};1+\frac{2}{\alpha},1+\frac{5}{2\alpha};\frac{-3k^2\epsilon(\eta)}{\alpha^2}\right)\nn\\
 &+& B \, e^{-\frac32\eta}\, {}_1 F_2\left(\frac{-7+2\alpha}{6\alpha};1-\frac{1}{2\alpha},1-\frac{5}{2\alpha};\frac{-3k^2\epsilon(\eta)}{\alpha^2}\right)\nn\\
 &+& C \, e^{-\eta}\, {}_1 F_2\left(\frac{-2+\alpha}{3\alpha};1-\frac{2}{\alpha},1+\frac{1}{2\alpha};\frac{-3k^2\epsilon(\eta)}{\alpha^2}\right)\,,\nn\\
\eea
where $A,B,C$ are free coefficients. For $\epsilon k^2\ll 1$, all generalized hypergeometric functions approach unity, and one recovers the usual growing
and decaying modes, plus an extra decaying mode arising from the $g_k$ perturbation of the velocity dispersion tensor. The solution given in Eq.~\eqref{eq:lincum2F1delta}
corresponds to $B=C=0$. Assuming $\epsilon$ grows with time, it is justified to assume these generalized growing-mode initial conditions, which we shall
do from now on. Nevertheless, the general solution can be used to obtain an analytic expression for the linear propagator that generalizes the well-known linear propagator from SPT to include dispersion (see App.\,\ref{app:lin}).

\subsection{Impact of third and fourth cumulant}\label{sec:fourthcumlin}

We consider the evolution of perturbations when taking also the third and fourth cumulant into account (see Sec.\,\ref{sec:pi}).
The set of differential equations in linear approximation is given in Eq.~\eqref{eq:cum4lin}.
We also take the expectation value $\omega(\eta)$ of the fourth cumulant into account, in addition to $\epsilon(\eta)$.
For illustration, we assume a constant dimensionless ratio
\be
  \bar \omega \equiv \frac{\omega(\eta)}{\epsilon(\eta)^2}\,.
\ee
This choice is also motivated by the scaling solutions considered in Sec.\,\ref{sec:powerlaw}

It is convenient to use the dimensionless variables
\be
  \bar g\equiv\frac{g}{\epsilon},\
  \delta\bar \epsilon \equiv\frac{\delta \epsilon}{\epsilon},\
  \bar \pi \equiv\frac{\pi}{\epsilon},\
  \bar \chi \equiv\frac{\chi}{\epsilon},\
  \bar \kappa \equiv\frac{\kappa}{\epsilon^2},\
  \bar \xi \equiv\frac{\xi}{\epsilon^2},\
  \bar \psi \equiv\frac{\psi}{\epsilon^2},
\ee
in terms of which the linear evolution equations for the Fourier mode $k$
read (approximating $\Omega_m/f^2\to 1$ and using $A_k=\delta_k$ at linear level)
\bea\label{eq:cum4linrescale}
  \partial_\eta \delta_k &=& \theta_k\,, \nn\\
  \left[\partial_\eta+\frac12\right]\theta_k &=& \frac32\delta_k -\epsilon k^2 (\delta_k+\bar g_k+\delta\bar\epsilon_k)\,,\nn\\
  \left[\partial_\eta+1+\alpha\right]\bar g_k &=& 2\theta_k-\bar\pi_k+\frac35\bar\chi_k\,,\nn\\
  \left[\partial_\eta+1+\alpha\right]\delta\bar\epsilon_k &=& -\frac15\bar\chi_k\,,\nn\\
  \left[\partial_\eta+\frac32+\alpha\right]\bar\pi_k &=& \epsilon k^2(\bar\omega \delta_k+3\bar g_k+5\delta\bar\epsilon_k+\bar\xi_k)\,,\nn\\
  \left[\partial_\eta+\frac32+\alpha\right]\bar\chi_k &=& \epsilon k^2\big( \bar\omega \delta_k+5\delta\bar\epsilon_k\,,\nn\\
  && {} +\frac12 (5\bar\kappa_k-5\bar\xi_k-8\bar\psi_k)\big)\,,\nn\\
  \left[\partial_\eta+2+2\alpha\right]\bar\kappa_k &=& 4\bar\omega \theta_k -4\bar\pi_k\,,\nn\\
  \left[\partial_\eta+2+2\alpha\right]\bar\xi_k &=& \frac{16}{5}\bar\omega \theta_k -4\bar\pi_k+\frac45\bar\chi_k\,,\nn\\
  \left[\partial_\eta+2+2\alpha\right]\bar\psi_k &=& 0\,.
\eea

We initialize the perturbations in the growing mode of the linear set of equations Eq.~\eqref{eq:cum4linrescale}
in the limit $k^2\epsilon\to 0$, given by
\bea
 \bar\psi &\equiv& (\delta_k,\theta_k,\bar g_k,\delta\bar\epsilon_k,\bar\pi_k,\bar\chi_k,\bar\kappa_k,\bar\xi_k,\bar\psi_k) \nn\\
 &\to& e^\eta\left(1,1,\frac{2}{2+\alpha},0,0,0,\frac{4 \bar\omega}{3+2\alpha},\frac{16 \bar\omega}{5(3+2\alpha)},0\right)\delta_{k0}\,.\nn\\
\eea
Note that also the $\delta\bar\epsilon_k$, $\bar\pi_k$ and $\bar\chi_k$ modes are generated in the time-evolution due to the terms
proportional to $k^2\epsilon$ in the evolution equations Eq.~\eqref{eq:cum4linrescale}. We define linear kernels $F_{1,a}$ for all perturbation variables by
\be\label{eq:Ta}
  \bar\psi_a \equiv F_{1,a}(k,\eta)\,e^\eta\delta_{k0}\,.
\ee

\begin{figure}[t]
  \begin{center}
  \includegraphics[width=\columnwidth]{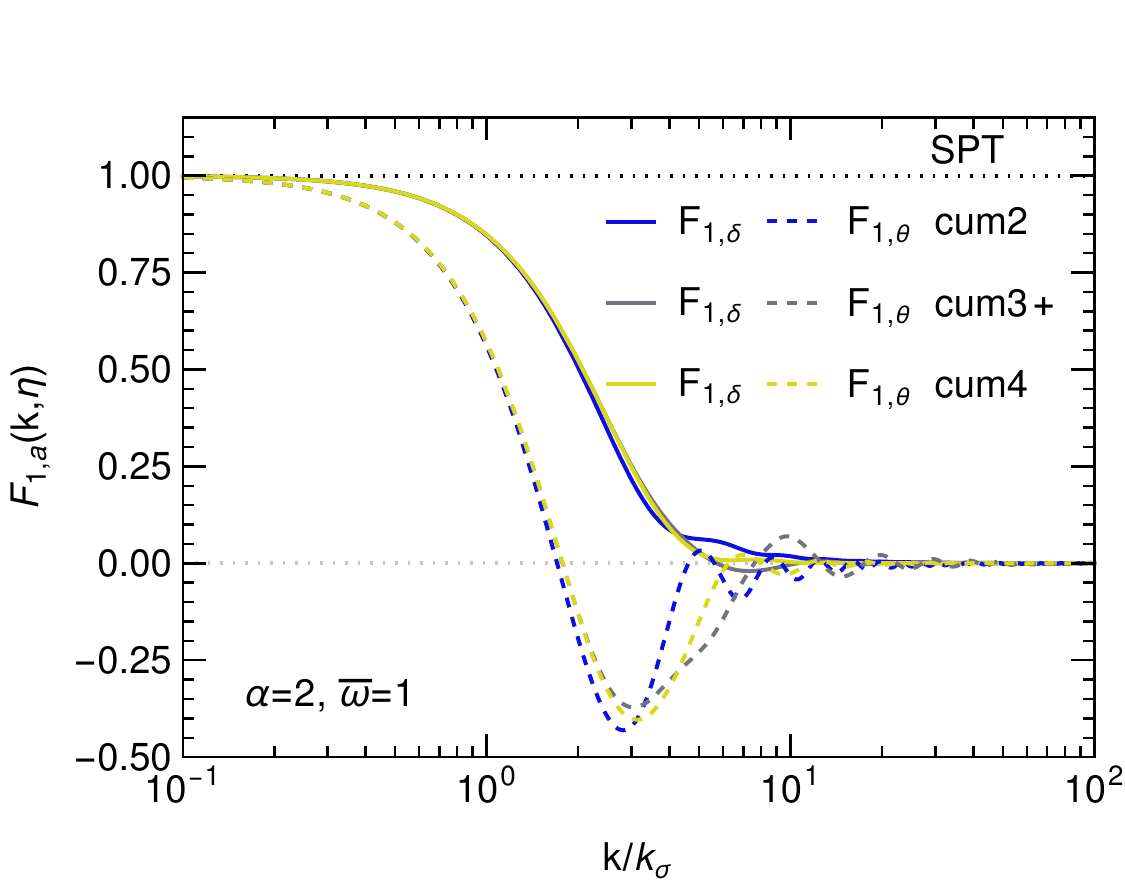}\\
  \includegraphics[width=\columnwidth]{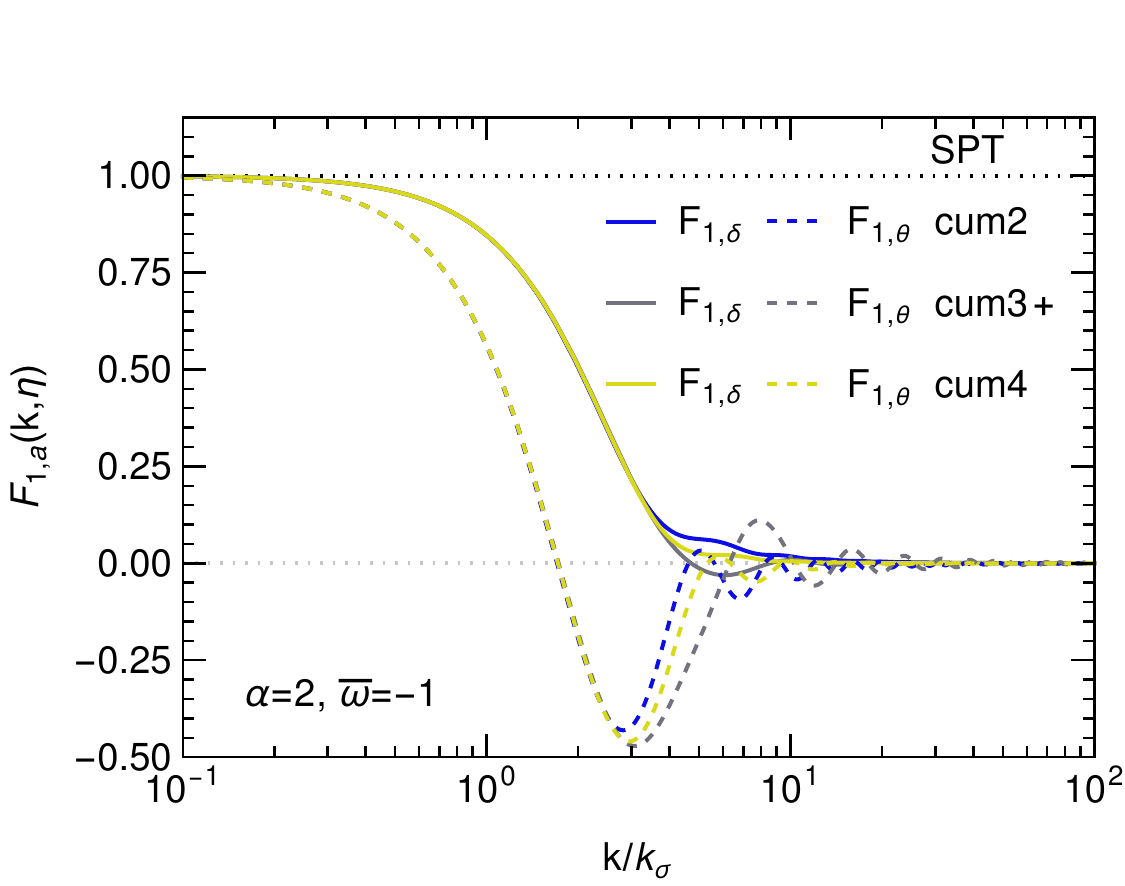}
  \end{center}
  \caption{\label{fig:Tcum}
  Linear \vpt~kernels $F_{1,a}(k,\eta)$ when taking the second, third and fourth cumulant into account, respectively. The upper panel shows the case $\bar\omega=1$, and the lower $\bar\omega=-1$, for $\delta$ and $\theta$. Furthermore, $\alpha=2$ and $k_\sigma=1/\sqrt{\epsilon(\eta)}$.}
\end{figure}

\begin{figure}[t]
  \begin{center}
  \includegraphics[width=\columnwidth]{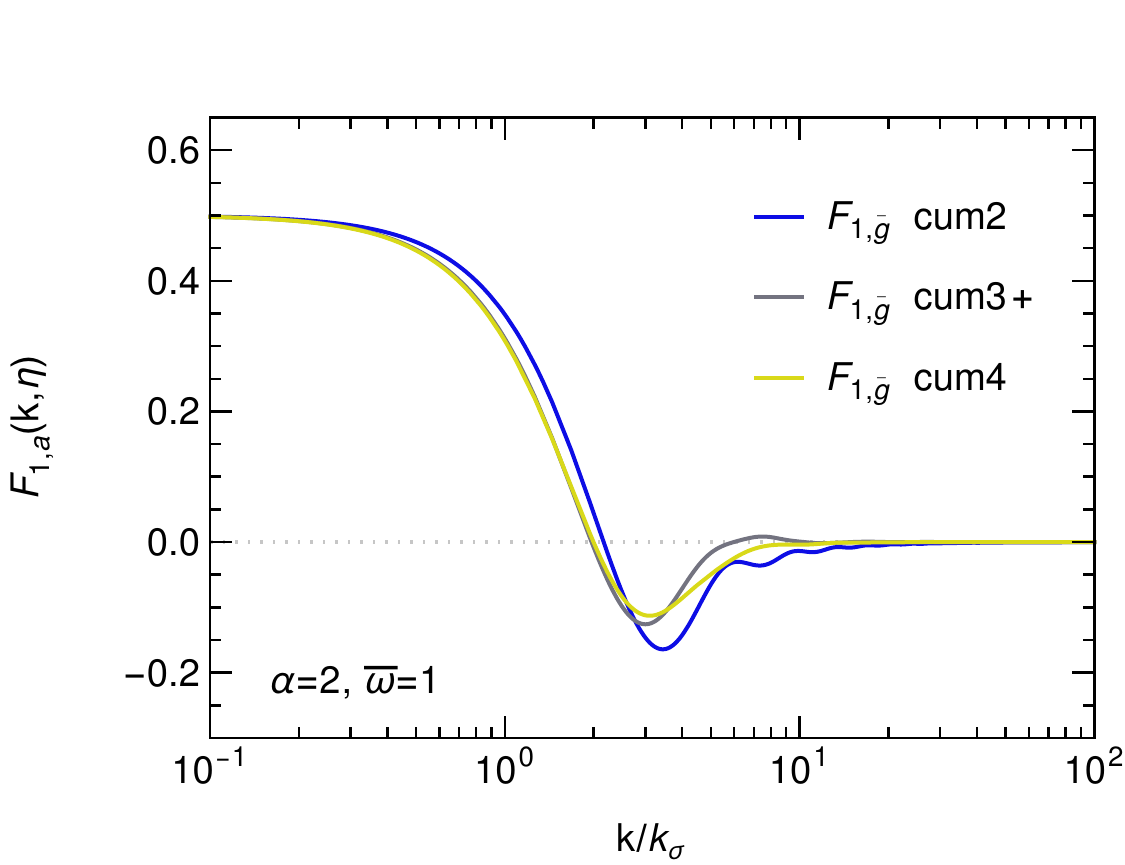}
  \end{center}
  \caption{\label{fig:Tcum_gbar}
  Linear \vpt~kernel $F_{1,\bar g}(k,\eta)$ when taking the second, third and fourth cumulant into account, respectively, for $\bar\omega=1$ and $\alpha=2$.}
\end{figure}

Let us now compare three approximations for the perturbation modes:
\begin{description}
  \item[cum2] second cumulant approximation ($\delta_k,\theta_k,\bar g_k,\delta\bar\epsilon_k$ and background dispersion $\epsilon(\eta)$)\,,
  \item[cum3+] third cumulant approximation for perturbation modes (+ $\bar\pi_k,\bar\chi_k$), and fourth cumulant approximation for expectation values (+ $\omega(\eta)$)\,,
  \item[cum4] fourth cumulant approximation (+ $\bar\kappa_k,\bar\xi_k,\bar\psi_k$)\,,
\end{description}
where in parenthesis we indicated the modes taken into account. Here (cum2) corresponds to the analytical result from Sec.\,\ref{sec:linanalyt}, and (cum3+) to 
neglecting $\bar\kappa_k,\bar\xi_k,\bar\psi_k$ in Eq.~\eqref{eq:cum4linrescale}, but keeping the expectation value $\bar\omega$ of the fourth cumulant.
Finally, (cum4) comprises the complete set Eq.~\eqref{eq:cum4linrescale}. In practice one can disregard $\bar\psi_k$ for (cum4)
in linear approximation, since its evolution equation is decoupled and this mode decays at all times. 
The same is true for $\delta\bar\epsilon_k$ for (cum2) as discussed in Sec.\,\ref{sec:linanalyt}.
Note that $\delta\bar\epsilon_k$ has to be included even at linear level for (cum3+) and (cum4) since this
mode of the dispersion tensor is sourced by the $\bar\chi_k$ perturbation of the third cumulant, see Eq.~\eqref{eq:cum4linrescale}.

As opposed to the second cumulant approximation (cum2), we solve the equations numerically when taking higher cumulants into account.
We compare the three linear approximations for the linear kernels $F_{1,\delta}$ and $F_{1,\theta}$ in Fig.~\ref{fig:Tcum}, for two values of $\bar\omega=\pm 1$.
The suppression on scales $k\lesssim 3k_\sigma$ is only weakly dependent on the higher cumulant perturbations, especially for negative $\bar\omega$,
with somewhat larger differences occurring in the damping tail. 
The numerical finding is also supported by an approximate analytical solution for small $k^2\epsilon$. We find that up to linear order
in $k^2\epsilon$ the linear kernel is not affected by higher cumulant perturbations, i.e. identical to Eq.~\eqref{eq:Tsmall} for
(cum2), (cum3+) and (cum4) for both $F_{1,\delta}$ and $F_{1,\theta}$.

For $F_{1,\bar g}$ (Fig.~\ref{fig:Tcum_gbar}) we find a mild shift when including the third cumulant, and a smaller difference between (cum3+) and (cum4).
Analytically, we find for small $k^2\epsilon$ (setting $\alpha=2$ for illustration)
\be
  F_{1,\bar g}\to \frac{1}{2}- k^2\epsilon\times \left\{\begin{array}{ll}
    \frac16 & \text{cum2}\,, \\
    \frac16 + \frac{1+4\bar\omega/15}{26} & \text{cum3+}\,, \\
    \frac16 + \frac{1+4\bar\omega/15}{26} + \frac{2\bar\omega}{273} & \text{cum4}\,, \\
  \end{array}\right.
\ee
with no further changes when including even higher (i.e. fifth or more) cumulants.

Altogether, even though the hierarchy of perturbation equations is coupled, we observe that for $\delta$ and $\theta$ the
transition region between the ideal fluid behavior for $k\ll k_\sigma$ and the strongly damped regime for $k\gg k_\sigma$ is only mildly dependent on
contributions from the third and fourth cumulant.

\section{Full hierarchy of cumulants}
\label{sec:hierarchy}

The formulation of perturbation theory up to the fourth cumulant presented in Secs.\,\ref{sec:sigma} and \ref{sec:pi} is suitable for a non-linear perturbative analysis.
Here we introduce an alternative formulation that will allow us to include an in principle arbitrary number of higher cumulants beyond fourth order, but is restricted to the linear approximation.
We use this approach in the following to study the convergence of the cumulant expansion at linear level.

For that purpose it is convenient to consider the Fourier decomposition of the cumulant generating function Eq.~\eqref{eq:delC},
\be
  \delta\bar{\cal C}(\eta,{\bm x},{\bm L}) = \int d^3k\, e^{i{\bm k}\cdot{\bm x}}\, \delta\bar{\cal C}(\eta,{\bm k},{\bm L})\,.
\ee
In the following we assume adiabatic cold dark matter initial conditions, being growing mode initial conditions for the density contrast and velocity divergence, and
vorticity and higher cumulants that vanish relative to the density contrast at early times.
Then, in linear approximation, all perturbation modes are proportional to the initial, linear density field $\delta_{k0}$.
In turn, this implies $\delta\bar{\cal C}(\eta,{\bm k},{\bm L})\propto \delta_{k0}$ for the generating function. The proportionality factor is described by a deterministic linear kernel that is given by the linearized equation of motion Eq.~\eqref{eq:dCbar}, while the dependence on the stochastic initial density field factors out. Due to rotational invariance, the linear kernel depends on the wavevector $\bm{k}$ and the auxiliary vector $\bm{L}$ only via their magnitudes $k$ and $L$ as well as the scalar projection $\mu\equiv {\bm k}\cdot{\bm L}/kL$.
We introduce the multipole decomposition
\be
  \delta\bar{\cal C}(\eta,{\bm k},{\bm L}) = \sum_\ell i^{-\ell}(2\ell+1){\cal C}_\ell(\eta,k,L) P_\ell(\mu)e^\eta\delta_{k0}\,,
\ee
where $P_\ell(\mu)$ are Legendre polynomials and
\be
  {\cal C}_\ell(\eta,k,L) \equiv i^\ell \int_{-1}^1\frac{d\mu}{2}\delta\bar{\cal C}(\eta,{\bm k},{\bm L})P_\ell(\mu)/(e^\eta\delta_{k0})\,.
\ee
One can express the first four multipoles in terms of the linear kernels of the scalar modes for the first four cumulants introduced above, see Eq.~\eqref{eq:Ta},
\bea\label{eq:Celldecomposition}
{\cal C}_0 &=& F_{1,A} +\frac{\epsilon L^2}{6}(F_{1,\bar g}+3F_{1,\delta\bar\epsilon})+\frac{\epsilon^2L^4}{5!}F_{1,\bar \kappa}+{\cal O}(L^6)\,,\nn\\
{\cal C}_1 &=& \frac{L}{k}\Bigg( \frac{F_{1,\theta}}{3} - \frac{\epsilon L^2}{30}F_{1,\bar \pi} +{\cal O}(L^4)\Bigg)\,,\nn\\
{\cal C}_2 &=& -\frac{\epsilon L^2}{15}F_{1,\bar g} +\frac{2\epsilon^2 L^4}{7\cdot 5!}(F_{1,\bar \kappa}-3F_{1,\bar \xi})+{\cal O}(L^6)\,,\nn\\
{\cal C}_3 &=& \frac{L}{k}\Bigg(   \frac{\epsilon L^2}{105}(F_{1,\bar \pi}-F_{1,\bar \chi}) +{\cal O}(L^4)\Bigg)\,,\nn\\
{\cal C}_4 &=& \frac{\epsilon^2 L^4}{9\cdot 7\cdot 5\cdot 3}(-4F_{1,\bar \kappa}+5F_{1,\bar \xi}+7F_{1,\bar \psi})+{\cal O}(L^6)\,.\nn\\
\eea
Each multipole contains a tower of higher cumulants multiplied by powers of $L^2$, with the lowest power being ${\cal C}_\ell\propto L^\ell$.
Note that the decomposition is limited to the linear approximation, for which only scalar modes contribute. Furthermore, we can replace the linear kernel of the log-density field via $F_{1,A}\to F_{1,\delta}$ in ${\cal C}_0$, as appropriate at linear order.

From the equation of motion Eq.~\eqref{eq:dCbar} of the cumulant generating function we obtain in the linear approximation (i.e. neglecting the quadratic term in $\delta\bar{\cal C}$)
\bea\label{eq:eomCell}
  \lefteqn{ \left[\partial_\eta+1+\left(\frac32\frac{\Omega_m}{f^2}-1\right)(L\cdot\partial_L)\right]{\cal C}_\ell }\nn\\
  &=&  \frac{k}{2\ell+1}\left[ 2\frac{\partial{\cal E}}{\partial L^2}L+\partial_L\right]\left((\ell+1){\cal C}_{\ell+1}-\ell{\cal C}_{\ell-1}\right)\nn\\
  && + \frac{k}{2\ell+1}\frac{1}{L}\left((\ell+1)(\ell+2)){\cal C}_{\ell+1}+\ell(\ell-1){\cal C}_{\ell-1}\right)\nn\\
  && - \frac{k}{3}LF_{1,\hat\Phi}\delta_{\ell 1}\,,
\eea
where 
\be
F_{1,\hat\Phi}\equiv \hat\Phi_k/(e^\eta\delta_{k0})=\frac32\frac{\Omega_m}{f^2}\frac{F_{1,\delta}}{k^2}=\frac32\frac{\Omega_m}{f^2}\frac{{\cal C}_0}{k^2}\Big|_{L=0}\,.
\ee
The background values of all cumulants enter via the function ${\cal E}(\eta,L^2)=\langle\bar{\cal C}\rangle$. The source term $Q_{\cal E}$ that enters its equation of
motion Eq.~\eqref{eq:E} can also be expressed in terms of the multipole moments,
\bea\label{eq:Qe}
  Q_{\cal E} &=& 4\pi \int_0^\infty dk\, k^3e^{2\eta}P_0(k)\,\sum_\ell (\ell+1)\Big(  {\cal C}_{\ell+1}\partial_L{\cal C}_\ell \nn\\
  && - {\cal C}_{\ell}\partial_L{\cal C}_{\ell+1}-\frac{2(\ell+1)}{L}{\cal C}_{\ell+1}{\cal C}_\ell\Big)\,,
\eea
where $P_0(k)$ is the conventional linear input power spectrum, and the impact of the Vlasov dynamics is encapsulated in the cumulants contained in the ${\cal C}_\ell$.
By inserting the decomposition Eq.~\eqref{eq:Celldecomposition} into Eq.~\eqref{eq:eomCell} and Taylor expanding up to the fourth order
in $L$ we recover the evolution equations Eq.~\eqref{eq:cum4lin} for the scalar perturbation modes of up to the fourth cumulant (when written in terms of the linear kernels from Eq.~\eqref{eq:Ta}). Similarly,
one recovers the equations Eq.~\eqref{eq:eomomegaepsilon} for the background dispersion $\epsilon$ and the
fourth cumulant background value $\omega$ by expanding Eq.~\eqref{eq:Qe} in $L$ and using Eqs.~(\ref{eq:E},~\ref{eq:Eexpansion},~\ref{eq:Plinab}) and Eq.~\eqref{eq:Celldecomposition}.

\subsection{Linear kernels beyond fourth cumulant order}

The above formulation allows us to include also higher cumulants beyond the fourth order.
It is convenient to scale out the leading $L^\ell$ dependence of ${\cal C}_\ell$, and a factor $1/k$ for odd $\ell$,
\be
  {\cal C}_\ell(\eta,k,L) = L^\ell \times \left\{\begin{array}{ll} \hat{\cal C}_\ell(\eta,k,L) & \ell\ \text{even}\,,\\
  \hat{\cal C}_\ell(\eta,k,L)/k & \ell\ \text{odd}\,.
  \end{array}\right.
\ee
To extract evolution equations up to a given order in the cumulant expansion, we expand
\be
  \hat{\cal C}_\ell(\eta,k,L) = \sum_n \frac{L^{2n}}{(2n)!}{\cal C}_{\ell, 2n}(\eta,k)\,,
\ee
where we used that due to the symmetry of the Legendre decomposition only even powers of $L$ appear.
The relation to the previous notation can be obtained using Eq.~\eqref{eq:Celldecomposition}, e.g.
\be
  {\cal C}_{0, 0}=F_{1,\delta},\quad {\cal C}_{1, 0}=F_{1,\theta}/3\,.
\ee
Similarly, as introduced already above in Eq.\,\eqref{Ecalcosmo}, we Taylor expand the ensemble-averaged values of the cumulants,
\be
  {\cal E}(\eta,L^2) = \sum_n \frac{L^{2n}}{(2n)!} {\cal E}_{2n}(\eta)\,,
\ee
such that the previously introduced background values of the $0$th, $2$nd and $4$th cumulant are given by
\be
  {\cal E}_0={\cal A},\ {\cal E}_2=\epsilon,\ {\cal E}_4=\frac{3\omega}{5}\,,
\ee
respectively. In general ${\cal E}_{2n}(\eta)$ denotes the ensemble expectation value of the $2n$th cumulant.
Note that due to statistical isotropy, only even cumulants can possess a non-zero expectation value.

\begin{figure}[t]
  \begin{center}
  \includegraphics[width=\columnwidth]{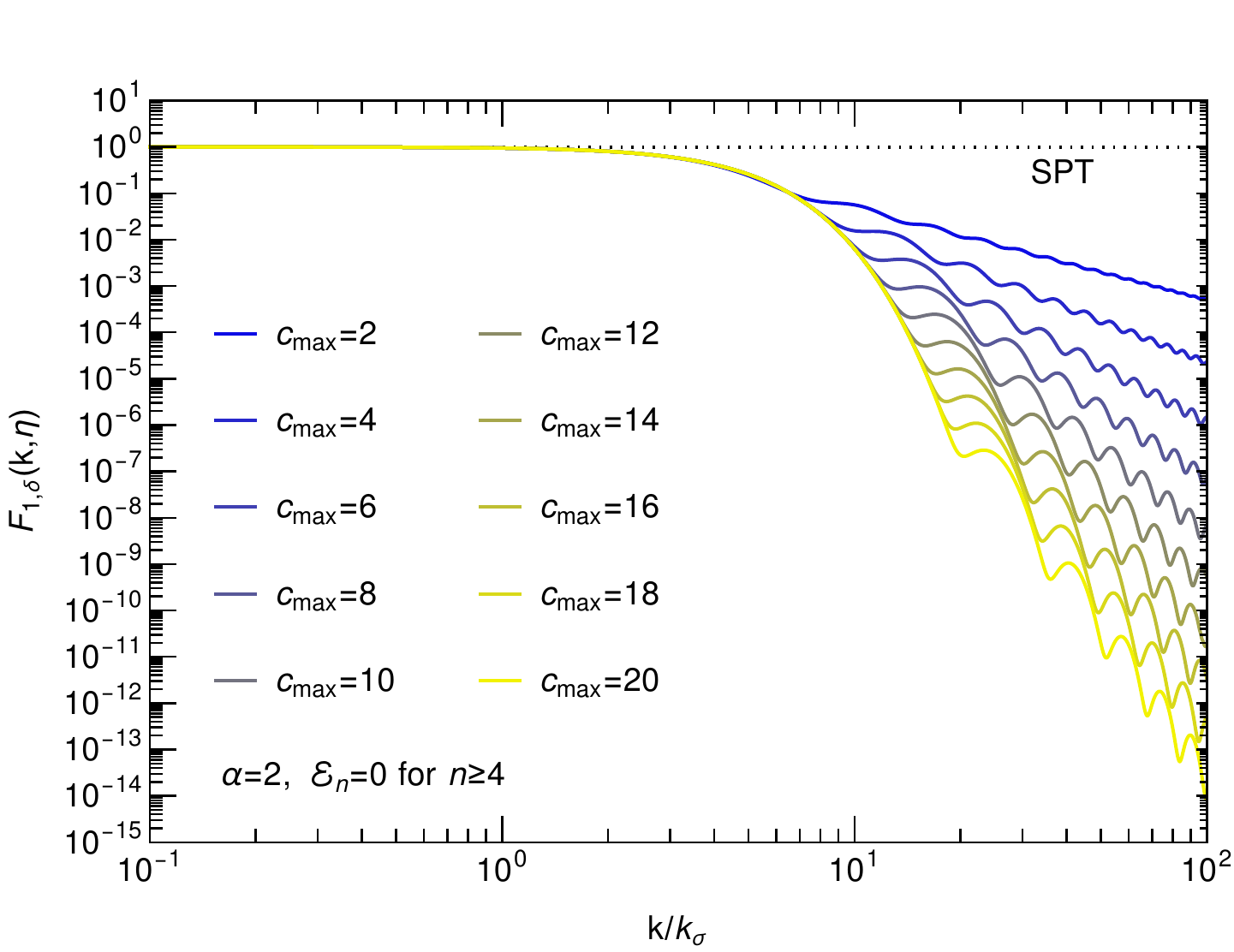}
  \end{center}
  \caption{\label{fig:Tcum_a2}
  Linear \vpt~kernel $F_{1,\delta}(k,\eta)={\cal C}_{0,0}$ when taking perturbation modes ${\cal C}_{\ell,2n}$ of cumulants up to $\ell+2n\leq c_\text{max}$
  into account. For this figure we set $\alpha=2$ and ${\cal E}_{2n}=0$ with $2n\geq 4$. As before, the time-dependence of the linear kernel is scaled out by normalizing the wavenumber to $k_\sigma=1/\sqrt{\epsilon}$. Note how the inclusion of higher cumulants only enhances the suppression of UV modes.}
\end{figure}

Inserting these expansions into the equation of motion Eq.~\eqref{eq:eomCell}
 one obtains a system of coupled, ordinary differential equations for
the ${\cal C}_{\ell, 2n}$, that is given explicitly in App.\,\ref{app:hierarchy}.

In the following we consider truncations that include perturbation modes up to a certain maximal cumulant order $c_\text{max}$, that is we include all
${\cal C}_{\ell, 2n}$ with 
\be
  \ell+2n\leq c_\text{max}\,,
\ee
and set those with higher values $\ell+2n>c_\text{max}$ to zero.
For truncations with $c_\text{max}\leq 4$, this corresponds to the approximation schemes considered in Sec.\,\ref{sec:fourthcumlin} as follows:
\be\begin{array}{ll}
  \text{(cum2)} & c_\text{max}=2\,,\nn\\
  \text{(cum3+)} & c_\text{max}=3\,,\nn\\
  \text{(cum4)} & c_\text{max}=4\,.
\end{array}\ee
Note that the evolution equations depend
on the background values ${\cal E}_{2n}$ with $2n\leq c_\text{max}+1$ for odd $c_\text{max}$, and $2n\leq c_\text{max}$ for even $c_\text{max}$.
For example, as noted previously, for $c_\text{max}=3$ the expectation value of the $4$th cumulant enters the
evolution equations of the $3$rd cumulant perturbation modes. Here we see that this pattern extends to higher cumulant orders correspondingly. 

Evolution equations for the ${\cal E}_{2n}$ can be obtained by Taylor expanding the source term given in Eq.~\eqref{eq:Qe} in powers of $L$.
They are also given in App.\,\ref{app:hierarchy}.

\medskip

As an illustrative example, we consider the solutions for the perturbation modes obtained when assuming that the expectation
values ${\cal E}_{2n}$ are taken as external input, while postponing self-consistent solutions of perturbations and background values to Sec.\,\ref{sec:powerlaw}.
Specifically, for concreteness we show in Fig.~\ref{fig:Tcum_a2} the linear kernel $F_{1,\delta}={\cal C}_{0, 0}$ for various $c_\text{max}$ when assuming ${\cal E}_2\equiv \epsilon=\epsilon_0\,e^{\alpha\eta}$
as previously, and in addition vanishing values for all ${\cal E}_{2n}$ with $2n\geq 4$. We checked that the solutions agree with (cum2), (cum3+) and (cum4) obtained in Sec.\,\ref{sec:linear} for $c_\text{max}=2,3,4$, respectively.

We observe that for any given wavenumber $k$, the linear \vpt~kernel for the density contrast converges to a common limit when increasing $c_\text{max}$. However, for higher wavenumbers a larger value of $c_\text{max}$ is required. The second cumulant approximation $c_\text{max}=2$ (denoted by (cum2) previously) is sufficient for $k\lesssim 7k_\sigma$, at which point the linear kernel is already suppressed by about a factor $10$ relative to its SPT value (being equal to unity). For higher wavenumbers, the linear kernel quickly drops. The $4$th order (cum4) approximation is close to the limiting value for $k\lesssim 9k_\sigma$, with a suppression of already around two orders of magnitude relative to SPT. For the highest order we consider, $c_\text{max}=20$, the linear kernel is converged for $k\lesssim 18k_\sigma$, corresponding to a suppression of $10^{-7}$. Therefore, while higher cumulant perturbations are important to capture the behavior for very large $k$, they only mildly affect the transition region between the ideal fluid regime and the onset of suppression within the linear approximation.

\subsection{Rescaling to dimensionless variables}
\label{subsec:rescaling}

When including background values ${\cal E}_{2n}$ of higher cumulants,  it is convenient to consider again the dimensionless quantities 
\be\label{eq:E2nbar}
  \bar{\cal E}_{2n}={\cal E}_{2n}/\epsilon^n={\cal E}_{2n}/{\cal E}_2^n\,,
\ee
and define dimensionless linear kernels 
\be\label{eq:Tl2n}
  T_{\ell,2n}={\cal C}_{\ell,2n}/\epsilon^{n+[\ell/2]}\,,
\ee
where $[\ell/2]=\ell/2$ for even $\ell$, and $(\ell-1)/2$ for odd $\ell$. The evolution equations for these rescaled variables are given in App.\,\ref{app:hierarchy}. 

Collecting all perturbation variables $T_{\ell,2n}$ with $\ell+2n\leq c_\text{max}$ into a single vector $\bar\psi$, the evolution equations may be brought into the form
\be\label{eq:psibareomhierarchy}
  \bar\psi'+(\Omega_0+\epsilon k^2\Omega_1)\bar\psi=0\,,
\ee
that is formally analogous to Eq.~\eqref{eq:psibareom}. The number of rows and columns of the matrices $\Omega_0$ and $\Omega_1$ equals the number of all scalar perturbation modes for a given $c_\text{max}$, being given by $4,6,9,12,16$ for $c_\text{max}=2,3,4,5,6$, respectively. We emphasize that $c_\text{max}$ denotes the truncation order for the cumulant expansion of perturbation modes collected in $\bar\psi$. The background values of higher cumulants $\bar{\cal E}_{2n}$ with $2n\geq 4$ enter the matrices $\Omega_0$ and $\Omega_1$. As mentioned above, for even $c_\text{max}$, the background values $\bar{\cal E}_{4}, \bar{\cal E}_{6},\dots\bar{\cal E}_{c_\text{max}}$ contribute, while for odd $c_\text{max}$, the equations for the perturbation modes depend on $\bar{\cal E}_{4}, \bar{\cal E}_{6},\dots\bar{\cal E}_{c_\text{max}+1}$.

Let us assume again that the background dispersion has a power-law dependence on the linear growth factor, with constant $\alpha=\partial_\eta\ln\epsilon$, and set $\Omega_m/f^2\to 1$. Furthermore, we assume for definiteness that the ratios $\bar{\cal E}_{2n}$ are constant in time. As will be seen in Sec.\,\ref{sec:powerlaw}, this assumption is consistent for a scaling universe, and may serve as a basis for a more general treatment in the future. In that case all entries of $\Omega_0$ and $\Omega_1$ are constant in time, and the time-dependence is entirely given by the factor $\epsilon k^2$. 

\subsection{Scaling in the limit $\epsilon\to 0$}
\label{subsec:epsscaling}

Since we assume initial conditions for the perturbation modes with vanishing second and higher cumulants, all higher cumulant modes can only be generated due to the
presence of the background dispersion $\epsilon$, as well as the background values of higher cumulants. Therefore, one expects the higher cumulant perturbations to
vanish with a certain power of $\epsilon$ in the limit $\epsilon\to 0$, and when assuming the dimensionless ratios $\bar{\cal E}_{2n}$ to remain finite.
Indeed, it turns out that as expected higher cumulants are more strongly suppressed for small $\epsilon$. In order to see this, we consider
the solutions of Eq.~\eqref{eq:psibareomhierarchy} in the limit $\epsilon k^2\ll 1$.
They are determined by the eigenmodes of $\Omega_0$. For any $c_\text{max}\geq 1$ the eigenmodes comprise the usual growing and decaying mode familiar from SPT, as well as further decaying modes for $c_\text{max}\geq 2$. 
Inspecting the evolution equation Eq.~\eqref{eq:eomTl2n} one finds that all $T_{\ell,2n}$ with even $\ell$ have a Taylor expansion in powers of $\epsilon k^2$ that starts with a constant term, while those with odd $\ell$ involve at least one factor of $\epsilon k^2$.
The only exception is $\ell=1, n=0$, related to the velocity divergence $T_{1,0}=F_{1,\theta}/3=1/3+{\cal O}(\epsilon k^2)$. The reason is the extra term in its evolution equation corresponding to the gravitational force in the Euler equation.
Together with Eq.~\eqref{eq:Tl2n} this implies the counting
\be
  {\cal C}_{\ell,2n} \propto \left\{\begin{array}{ll}
    \epsilon^{(\ell+2n)/2}[1+{\cal O}(\epsilon k^2)+\dots] & \ell\ \text{even}\,,\\
    \epsilon^{0}[1+{\cal O}(\epsilon k^2)+\dots] & \ell=1, n=0\,,\\
    \epsilon^{(\ell-1+2n)/2}[{\cal O}(\epsilon k^2)+\dots] & \ell\ \text{odd}, \ell+2n\geq 3\,,\\
  \end{array}\right.
\ee
which shows that higher cumulants of order $c=\ell+2n$ are suppressed by higher powers of the background dispersion $\epsilon$
in the limit $\epsilon\to 0$. The counting assumes that the $\bar{\cal E}_{2n}$ are parametrically of order unity in this limit,
which implies ${\cal E}_{2n}\propto \epsilon^n$ for the background values of cumulant order $2n$. This is consistent with the scaling of
the perturbation modes ${\cal C}_{\ell,2n}$ of the same cumulant order. By inspecting the evolution equation Eq.~\eqref{eq:dCbar} of the cumulant
generating function, we find that this result can be generalized to the following scaling of the leading contribution in the limit $\epsilon\to 0$:
\bea
  \delta,A,\theta &\propto& {\cal O}(\epsilon^0)\,,\nn\\
  w_i=(\vec\nabla\times \vec u)_i &\propto& {\cal O}(\epsilon^1)\,,\nn\\
  \epsilon_{ij} &\propto& {\cal O}(\epsilon^1)\,,\nn\\
  {\cal C}_{ijk} &\propto& {\cal O}(\epsilon^2)\,,\nn\\
  {\cal C}_{ijkl} &\propto& {\cal O}(\epsilon^2)\,,\nn\\
  {\cal C}_{ijklm} &\propto& {\cal O}(\epsilon^3)\,,\nn\\
  {\cal C}_{ijklmn} &\propto& {\cal O}(\epsilon^3)\,,
\eea
and so on for cumulants of order $0,1,2,3,4,5,6,\dots$.
Here the only exception is $\theta$ that is contributing already at zeroth order in $\epsilon$ as discussed above, while being of cumulant order one.
Note that the remaining part of the first cumulant, i.e. the vorticity, has the ``generic'' scaling since the gravitational
force does not contribute to the vorticity equation. 

Since the equations of motion couple the various cumulants, the self-consistency of this hierarchy is not obvious, but, as can be checked using Eq.~\eqref{eq:dCbar}, indeed holds in general. In particular, it remains valid beyond the linear approximation, and holds also for the fully non-linear system, i.e. the complete Vlasov hierarchy.

\subsection{Stability conditions}
\label{subsec:stability}

Within collisional fluid dynamics, fluid perturbations are damped due to microscopic pressure and viscosity. In contrast, for the collisionless Vlasov system underlying the cumulant evolution equations derived above this is in general not guaranteed. As we shall show below, the linearized system of coupled cumulants indeed may develop instabilities. In the following we show under which circumstances this occurs, and derive stability conditions.

When including background values ${\cal E}_{2n}$ of higher cumulants, the linear kernels remain qualitatively similar to those shown in Fig.~\ref{fig:Tcum_a2} provided the dimensionless quantities $\bar {\cal E}_{2n}$ (see Eq.~\ref{eq:E2nbar}) are not too large in magnitude. However, for sizeable values of the $\bar {\cal E}_{2n}$, the linear kernels develop an exponential instability for large $\epsilon k^2$. Even though this behavior may in principle be cured when including non-linearities, the resulting dynamics would be outside the realm of perturbation theory. We therefore require that no such exponential growth occurs. This imposes restrictions on the magnitude of the higher cumulant expectation values.

To make this statement quantitative, we investigate the solutions of Eq.~\eqref{eq:psibareomhierarchy} in the limit $\epsilon k^2\gg 1$.
The asymptotic behavior can be obtained by taking a further $\eta$ derivative and using that $\Omega_1$ is a nilpotent matrix, with $\Omega_1\cdot\Omega_1=0$ in the matrix sense, giving $\bar\psi''-[\Omega_0\cdot\Omega_0+\epsilon k^2(\Omega_0\cdot\Omega_1+\Omega_1\cdot\Omega_0-\alpha\Omega_1)]\bar\psi=0$. Furthermore we switch variables from $\eta$ to 
\be
  s_k\equiv \sqrt{3\epsilon(\eta)k^2}\,,
\ee
using $\alpha=\partial_\eta\ln\epsilon=$const., giving
\be\label{detads}
\partial_\eta = \frac12(\partial_\eta\ln\epsilon)s_k\partial_{s_k}=\frac{\alpha}{2}s_k\partial_{s_k}\,.
\ee
Altogether, Eq.~\eqref{eq:psibareomhierarchy} can be rewritten as
\bea
  \Bigg[\frac{\alpha^2}{4}s_k^2\partial^2_{s_k}-\Omega_0\cdot\Omega_0-\frac{\alpha}{2}\Omega_0 && \nn\\
  -\frac{s_k^2}{3}\left(\Omega_0\cdot\Omega_1+\Omega_1\cdot\Omega_0-\frac{\alpha}{2}\Omega_1\right)\Bigg]\bar\psi &=& 0\,.
\eea
For $s_k\gg 1$, the solution is given by a linear combination of eigenmodes with time-dependence given by 
\be
  T_{\ell,2n} \propto e^{\pm 2\sqrt{\lambda}s_k/\alpha}\qquad \text{for}\ s_k\gg 1\,,
\ee
where $\lambda$ are the eigenvalues of the matrix
\be
  M_{c_\text{max}} \equiv \frac{1}{3}\left(\Omega_0\cdot\Omega_1+\Omega_1\cdot\Omega_0-\frac{\alpha}{2}\Omega_1\right)\,.
\ee
The solutions therefore possess an exponential instability if $\sqrt{\lambda}$ has a non-zero real part for any of the eigenvalues.
The absence of this instability requires that all eigenvalues are real and smaller or equal to zero, 
\be
  \text{Im}(\lambda)=0,\quad \text{Re}(\lambda)\leq 0\,.
\ee
This statement is equivalent to the condition that all roots of the characteristic polynomial
\be
  p_{c_\text{max}}(\lambda)\equiv\text{det}(\lambda\, {\bm 1}-M_{c_\text{max}})\,,
\ee
are zero or lie on the negative real axis. For the truncations up to the fourth cumulant we find
\bea
  p_1 &=& (1/3+\lambda)^2\,,\nn\\
  p_2 &=& \lambda^2(1+\lambda)^2\,,\nn\\
  p_3 &=& (1/3+\lambda)^2\left(\lambda^2+2\lambda+\frac19(3-\bar{\cal E}_4)\right)^2\,,\nn\\
  p_4 &=& \lambda^3(1+\lambda)^2\left(\lambda^2+\frac{10}{3}\lambda+\frac59(3-\bar{\cal E}_4)\right)^2\,.
\eea
Remarkably, the characteristic polynomials are independent of $\alpha=\partial_\eta\ln\epsilon$. We find that this property extends also to higher $c_\text{max}$.
This implies that any constraints from stability are insensitive to the time-dependence of the background dispersion. In addition, there is no restriction
from stability on the size of the background dispersion $\epsilon\equiv {\cal E}_2$ itself. Thus, only fourth and higher cumulant expectation values are subject to stability conditions.

For $c_\text{max}=1,2$ the stability condition is always satisfied. 
We note that for $c_\text{max}=2$ (equivalent to (cum2)) the two solutions $\lambda=0,-1$
precisely correspond to the exponential factors in the time-dependence found in the asymptotic limit given in Eq.~\eqref{eq:Tlarge2} of the analytical solution of the linear kernel.
In addition there is a power-law dependence on $s_k$ that is not captured by the leading asymptotic solution considered above.
Note that for $c_\text{max}=1$ a Jeans-like term proportional to
$\epsilon$ is contained in the perturbation equations, leading to a non-trivial linear kernel even in that case. 
The negative root $\lambda=-1/3$ implies that the linear kernels exhibit oscillations in the limit $\epsilon k^2\gg 1$ for this most restrictive truncation.

For $c_\text{max}=3,4$ (equivalent to (cum3+) and (cum4), respectively), the expectation value $\bar{\cal E}_4=3\bar\omega/5$ of the $4$th cumulant enters the equations of motion.
The condition that all roots are real and negative or zero imposes a restriction on the size of $\bar{\cal E}_4$,
\bea\label{eq:E4condition}
  c_\text{max}=3: &\qquad& -6\leq \bar{\cal E}_4 \leq 3\,,\nn\\
  c_\text{max}=4: &\qquad& -2\leq \bar{\cal E}_4 \leq 3\,.
\eea
When including even higher cumulant perturbations, we find that the characteristic polynomials possess a recursive structure,
\be
  p_{c_\text{max}}(\lambda) = \left\{\begin{array}{ll}
  \lambda p_{c_\text{max}-2}(\lambda)\,q_{c_\text{max}}(\lambda)^2 & c_\text{max}\ \text{even}\,,\\
  p_{c_\text{max}-2}(\lambda)\,q_{c_\text{max}}(\lambda)^2 & c_\text{max}\ \text{odd}\,.
  \end{array}\right.
\ee
This implies that the roots of $p_{c_\text{max}-2}$ are also roots of $p_{c_\text{max}}$ for any $c_\text{max}>2$.
In addition, for even $c_\text{max}$, the solution $\lambda=0$ corresponds to a root with
higher multiplicity. The new roots that appear when increasing $c_\text{max}$ by two units are described by an additional
factor, that can be written as the square of a polynomial $q_{c_\text{max}}(\lambda)$ given in Table~\ref{tab:Tasymptotichierarchy} in App.\,\ref{app:hierarchy} up to $c_\text{max}=12$.
They are of order $N_{c_\text{max}}=2,2,3,3,4,4,\dots$ for $c_\text{max}=3,4,5,6,7,8,\dots$.

The recursive structure implies for the set of eigenvalues that appear at a given truncation order
\be
  \{\lambda\}_{c_\text{max}-2}\subset\{\lambda\}_{c_\text{max}}\subset\{\lambda\}_{c_\text{max}+2} \cdots \,.
\ee
This means that the stability conditions of the linear solution obtained for  given even (odd) $c_\text{max}$ continue to hold at all higher even (odd) values of $c_\text{max}$, with additional
conditions arising from the additional eigenvalues given by the roots of $q_{c_\text{max}}(\lambda)$.

A \emph{necessary} condition for stability is that no real and positive roots exist. This is ensured if all coefficients of $q_{c_\text{max}}(\lambda)$
have the same sign, according to the Descartes sign rule (which can easily be proven by contradiction in that case). Inspecting Table~\ref{tab:Tasymptotichierarchy}, this leads to the conditions
\bea\label{eq:E2ncond}
  \bar{\cal E}_4 &\leq& 3\,,\nn\\
  \bar{\cal E}_6 &\geq& 15(\bar{\cal E}_4-1)\,,\\
  \bar{\cal E}_8 &\leq& 105 - 210 \bar{\cal E}_4 + 35 \bar{\cal E}_4^2 + 28 \bar{\cal E}_6 \,,\nn\\
  \bar{\cal E}_{10} &\geq& 45 \bar{\cal E}_8  + 210 \bar{\cal E}_4 (15 + \bar{\cal E}_6) - 945 - 1575 \bar{\cal E}_4^2 - 630 \bar{\cal E}_6 \,, \nn\\
  \bar{\cal E}_{12} &\leq& 10395 + 51975 \bar{\cal E}_4^2 - 5775 \bar{\cal E}_4^3 + 13860 \bar{\cal E}_6 + 462 \bar{\cal E}_6^2 \nn\\
  && -  495 \bar{\cal E}_4 (105 + 28 \bar{\cal E}_6 - \bar{\cal E}_8) - 1485 \bar{\cal E}_8 + 66 \bar{\cal E}_{10} \,,\nn
\eea
for all values of $c_\text{max}$ for which the corresponding expectation values enter, being $c_\text{max}\geq 3,5,7,9,11$ for the five inequalities, respectively.
The first condition is consistent with Eq.~\eqref{eq:E4condition}. These conditions are however not sufficient.

A \emph{sufficient} set of stability conditions can be obtained by applying an algorithm known as Sturm chain.
To that end we define a set of polynomials $P_n(\lambda)$, starting from 
\be
  P_0(\lambda)=q_{c_\text{max}}(\lambda),\quad P_1(\lambda)=dq_{c_\text{max}}/d\lambda\,.
\ee
Then, we recursively compute the polynomial quotient $Q_n(\lambda)$ from polynomial division of $P_n$ by $P_{n+1}$, and define 
\be
  P_{n+2}=Q_nP_{n+1}-P_n\,,
\ee
being the rest term up to an overall sign.
The order of the polynomials decreases with increasing $n$, and at some point one obtains a constant, where the chain is terminated.
For all $c_\text{max}$ we considered, this is the case for $n$ being equal to the order of the polynomial $q_{c_\text{max}}(\lambda)$, being
\be
  N_{c_\text{max}}\equiv [(c_\text{max}+1)/2]\,.
\ee
Here the square bracket denotes the integer part.
The number of roots of $q_{c_\text{max}}(\lambda)$ in the intervall $a<\lambda\leq b$ for some real values $a<b$ is then given by $\sigma(a)-\sigma(b)$,
where $\sigma(\lambda)$ is the number of sign changes in the series $P_0(\lambda),P_1(\lambda),\dots,P_{N_{c_\text{max}}}(\lambda)$.

Stability requires that all $N_{c_\text{max}}$ roots of $q_{c_\text{max}}(\lambda)$ lie in the intervall $-\infty<\lambda\leq 0$. 
We therefore consider the choice $a\to -\infty$ and $b=0$, and require $\sigma(a)-\sigma(b)=N_{c_\text{max}}$. The only way how this condition can be
satisfied is if $\sigma(a)=N_{c_\text{max}}$ and $\sigma(b)=0$. For $\lambda\to-\infty$ the coefficient of the highest monomial in each $P_n$ determines
its sign, and for $\lambda=0$ the constant term. We write the polynomials in the form 
\be
  P_n(\lambda)=C_n\lambda^{N_{c_\text{max}}-n}+\dots+D_n\,,  
\ee
where the ellipsis denotes summands with powers $\lambda^m$ with $0<m<N_{c_\text{max}}-n$, that are irrelevant here. Due to the alternating sign of $\lambda^{N_{c_\text{max}}-n}$ for $n=0,\dots,N_{c_\text{max}}$ and $\lambda<0$, all coefficients
$C_n$ are required to have the same sign in order to satisfy $\sigma(-\infty)=N_{c_\text{max}}$. Furthermore, by definition of the characteristic polynomial, we have $C_0=+1$. This implies the conditions that $C_n\geq 0$ for all $n=0,\dots,N_{c_\text{max}}$. Similarly, the constraint $\sigma(0)=0$ implies that all $D_n$ have to have the same sign. In principle they could all be positive or all be negative. However, the latter can be excluded by the following argument: For the constant polynomial with $n=N_{c_\text{max}}$, the term with highest and lowest power of $\lambda$ are trivially identical, i.e. $C_{N_{c_\text{max}}}=D_{N_{c_\text{max}}}$. Since we already obtained the condition that all $C_n$ need to be non-negative, this implies that also all $D_n$ have to be non-negative for stability to hold,
\be\label{eq:stability}
  C_n\geq 0,\quad D_n\geq 0,\qquad 0\leq n\leq N_{c_\text{max}}\,.
\ee
As mentioned above, the condition is trivially satisfied for $C_0=1$ and degenerate for $C_n=D_n$ for $n=N_{c_\text{max}}$.
There are therefore in general $2N_{c_\text{max}}$ distinct conditions. However, some of them are either trivially satisfied or equivalent.
For example, for $c_\text{max}=4$ one has $N_{c_\text{max}}=2$ and we find $C_1=2, C_2=D_2=\frac59(2+\bar{\cal E}_4)$ and $D_0=\frac59(3-\bar{\cal E}_4), D_1=\frac{10}{3}$.
The stability conditions Eq.~\eqref{eq:stability} therefore precisely yield the constraint Eq.~\eqref{eq:E4condition} obtained previously for $c_\text{max}=4$.
The same can be checked for $c_\text{max}=3$.

In addition, the Sturm chain algorithm allows us to obtain stability conditions for arbitrary truncation order $c_\text{max}$.
We observe a number of general patterns: We find that $C_1=N_{c_\text{max}}$, such that the condition $C_1\geq 0$ is trivially satisfied. From the condition $C_2\geq 0$ we obtain a lower bound
on $\bar{\cal E}_4$ for all $c_\text{max}\geq 3$,
\be\label{eq:E4lowerbound}
  \bar{\cal E}_4 \geq \left\{\begin{array}{l@\qquad l}
  -\frac{6}{c_\text{max}-2} & c_\text{max}\ \text{odd}\,,\\[1.5ex]
  -\frac{6}{c_\text{max}-1} & c_\text{max}\ \text{even}\,.\\
  \end{array}\right.
\ee
This is a generalization of the lower bound obtained in Eq.~\eqref{eq:E4condition}.\footnote{The same condition can alternatively be derived from the Laguerre-Samuelson rule, being that the generalized discriminant is positive,
$q_{c_\text{max}}^{(2)}q_{c_\text{max}}^{(0)}\frac{2N_{c_\text{max}}}{N_{c_\text{max}}-1}\leq \left( q_{c_\text{max}}^{(1)} \right)^2$, where $q_{c_\text{max}}(\lambda) = \sum_n q_{c_\text{max}}^{(n)} \lambda^{N_{c_\text{max}}-n}$.
However, the Laguerre-Samuelson rule is necessary but not sufficient for stability, while the full set of conditions derived from the Sturm chain are sufficient. We checked Eq.~\eqref{eq:E4lowerbound} up to $c_\text{max}=12$.}

Furthermore, the condition obtained from $D_0\geq 0$ is equivalent to the first line in Eq.~\eqref{eq:E2ncond} for $c_\text{max}=3,4$, to the second line for $c_\text{max}=5,6$, the third line for $c_\text{max}=7,8$, and so on.
Similarly, $D_1\geq 0$ yields the first line in Eq.~\eqref{eq:E2ncond} for $c_\text{max}=5,6$, the second line for $c_\text{max}=7,8$, etc.

\begin{figure}[t]
  \begin{center}
  \includegraphics[width=\columnwidth]{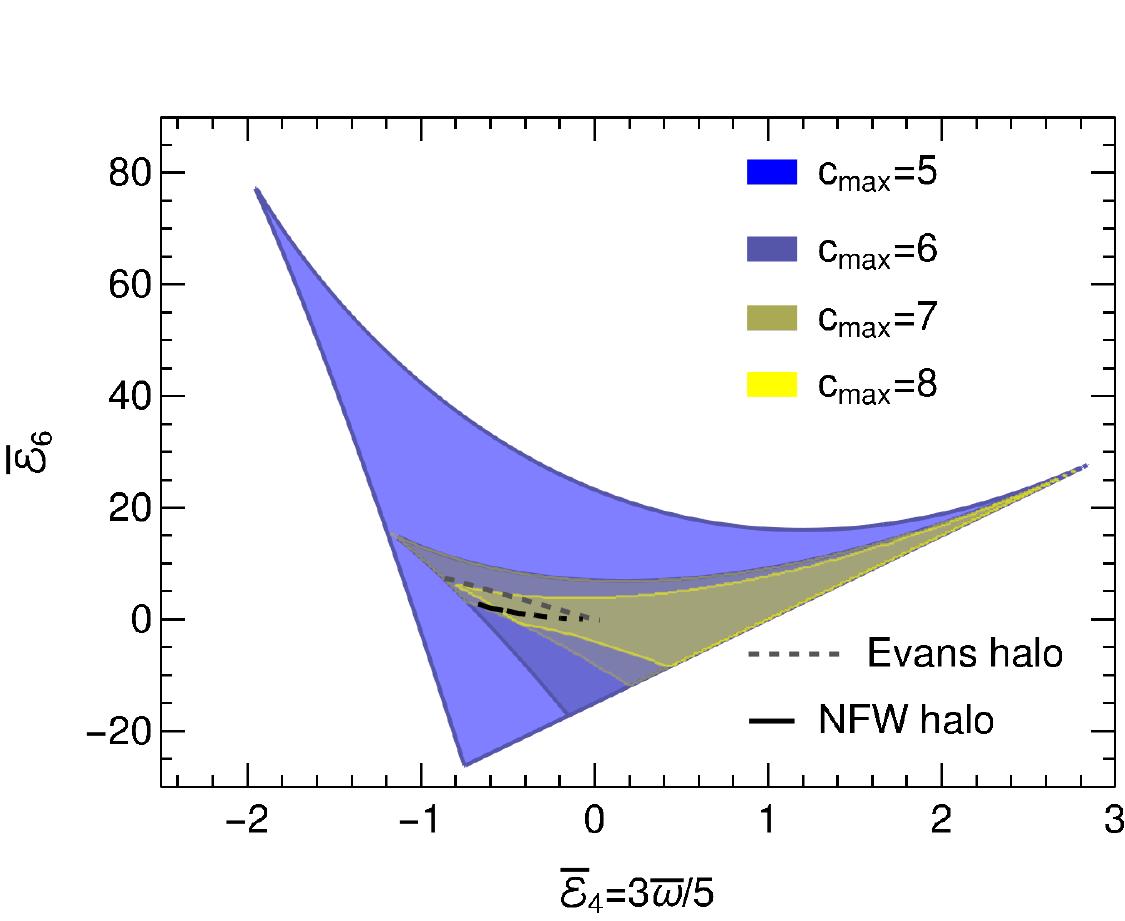}
  \\
  \includegraphics[width=\columnwidth]{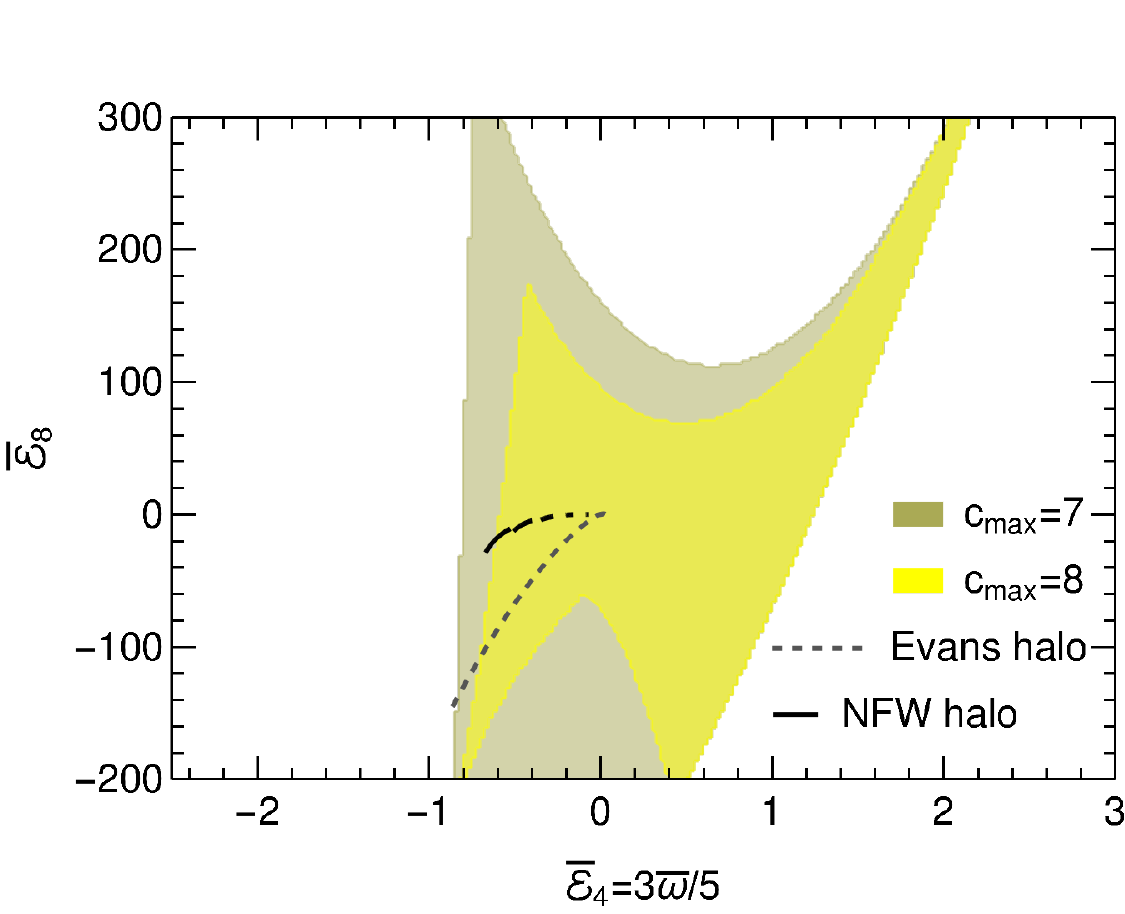}
  \end{center}
  \caption{\label{fig:cum4vs6} Constraints on the non-Gaussianity of the distribution function from requiring stability of the linear solutions. The top panel shows the parameter region allowed by stability within the plane spanned by the dimensionless $4$th and $6$th cumulant expectation values $(\bar{\cal E}_4,\bar{\cal E}_6)$,   while the bottom panel shows constraints on the $4$th and $8$th cumulant expectation values $(\bar{\cal E}_4,\bar{\cal E}_8)$. In each case, these are obtained from the evolution equations for perturbations modes of cumulants up to $c_\text{max}$ as given in the legend. Note that  $\bar{\cal E}_6$ is relevant for $c_\text{max}\geq 5$, and $\bar{\cal E}_8$ for $c_\text{max}\geq 7$. The grey dashed and black lines show the expectation from the Evans and NFW halo  models, see Sec.\,\ref{sec:halo}. The fact that  the backreaction on linear modes from dispersion and higher cumulants expected from halos is broadly stable is reassuring, making linearized \vpt~a good starting point for a perturbative expansion.
}      
\end{figure}

The complete set of stability conditions obtained from the Sturm chain algorithm for $c_\text{max}=5$ reads
\bea\label{eq:cmax5}
  -2 &\leq& \bar{\cal E}_4 \leq 3\,,\nn\\
  15(\bar{\cal E}_4-1) &\leq& \bar{\cal E}_6 \leq 10(6-\bar{\cal E}_4)\,,\\
  0 &\leq& 100(24 +12\bar{\cal E}_4 -6 \bar{\cal E}_4^2 + 5 \bar{\cal E}_4^3) \nn\\
  && - 40 \bar{\cal E}_6 (2+3\bar{\cal E}_4) - \bar{\cal E}_6^2\,.\nn
\eea
For $c_\text{max}=6$ we obtain
\bea\label{eq:cmax6}
  -\frac65 &\leq& \bar{\cal E}_4 \leq 3\,,\nn\\
  15(\bar{\cal E}_4-1) &\leq& \bar{\cal E}_6 \leq 10(2+\bar{\cal E}_4/3)\,,\\
  0 &\leq& 20(216+ 324\bar{\cal E}_4 + 90\bar{\cal E}_4^2 + 175\bar{\cal E}_4^3) \nn\\
  && - 108 \bar{\cal E}_6 (4+ 10\bar{\cal E}_4)  -  27 \bar{\cal E}_6^2\,.\nn
\eea
The conditions for $c_\text{max}=7,8$ are given in App.\,\ref{app:hierarchy}. 
We checked that vanishing expectation values $\bar{\cal E}_{2n}=0$ for $2n\geq 4$ do satisfy
the stability conditions in all cases. Therefore, in general, stability sets an upper limit on the magnitude of the $\bar{\cal E}_{2n}$, i.e. how strongly non-Gaussian the distribution function can be (on average).

Let us now discuss the impact of stability constraints. For $c_\text{max}=1,2$ the stability conditions are trivially satisfied, and for $c_\text{max}=3,4$ they amount to the constraint
on the size of $\bar{\cal E}_4$ given in Eq.~\eqref{eq:E4condition}. 
For $c_\text{max}=5, 6$ the perturbation equations depend on $\bar{\cal E}_4$ and $\bar{\cal E}_6$, and the stability constraints given in Eqs.~(\ref{eq:cmax5},~\ref{eq:cmax6}) are satisfied within a finite region in the two-dimensional $(\bar{\cal E}_4,\bar{\cal E}_6)$ parameter space, shown in Fig.~\ref{fig:cum4vs6} (top panel). We see that the point
$\bar{\cal E}_{4}=\bar{\cal E}_{6}=0$ lies within the stable region, as claimed aboved. Furthermore, the allowed region is more restricted for $c_\text{max}=6$ compared to $c_\text{max}=5$.

For $c_\text{max}=7, 8$, also $\bar{\cal E}_8$ enters in the perturbation equations. Stability yields an allowed region in the three-dimensional parameter space $(\bar{\cal E}_4,\bar{\cal E}_6,\bar{\cal E}_8)$, that contains the origin. In Fig.~\ref{fig:cum4vs6} we show the projection of this region on the $(\bar{\cal E}_4,\bar{\cal E}_6)$ as well as $(\bar{\cal E}_4,\bar{\cal E}_8)$ plane in the top and bottom panel, respectively. As expected, the allowed regions for higher $c_\text{max}$ are contained in those for lower values of $c_\text{max}$.
Generically, one may expect the normalized quantities $\bar{\cal E}_{2n}$ to be broadly of order unity, which is allowed by stability. Nevertheless, the stability conditions impose relevant constraints, in particular for $\bar{\cal E}_{4}$, as well as correlations among the relative size of the $\bar{\cal E}_{2n}$.

It is instructive to compare the stability regions of the cumulant expectation values
with those obtained within the Evans and NFW halo models discussed in Sec.\,\ref{sec:halo}, that are shown as grey dashed and black lines in Fig.~\ref{fig:cum4vs6}, respectively.  Since the stability conditions for the linearized Vlasov hierarchy and the averaged cumulants obtained from virialized halos are based on  quite distinct physical situations, it is remarkable that the latter are largely contained within the stable regions for a wide range of halo concentrations (NFW halos) and different halo shapes (Evans halos). Note the Evans halos hit the stability boundary when the shape parameter $q$ approaches the unphysical limit of extreme oblateness $q=1/\sqrt{2}$, where the density ceases to be positive definite. For NFW halos, the stability boundary is hit for high mass halos, corresponding to high concentrations which are the most non-Gaussian (See Fig.~\ref{fig:E4E6E8NFW}). But this is unrealistic since in practice we have a spectrum of halos, and the abundance of high-mass halos is exponentially suppressed, therefore one must integrate over the mass function to compare properly. The advantage of using the results of individual halos in Fig.~\ref{fig:cum4vs6} is that we are insensitive to the initial power spectrum shape, therefore these results apply broadly to scale-free power spectra as well as CDM spectra. The dependence on the initial spectrum enters only through the weight given to different halo masses (or concentrations) by the mass function. That the constrains on non-Gaussianity of the distribution function are satisfied by halo estimates 
 implies that the linear solutions within \vpt~can be considered as a good starting point for a perturbative analysis for realistic values of the higher cumulant expectation values.

\section{Dispersion  in a scaling universe}
\label{sec:powerlaw}

So far, we treated the background dispersion $\epsilon(\eta)={\cal E}_2$ as well as the
expectation value $\omega(\eta)=5{\cal E}_4/3$ of the fourth cumulant, and those of yet higher cumulants (${\cal E}_{2n}$), as external inputs for solving the
equations for perturbation modes up to a certain cumulant order. In this section we return to the
Eqs.~(\ref{eq:epseom},~\ref{eq:eomcumexpectation},~\ref{eq:E},~\ref{eq:Qe}) for the background quantities themselves,
that are in turn sourced by the fluctuations of the perturbation modes, and present self-consistent solutions of the perturbation and background equations in various approximations.
For illustration, we restrict ourselves to a scaling universe in this work, for which the differential equations
for the background values turn into algebraic equations, as we shall see. This makes the problem tractable and allows us to study the dependence
on the truncation of the cumulant expansion.

A scaling universe is characterized by a power-law initial spectrum,
\be
  P_0(k) = A\,k^{n_s}\,,
\ee
with spectral index $n_s$, and an EdS background ($\Omega_m=1$). The linear power spectrum in the SPT approximation is given by
\be
  P^\text{lin}_\text{SPT}(k,\eta)=e^{2\eta} P_0(k)\,,
\ee
where $e^{2\eta}=D^2$ is the square of the conventional linear growth factor, which in turn equals the scale-factor within EdS.
For the dimensionless power spectrum $\Delta\equiv 4\pi k^3P$ this means
\be
 \Delta^\text{lin}_\text{SPT}(k,\eta)=4\pi e^{2\eta}A\,k^{n_s+3}\equiv \left(\frac{k}{k_{nl}(\eta)}\right)^{n_s+3}\,,
\ee
where we introduced the usual non-linear scale
\be
  k_{nl}(\eta) = k_{nl}\,e^{-2\eta/(n_s+3)}\,,
\ee
with $k_{nl}=(4\pi A)^{-1/(n_s+3)}$ being the non-linear scale today ($\eta=0$). 
The power spectrum obeys a scaling symmetry (for any $r>0$)
\be
  k \to r k,\qquad e^{\eta}\to r^{-(n_s+3)/2} e^{\eta}\,,
\ee
that suggests that the non-linear power spectrum (of any dimensionless variable) is a function of the ratio
\be
  \Delta(k,\eta) = \Delta(k/k_{nl}(\eta))\,.
\ee
The background value $\epsilon(\eta)$ of the velocity dispersion defines the scale
\be
  k_\sigma(\eta) \equiv \frac{1}{\sqrt{\epsilon(\eta)}}\,.
\ee
The scaling symmetry suggests that 
$ k_\sigma(\eta)/k_{nl}(\eta)$
 is constant, implying that
\be\label{eq:epspowerlaw}
\epsilon=\epsilon_0 e^{\alpha\eta}\,,
\ee
follows a power-law with exponent 
\be
\alpha=4/(n_s+3)\,.
\ee
We denote the value today by 
\be
  k_\sigma\equiv 1/\sqrt{\epsilon_0}\,,
\ee
without time-argument. In addition, when taking higher cumulants into account, the dimensionless ratio $\bar{\cal E}_4=3\bar\omega/5$ as well as in general all $\bar{\cal E}_{2n}$
are constant in time. As anticipated, the scaling universe therefore provides an example for which the assumptions on the time-dependence of $\epsilon(\eta)$ and higher cumulant expectation values
taken in Secs.~\ref{sec:linear} and~\ref{sec:hierarchy} are satisfied exactly, and the corresponding linear kernels for the perturbation variables can be used.

The linear matter power spectrum within \vpt~is given by
\be\label{eq:PlinP0}
  P_{\delta\delta}^\text{lin}(k,\eta) = F_{1,\delta}(k,\eta)^2\,e^{2\eta}P_0(k)\,,
\ee
where $F_{1,\delta}$ is the linear kernel obtained from a solution of the appropriate linearized perturbation equations.
The (cross-)power spectra for any pair $a,b=\delta,\theta,\bar g, \delta\bar\epsilon,\dots$ of dimensionless perturbation modes
in \vpt~is analogously given by
\be\label{eq:Plinab}
  P_{ab}^\text{lin}(k,\eta) = F_{1,a}(k,\eta)F_{1,b}(k,\eta)\,e^{2\eta}P_0(k)\,.
\ee
As shown in Secs.~\ref{sec:linear} and~\ref{sec:hierarchy}, for any truncation order of the cumulant expansion the linear kernels $F_{1,a}(k,\eta)=F_{1,a}(s)$ depend on time and scale only via the dimensionless variable 
\be
  s \equiv s_k(\eta)= \sqrt{3\epsilon(\eta)k^2}\,.
\ee
Therefore we may write the dimensionless power spectrum $\Delta_{ab}^\text{lin} = 4\pi k^3 P_{ab}^\text{lin}$ as
\be
  \Delta_{ab}^\text{lin}(k,\eta) = x \times \hat\Delta_{ab}^\text{lin}(s)\,,
\ee
where 
\bea\label{eq:xdef}
  x &\equiv&  \left(\frac{k_\sigma}{\sqrt{3}k_{nl}}\right)^{n_s+3} = \left(\frac{1}{3\epsilon_0 k_{nl}^2}\right)^{\frac{n_s+3}{2}}\,,
\eea
is time-independent, and we defined
\be\label{eq:DeltalinLloop}
 \hat\Delta_{ab}^\text{lin}(s) \equiv F_{1,a}(s)F_{1,b}(s)s^{n_s+3}\,.
\ee
A similar relation holds for the full non-linear power spectrum, computed within \vpt.
As in SPT, one can use the perturbative solutions to obtain a loop expansion,
\be
  \Delta_{ab}(k,\eta)=\sum_{L\geq 0} \Delta_{ab}^{L-\text{loop}}(k,\eta)\,,
\ee
with $L=0$ being the linear solution. For $ab=\delta\delta,\delta\theta,\theta\theta$, the loop corrections formally take a similar  form as in SPT,
but with non-linear kernels  computed based on the perturbation modes and their non-linear vertices presented here.
For a detailed discussion of loop corrections within \vpt~we refer to paper II~\cite{cumPT2}. Here we restrict ourselves to the general structure of loop corrections.
Since the $L$-loop contribution encompasses $L+1$ factors of $P_0\propto A\propto x$, one has
\be\label{eq:DeltaLloop}
  \Delta_{ab}^{L-\text{loop}}(k,\eta) = x^{L+1} \times \hat\Delta_{ab}^{L-\text{loop}}(s)\,,
\ee
with the second factor involving loop integrals for $L\geq 1$.

Let us now discuss how to obtain a self-consistent solution for the background dispersion  $\epsilon(\eta)$. Its equation of motion is given in Eq.~\eqref{eq:epseom}, with source term from Eq.~\eqref{eq:eomomegaepsilon}. It can be written in terms of dimensionless quantities as
\be\label{eq:epseom2nd}
  \frac{\epsilon'(\eta)}{\epsilon(\eta)}+1 = \frac13 \int_0^\infty\frac{dk}{k}\,\left( \Delta_{\theta\bar g}-\Delta_{\theta\delta\bar \epsilon}+2\Delta_{w_i\bar\nu_i}+\Delta_{A\bar\pi}\right)\,,
\ee
where $\bar g=g/\epsilon$, $\delta\bar\epsilon=\delta\epsilon/\epsilon$ and $\bar\nu_i=\nu_i/\epsilon$ are the
dimensionless scalar and vector perturbation modes of the velocity dispersion tensor, $\bar\pi=\pi/\epsilon$ is a scalar mode of the third cumulant,
and we have set $\Omega_m/f^2\mapsto 1$.

For the time-dependence Eq.~\eqref{eq:epspowerlaw} expected from scaling symmetry, the left-hand side of this equation is constant, and equal to $\alpha+1=(n_s+7)/(n_s+3)$.
Therefore, the ansatz Eq.~\eqref{eq:epspowerlaw} is consistent if also the right-hand side is time-independent.
Using Eq.~\eqref{eq:DeltalinLloop} in linear approximation and Eq.~\eqref{eq:DeltaLloop} in general, we see that this is indeed the case, since the variable $x$ is time-independent
and the power spectrum integrated over $s\propto k$ as well. Therefore, Eq.~\eqref{eq:epseom2nd} turns into an algebraic equation for $x$ given by
\be\label{eq:eps0selfconsistent}
  \frac{n_s+7}{n_s+3} = \frac13\sum_{L\geq 0} x^{L+1} I^{L-\text{loop}}(n_s)\,,
\ee
where
\bea\label{eq:ILloop}
  I^{L-\text{loop}}(n_s) &\equiv& \int_0^\infty\frac{ds}{s}\Big( \hat\Delta_{\theta\bar g}(s)-\hat\Delta_{\theta\delta\bar \epsilon}(s)\nn\\
  && +2\hat\Delta_{w_i\bar\nu_i}(s)+\hat\Delta_{A\bar\pi}(s)\Big)^{L-\text{loop}}\,.
\eea
In particular, the linear contribution ($L=0$) reads
\bea\label{eq:Ilin}
  I^{\text{lin}}(n_s) &=& \int_0^\infty ds\,s^{n_s+2}\Big(F_{1,\theta}(s)(F_{1,\bar g}(s)-F_{1,\delta\bar \epsilon}(s))\nn\\
  && +F_{1,\delta}(s)F_{1,\bar\pi}(s)\Big)\,,
\eea
where we used that vorticity and vector modes contribute only starting at one-loop, and that $A\mapsto\delta$ in linear approximation.
When including terms up to a given loop order $L$, Eq.~\eqref{eq:eps0selfconsistent} is a polynomial equation for $x$ of degree $L+1$, the solution(s) $x_*$ of which determine
the ratio of scales
\be\label{eq:ksigknl}
  k_\sigma/k_{nl}=\sqrt{3}x_*^{1/(n_s+3)}\,,
\ee 
setting the overall magnitude of the background dispersion $\epsilon_0=1/k_\sigma^2$.
In linear approximation, the solution is given by
\be\label{eq:xstar}
  x_*^\text{lin}=3(n_s+7)/((n_s+3) I^\text{lin}(n_s))\,.
\ee
While significant changes from non-linear corrections can be expected,
we discuss self-consistent solutions in linear approximation as a proof-of-principle in this work. Non-linear corrections are presented in paper II~\cite{cumPT2}.
In particular, we are interested in the sensitivity to the truncation of the cumulant expansion, and start with the case of including velocity dispersion only.

\subsection{Self-consistent solution in second cumulant approximation}\label{sec:powerlawcum2}

\begin{figure}[t]
  \begin{center}
  \includegraphics[width=\columnwidth]{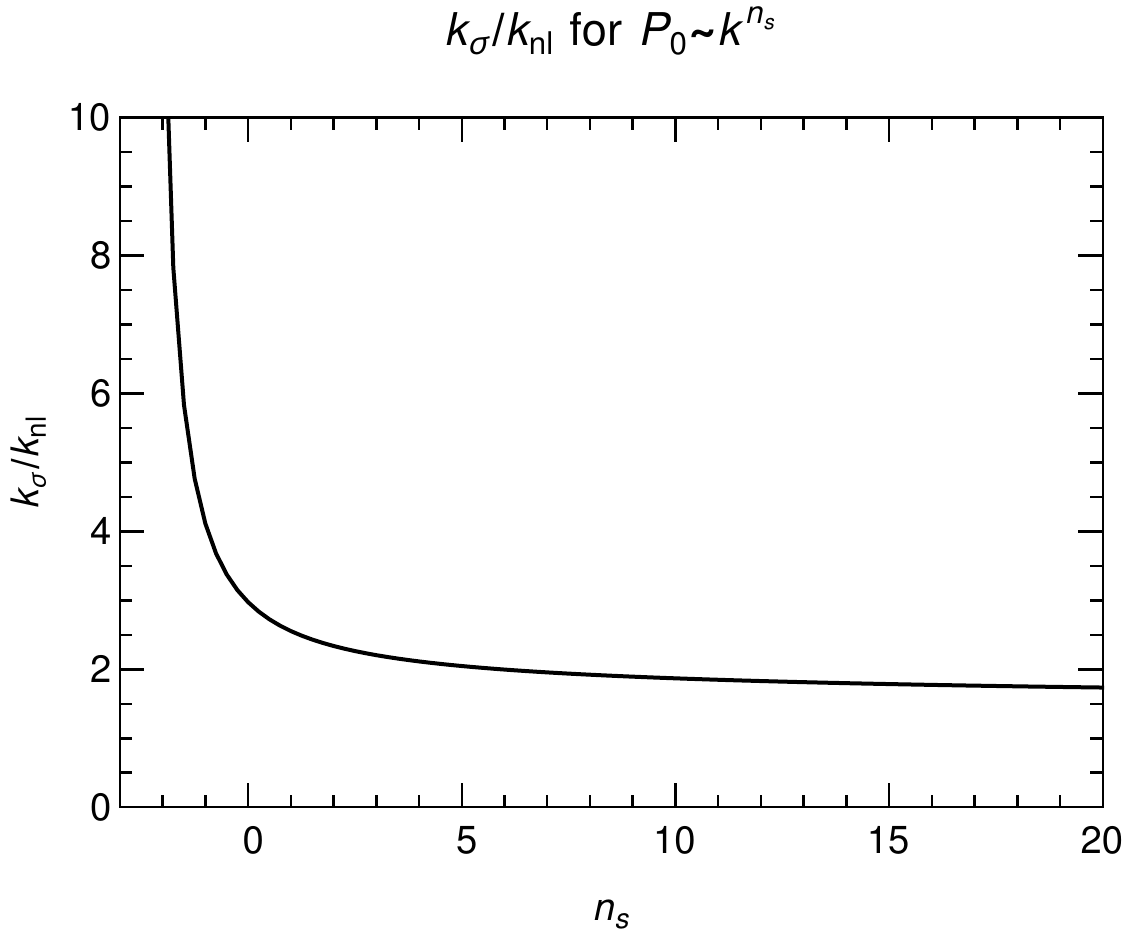}
  \end{center}
  \caption{\label{fig:ksigPowerLaw} Velocity dispersion scale $k_\sigma=\epsilon^{-1/2}$ relative to the non-linear scale $k_{nl}$ for power-law initial spectrum $P_0\propto k^{n_s}$, obtained when using the linear approximation for the source term for $\epsilon$, and neglecting third and higher cumulants.}
\end{figure}

When neglecting third and higher cumulants, we can obtain the self-consistent solution for the
background dispersion by computing the integral Eq.~\eqref{eq:Ilin} using the analytical linear kernels
for $F_{1,\theta}$ and $F_{1,\bar g}$ given in App.\,\ref{app:lin}, while $F_{1,\bar\pi}=F_{1,\delta\bar \epsilon}=0$.
The latter follows since the scalar mode $\delta\epsilon$ has no growing mode in linear approximation, and
at second cumulant order, and the former since the third cumulant perturbation $\pi$ is neglected presently (see below for the generalization to higher cumulants). 

The self-consistent solution exists provided that the integral $I^\text{lin}(n_s)$ converges. In the infrared limit, $k \propto s\to 0$, one has $F_{1,\theta}\to 1$
and $F_{1,\bar g}\to 2/(2+\alpha)$. Therefore the integral is infrared-finite provided that $n_s>-3$. In the ultraviolet limit $k \propto s \gg 1$, the linear kernels
have asymptotic form given in Eq.~\eqref{eq:Tlarge}. The integral is absolutely convergent if
\be
  d(n_s)\equiv 3+n_s-\min(d_\theta+d_g,d_\theta+e_g,e_\theta+d_g,e_\theta+e_g)
\ee
is less than zero. Using results from App.~\ref{app:lin}
one finds
\be
  d(n_s) = \left\{ \begin{array}{ccc}
    -(11n_s+53)/24 & \mbox{for} & n_s<1\,, \\
    -(n_s+7)/3 & \mbox{for} & n_s>1\,, \\
  \end{array}\right.
\ee
which is less than zero for $n_s>-53/11\approx -4.8$. Note that even for a very blue initial spectrum (large $n_s$), the damping due to velocity dispersion is strong enough to compensate the growth of power at large $k$ and make the integral converge. On the contrary, the sensitivity to short modes grows when decreasing $n_s$, e.g. $d(-1)\approx-1.8$, $d(-2)\approx-1.3$, $d(-3)\approx-0.8$. Overall, the whole integral is convergent for all $n_s>-3$. Note that the solution is not valid for $n_s<-3$, because then one would have $\alpha<0$, i.e. a velocity dispersion that was larger in the past and decays with time, rather than being generated, leading to a qualitatively different behavior. For the relevant case where $n_s\to -3$ from above, one has $\alpha\to \infty$, i.e. velocity dispersion grows very quickly.

In Fig.~\ref{fig:ksigPowerLaw}, the ratio of $k_\sigma=\epsilon_0^{-1/2}$ and $k_{nl}$, 
as determined by the linear approximation, is shown by the black line. 
The ratio is always larger than one, and becomes very large for $n_s\to -3$. 
This implies that non-linear corrections are expected to be more relevant the smaller $n_s$.

\subsection{Self-consistent solution in third and fourth cumulant approximation}\label{sec:powerlawcum4}

When including perturbation modes of the third and fourth cumulant, the background value $\omega(\eta)$ of the
fourth cumulant has to be taken into account. Its equation of motion when neglecting fifth cumulant perturbations
is given in Eq.~\eqref{eq:eomomegaepsilon}, and can (for $\Omega_m/f^2=1$) be rewritten as
\bea\label{eq:eomomegadimless}
  \frac{\omega'+2\omega}{\epsilon^2}
  &=& \frac13 \int_0^\infty\frac{dk}{k}\,\Big\{ 4\Delta_{\theta\bar\xi}-\Delta_{\theta\bar\kappa}+2\Delta_{\bar g\bar\pi}\nn\\
  && +6\Delta_{\delta\bar \epsilon\bar\pi} -\frac85\Delta_{\bar g\bar\chi}\Big\} \,,
\eea
where $\bar\pi, \bar\chi$ and $\bar\xi, \bar\kappa$ are the dimensionless perturbation modes of the third and fourth cumulant, respectively.
Let us show that for a scaling universe, solutions with time-dependence $\omega(\eta)\propto \epsilon(\eta)^2$ are consistent with this equation
of motion. For constant $\bar\omega=\omega/\epsilon^2$, the left-hand side of Eq.~\eqref{eq:eomomegadimless} is equal to $2(\alpha+1)\bar\omega=2(n_s+7)\bar\omega/(n_s+3)$,
and is itself time-independent. The dimensionless power spectra on the right-hand side can be decomposed in a sum over loop contributions.
Each of them satisfies a relation analogous to Eq.~\eqref{eq:DeltaLloop}, except that the $\hat\Delta_{ab}^{L-\text{loop}}$ can depend on $\bar\omega$
in addition to $s$. Therefore, also the right-hand side of Eq.~\eqref{eq:eomomegadimless} is constant after integration over $k\propto s$, if $\bar\omega$ is constant.
Thus, both the left- and right-hand side of Eq.~\eqref{eq:eomomegadimless} are time-independent for constant $\bar\omega$, implying that $\omega(\eta)\propto \epsilon(\eta)^2$ is a consistent ansatz.
What remains to be done is to find a solution for the constant values of $k_\sigma/k_{nl}$ and $\bar\omega$, or equivalently $x$ and $\bar\omega$.
For that purpose we rewrite Eq.~\eqref{eq:eps0selfconsistent} and Eq.~\eqref{eq:eomomegadimless} in the form
\bea\label{eq:eps0wbarselfconsistent}
  \frac{n_s+7}{n_s+3} &=& \frac13\sum_{L\geq 0} x^{L+1} I^{L-\text{loop}}(n_s,\bar\omega)\,,\nn\\
  2\frac{n_s+7}{n_s+3}\bar\omega &=& \frac13\sum_{L\geq 0} x^{L+1} J^{L-\text{loop}}(n_s,\bar\omega)\,,
\eea
yielding a coupled set of equations for the unknowns $x$ and $\bar\omega$, with a polynomial dependence on $x$ for any given loop order, and an implicit dependence on $\bar\omega$ that can in general only be determined numerically.
Here we defined
\bea\label{eq:JLloop}
  J^{L-\text{loop}}(n_s,\bar\omega) &\equiv& \int_0^\infty\frac{ds}{s}\Big( 4\hat\Delta_{\theta\bar\xi}(s)-\hat\Delta_{\theta\bar\kappa}(s)+2\hat\Delta_{\bar g\bar\pi}(s)\nn\\
  && +6\hat\Delta_{\delta\bar \epsilon\bar\pi}(s) -\frac85\hat\Delta_{\bar g\bar\chi}(s)\Big)^{L-\text{loop}}\,.
\eea
The integrals $I^{L-\text{loop}}$ are given by the same expression as in Eq.~\eqref{eq:ILloop}, but with power spectra computed based on kernels and non-linear vertices including third and fourth cumulants. In turn, this leads to an implicit dependence
on the background value of the fourth cumulant $\bar\omega$, as indicated in the arguments of $I^{L-\text{loop}}$ and $J^{L-\text{loop}}$.

In linear approximation, we can eliminate $x$ by taking the ratio of both equations in Eq.~\eqref{eq:eps0wbarselfconsistent}, giving an implicit equation for the fourth cumulant expectation value $\bar\omega$,
\be\label{eq:omegabarimplicit}
  \bar\omega = \frac12\frac{J^{\text{lin}}(n_s,\bar\omega)}{I^{\text{lin}}(n_s,\bar\omega)}\,.
\ee
The solution $\bar\omega_*^\text{lin}$ of this equation can be determined numerically using Eq.~\eqref{eq:Ilin} for $I^\text{lin}$ as well as an analogous expression for $J^{\text{lin}}\equiv J^{0-\text{loop}}$ and Eq.~\eqref{eq:JLloop}. The power spectra entering
both integrals can be expressed in terms of the linear kernels $F_{1,a}$ for the dimensionless perturbation variables via Eq.~\eqref{eq:DeltalinLloop}. The $F_{1,a}$ are obtained by numerically solving the linear evolutions
equations Eq.~\eqref{eq:cum4linrescale} and using Eq.~\eqref{eq:Ta}. Finally, the background dispersion scale $k_\sigma/k_{nl}$ can be obtained using Eq.~\eqref{eq:ksigknl} with the linear solution for $x$ given by Eq.~\eqref{eq:xstar}
with $I^\text{lin}=I^\text{lin}(n_s,\bar\omega_*^\text{lin})$ evaluated on the solution $\bar\omega=\bar\omega_*^\text{lin}$.

When truncating the perturbation modes at third cumulant order, corresponding to (cum3+) or equivalently $c_\text{max}=3$, we find that Eq.~\eqref{eq:omegabarimplicit} indeed has a solution in linear approximation.
The corresponding values of $\bar\omega$ as well as $k_\sigma/k_{nl}$ are given in Table~\ref{tab:selfconsistent} for various spectral indices $n_s$. We observe that the dimensionless fourth cumulant expectation value is of order
unity, indicating that higher cumulants are relevant quantitatively, but of the same order as the background dispersion. In addition, the shift in the value of $k_\sigma/k_{nl}$ compared to the second cumulant approximation is sizeable, while
the overall magnitude is comparable. This indicates that higher cumulants are important quantitatively, but do not invalidate the qualitative behavior of the second cumulant approximation. We find that no self-consistent solutions exist when
including fourth cumulant perturbations, which may be attributed to the shortcomings of the linear approximation. We investigate the impact of cumulants beyond the fourth order in the next section.

\begin{table*}
  \centering
  \caption{Self-consistent solutions within linear approximation for the velocity dispersion scale $k_\sigma=\epsilon^{-1/2}={\cal E}_2^{-1/2}$ relative to the non-linear scale, as well as the normalized expectation values $\bar{\cal E}_{2n}={\cal E}_{2n}/\epsilon^n$ of higher cumulants. We show results for scaling universes with spectral indices $n_s=-1,0,1,2$, and for various truncations of the cumulant expansion, with perturbation modes up to order $c_\text{max}$. The cases $c_\text{max}=2,3$ are equivalent to (cum2) and (cum3+), respectively, with $\bar{\cal E}_4=3\bar\omega/5$.}
  \begin{ruledtabular}
    \begin{tabular}{c|cccc|cccc|cccc|cccc} 
   &&&&&&&&&&&&&&\\[-2ex]
   & \multicolumn{4}{c|}{$n_s=-1$} &  \multicolumn{4}{c|}{$n_s=0$} &  \multicolumn{4}{c|}{$n_s=1$} &  \multicolumn{4}{c}{$n_s=2$}  \\[1.5ex] \hline &&&&&&&&&&&&&&\\[-1.ex]
   $c_\text{max}$ & $k_\sigma/k_{nl}$ & $\bar{\cal E}_4$ & $\bar{\cal E}_6$ & $\bar{\cal E}_8$  
   & $k_\sigma/k_{nl}$ & $\bar{\cal E}_4$ & $\bar{\cal E}_6$ & $\bar{\cal E}_8$ 
   & $k_\sigma/k_{nl}$ & $\bar{\cal E}_4$ & $\bar{\cal E}_6$ & $\bar{\cal E}_8$ 
   & $k_\sigma/k_{nl}$ & $\bar{\cal E}_4$ & $\bar{\cal E}_6$ & $\bar{\cal E}_8$ \\[1.5ex] \hline &&&&&&&&&&&&&&\\[-1.ex]
  $2$ & 4.1 &-&-&- & 3.0 &-&-&- & 2.6  &-&-&- & 2.3  &-&-&- 
  \\[1.5ex]
  $3$ & 3.4 & 0.45 &-&- & 2.5 & 0.40 &-&- & 2.2 & 0.37 &-&- & 2.0 & 0.35 &-&-
  \\[1.5ex]
  $6$ & 3.8 & 0.37 & 0.86 &- & 2.7 & 0.34 & 0.92 &- & 2.3 & 0.31 & 0.93 &- & 2.1 & 0.29 & 0.92
  \\[1.5ex]
  $7$ & 3.8 & 0.36 & 0.78 & 3.5 & 2.7 & 0.36 & 0.94 & 4.5 & 2.3 & 0.35 & 1.03 & 5.1 & 2.1 & 0.34 & 1.08  & 5.3 
  \\
    \end{tabular}
  \end{ruledtabular}
  \label{tab:selfconsistent}
\end{table*}

\subsection{Self-consistent solutions for the full cumulant hierarchy}\label{sec:powerlawhierarchy}

\begin{figure}[t]
  \begin{center}
  \includegraphics[width=\columnwidth]{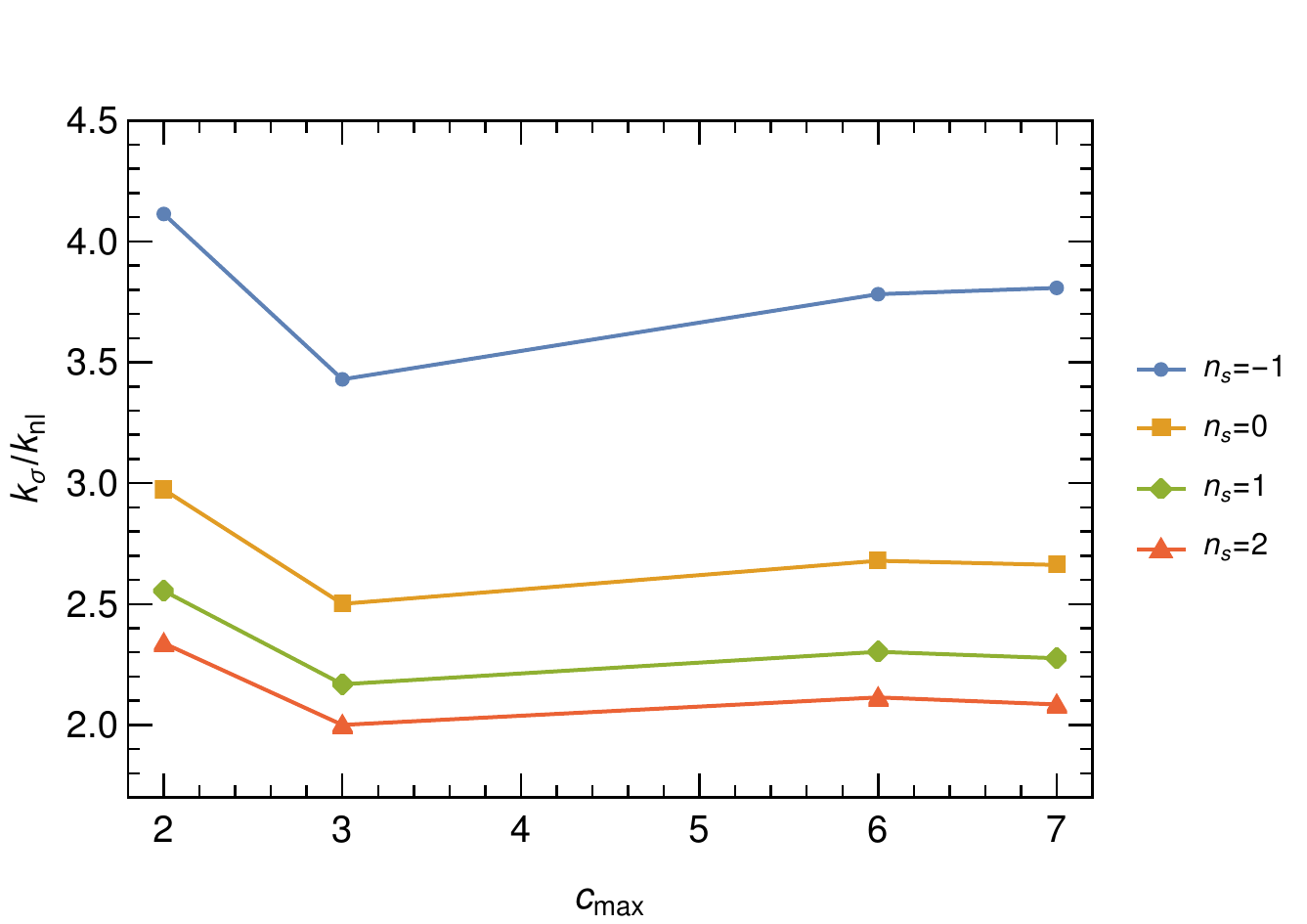}
  \end{center}
  \caption{\label{fig:ksigPowerLawCum} Velocity dispersion scale $k_\sigma=\epsilon^{-1/2}$ relative to the non-linear scale $k_{nl}$ for power-law initial spectrum $P_0\propto k^{n_s}$, obtained from a self-consistent solution when including cumulant perturbations up to order $c_\text{max}$, and for $n_s=-1,0,1,2$, respectively. The corresponding self-consistent solutions for the fourth, sixth and eighth cumulant expectation values are shown in Table~\ref{tab:selfconsistent}.}
\end{figure}

The self-consistent solutions of background values for a scaling universe can be extended to truncations including cumulants beyond the fourth order following Sec.\,\ref{sec:hierarchy}.
In linear approximation, the expectation values ${\cal E}_{2n}$ of the $2n$th cumulant satisfy equations of motion given in Eq.~\eqref{eq:eomE2n}. For $2n=2$ and $2n=4$ they agree with
those discussed above for $\epsilon={\cal E}_2$ and $\omega=5{\cal E}_4/3$, respectively. For a scaling universe these equations allow for constant values of the dimensionless
ratios $\bar{\cal E}_{2n}={\cal E}_{2n}/\epsilon^n$, determined by the set of implicit equations (in linear approximation)
\be\label{eq:selfconsistentE2n}
  n\frac{n_s+7}{n_s+3}\bar{\cal E}_{2n} = x \times I_{\bar{\cal E}_{2n}}^\text{lin}(n_s,\bar{\cal E}_{4},\bar{\cal E}_{6},\dots)\,,
\ee
with
\bea\label{eq:IbarE2n}
  I_{\bar {\cal E}_{2n}}&=& \int_0^\infty ds\,s^{n_s+2}\sum_{\ell=0}^n (\ell+1) \sum_{m_1,m_2=0}^{n-\ell}\delta_{m_1+m_2,n-\ell}\nn\\
  &&\frac{(2n)!(2(m_2-m_1-\ell)-3)}{(2m_1)!(2m_2)!}T_{\ell+1,2m_1}(s)T_{\ell,2m_2}(s)\,,\nn\\
\eea
where $\delta_{i,j}$ is the Kronecker symbol, and $T_{\ell,2m}(s)$ are the dimensionless linear kernels for perturbation modes of cumulant order $\ell+2m$ defined in Eq.~\eqref{eq:Tl2n}.
The linear kernels are given by numerical solutions of the equations of motion Eq.~\eqref{eq:eomTl2n}. They depend on time and scale via the single variable $s=\sqrt{3\epsilon(\eta) k^2}$, which can be
seen using Eq.~\eqref{detads} and that $\alpha=\partial_\eta\ln\epsilon$ is constant for a scaling universe.
In addition, the linear kernels depend parametrically on  the spectral index $n_s$ due to $\alpha=4/(3+n_s)$, as well as the backgound values $\bar{\cal E}_{4},\bar{\cal E}_{6},\dots$.
Therefore, Eq.~\eqref{eq:selfconsistentE2n} is a highly implicit and coupled set of equations for the self-consistent values of $\bar{\cal E}_{2n}$ with $2n\geq 4$ as well as the
overall magnitude of the background dispersion parameterized by the variable $x$ defined in Eq.~\eqref{eq:xdef}. By taking the ratio of Eq.~\eqref{eq:selfconsistentE2n} for $2n\geq 4$ and Eq.~\eqref{eq:selfconsistentE2n}
for $2n=2$, and using that $\bar{\cal E}_2=1$ by definition, we obtain a coupled set of equations for the cumulant expectation values of order $2n\geq 4$ that is independent of $x$,
\be\label{eq:E2nbarselfconsistent}
  \bar{\cal E}_{2n} =\frac{1}{n}\frac{I_{\bar{\cal E}_{2n}}^\text{lin}(n_s,\bar{\cal E}_{4},\bar{\cal E}_{6},\dots)}{I_{\bar{\cal E}_{2}}^\text{lin}(n_s,\bar{\cal E}_{4},\bar{\cal E}_{6},\dots)}\,.
\ee
These implicit equations can be viewed as a generalization of Eq.~\eqref{eq:omegabarimplicit} to beyond the fourth cumulant order. Once a solution to these equations is found, it can be inserted in Eq.~\eqref{eq:selfconsistentE2n}
for $n=1$, yielding the  solution for $x$ and thereby for the background dispersion scale
\be\label{eq:ksighierarchy}
  \frac{k_\sigma}{k_{nl}} = \sqrt{3}\left(\frac{n_s+7}{(n_s+3)I_{\bar {\cal E}_2}^\text{lin}}\right)^{1/(n_s+3)}\,.
\ee
For a given truncation of cumulant perturbations at order $c_\text{max}$, all $T_{\ell,2m}(s)$ with $\ell+2m>c_\text{max}$ are neglected. In this case the equations for the linear kernels and therefore
also $I_{\bar{\cal E}_{2n}}^\text{lin}$ depend only on the background values $\bar{\cal E}_{4},\dots,\bar{\cal E}_{c_\text{max}+1}$ for odd $c_\text{max}$, and $\bar{\cal E}_{4},\dots,\bar{\cal E}_{c_\text{max}}$ for even $c_\text{max}$.
For $c_\text{max}=2$ only the background dispersion enters and it is sufficient to solve Eq.~\eqref{eq:ksighierarchy}, with results identical to those from Sec.\,\ref{sec:powerlawcum2}. For $c_\text{max}=3,4$, Eq.~\eqref{eq:E2nbarselfconsistent}
reduces to a single equation and yields values for $\bar{\cal E}_{4}=3\bar\omega/5$ consistent with those obtained in Sec.\,\ref{sec:powerlawcum4}. For $c_\text{max}=5,6$, Eq.~\eqref{eq:E2nbarselfconsistent} yields a coupled set of equations
for $\bar{\cal E}_{4},\bar{\cal E}_{6}$. For $c_\text{max}=7,8$, one obtains three coupled equations for $\bar{\cal E}_{4},\bar{\cal E}_{6},\bar{\cal E}_{8}$. Up to eighth order, we find that within the linear approximation considered here
a joint self-consistent solution exists only for $c_\text{max}=2,3,6,7$, confirming the previous findings for $c_\text{max}=2,3$. The corresponding values are shown in Table~\ref{tab:selfconsistent} for various values of $n_s$. We observe that
the size of $\bar{\cal E}_4$ is comparable for $c_\text{max}=3,6,7$, and the one of $\bar{\cal E}_6$ for $c_\text{max}=6,7$, indicating that increasing the truncation order does not lead to dramatic changes.

The dependence of the dispersion scale $k_\sigma$ on the truncation order $c_\text{max}$ is shown in Fig.~\ref{fig:ksigPowerLawCum}. The largest shift occurs when going from $c_\text{max}=2$ to $c_\text{max}=3$, while even higher cumulants have only a minor impact. This indicates that the relevant contributions to the source terms Eq.~\eqref{eq:IbarE2n} arise from scales where the dependence of the linear kernels on $c_\text{max}$ is already converged.
In addition, we find that the impact of higher cumulants is smaller for larger $n_s$. While spectra with large values of $n_s$ do have a lot of power on small scales initially, it is more efficiently erased by the suppression due to the buildup of velocity dispersion, making the linear kernels drop faster, and hence the source terms less sensitive to the contribution from large wavenumbers.

\section{Conclusions}
\label{sec:conclusions}

In this work we discuss the extension of standard perturbation theory (SPT) to include higher cumulants of the phase-space distribution function, based on the underlying Vlasov-Poisson dynamics for collisionless matter, dubbed \emph{Vlasov Perturbation Theory} (\vpt). This takes into account that even for an initially perfectly cold dark matter distribution, orbit crossing generates velocity dispersion and higher cumulants. We provide the explicit form of non-linear evolution equations when taking up to the fourth cumulant into account, and derive evolution equations linear in perturbations up to arbitrary order in the cumulant expansion.

\vpt~splits cumulants into their average values and fluctuations around them. The evolution equations for the perturbations depend on the average values, and vice versa. Since in general the average values are sourced by fluctuations integrated over all scales, we argue that the background values of even cumulants should be treated as ``${\cal O}(1)$'' quantities. This leads to a consistent, systematic perturbative expansion scheme for the fluctuations, and allows us to describe the screening of UV modes crucial for improving the convergence of SPT, as we discuss in paper~II~\cite{cumPT2}. The resulting  \vpt~equations can be cast into a form that is formally analogous to SPT, but with an extended set of perturbation variables and non-linear interactions among them,  in presence of a background given by the average values of the even cumulants. 

Our main findings  are:

\begin{itemize}

\item[i)] 
Linear \vpt~is far richer compared to SPT. Even in the simplest approximation, where linear theory is truncated at the second cumulant, the effective description in terms of density and velocity divergence is non-local in time, which is key to satisfy the cosmic energy equation. When a given mode $k$ crosses the dispersion scale $k_\sigma$, its growth is suppressed. This  back reaction on modes from small-scale dispersion is completely absent in SPT. The suppression is only affected by higher cumulants when $k\gg k_\sigma$, and they make the screening mechanism even more efficient. For any given wavenumber the cumulant expansion converges. This motivates a further study of perturbation theory with a truncated cumulant expansion beyond the linear approximation. This is what we carry out in paper~II~\cite{cumPT2}.

\item[ii)] The UV screening mechanism is in principle not guaranteed, i.e. the complexity of the linear theory of collisionless dynamics allows exponential instabilities. Requiring that these be absent leads to stability conditions that we derive analytically up to cumulant of order eight. Remarkably, these stability conditions are independent of the value of the velocity dispersion and spectral index and only constrain the non-Gaussianity of the distribution function. The Gaussian case (vanishing average values of fourth and higher cumulants) is always within the stable domain.

\item[iii)] We therefore consider higher cumulants averaged over stationary dark matter haloes, finding that
they are generically of order unity when compared to an appropriate power of the second cumulant. While the halo analysis is not required for the development of the \vpt~framework, it serves as a useful benchmark and to gain some insight into the distribution function non-Gaussianity. Interestingly, we find that the cumulant expectation values obtained from the halo analysis satisfy the stability conditions. Altogether, this implies that the linear approximation within \vpt~is a good starting point for a perturbative analysis with realistic values of the cumulant expectation values.

\item[iv)] Finally, we determine self-consistent solutions of the coupled set of perturbation and background equations. We consider a scaling universe for this analysis, allowing us to transform the set of equations for the average values into coupled algebraic equations, that we solve up to cumulant of order eight. We find that the decoupling of UV modes is  more pronounced for larger spectral index $n_s$. This is because very blue initial spectra lead to pronounced orbit crossing on small scales and therefore quickly generate a large dispersion. Remarkably, this causes the  integral over the power spectra of cumulant perturbations that source the average values to converge even for arbitrarily large $n_s$. The resulting background values for the cumulants in this self-consistent approach also satisfy the stability conditions.

\end{itemize}

In summary, the \vpt~framework of perturbation theory for dark matter clustering laid out in this work is directly derived from the underlying fundamental collisionless Vlasov-Poisson equations. It can, from the conceptual point of view, be regarded as a straightforward extension of SPT by taking second and higher cumulants of the distribution function into account. Our results show that \vpt~captures physical effects that are neglected in SPT,  in particular shell crossing and the screening of UV modes. It therefore abandons a major shortcoming of SPT. Furthermore, the framework does by construction neither contain any ad hoc assumptions nor undetermined free fitting parameters, and therefore preserves the predictivity of the underlying Vlasov-Poisson equations. Nevertheless, it allows for systematic and tractable extensions of SPT. In an accompanying work, we show that the extended framework can be used beyond the linear approximation as well, and present detailed comparisons with N-body simulation results  (see paper II~\cite{cumPT2}).

\vspace*{2em}
\acknowledgments

We thank Michael Buehlmann, ChangHoon Hahn, Oliver Hahn and Lam Hui for useful discussions. MG and RS acknowledge support from the Munich Institute for Astro- and Particle Physics (MIAPP), and RS thanks the Theory Group at CERN for hospitality during the Fall of 2014, where this collaboration was started. We acknowledge support by the Excellence Cluster ORIGINS, which is funded by the Deutsche Forschungsgemeinschaft (DFG, German Research Foundation) under Germany’s Excellence Strategy - EXC-2094 - 390783311. 

\newpage

\begin{widetext}

\appendix

\section{Further Details on Halo Calculations}
\label{app:halo}
In this appendix we provide some extensions of the calculations presented in Sec.~\ref{sec:halo} for the cumulants of the phase-space distribution function and an alternative approach in the NFW case that gives the shape of the distribution function itself. 

\subsection{Evans Halos}

We first we provide some more details on the halo calculations presented in Sec.~\ref{subsec:Evans}. 
 For the distribution function given by Eq.~(\ref{fEvans2}), the expectation values of the sixth and eighth cumulants are, respectively
\beqa
7\, {\cal E}_6 &=& {315 \over 8} \lexp w_a \rexp + {105\over 8}  \lexp w_b \rexp  + 105 \lexp w_c \rexp + \frac{687}{4} \lexp w_a^3\rexp 
-\frac{1449}{8} \lexp w_a^2 \rexp -\frac{315}{2} \lexp w_a w_b \rexp  + \frac{987}{4} \lexp w_a^2 w_b \rexp  - \frac{315}{8} \lexp w_b^2\rexp \nonumber \\ & & 
+ \frac{525}{4} \lexp w_a w_b^2  \rexp  + \frac{105}{4} \lexp w_b^3 \rexp  - \frac{1785}{4} \lexp w_a w_c \rexp  + \frac{987}{2} \lexp w_a^2 w_c\rexp
  - \frac{945}{4} \lexp w_b w_c\rexp  + 525 \lexp w_a w_b w_c \rexp+ \frac{315}{2} \lexp w_b^2 w_c \rexp  \nonumber \\ & & - 315 \lexp w_c^2 \rexp + 
 525 \lexp w_a w_c^2 \rexp + 315 \lexp w_b w_c^2 \rexp + 210 \lexp w_c^3 \rexp \,,
\label{E6Evans}
\eeqa 
\beqa
9\, {\cal E}_8 &=& \frac{3465}{16} \lexp w_a\rexp + \frac{945}{16} \lexp w_b\rexp  + 945 \lexp w_c \rexp
- \frac{42399}{16} \lexp w_a^2\rexp+ \frac{28215}{4} \lexp w_a^3\rexp - \frac{42003}{8} \lexp w_a^4\rexp 
 - \frac{15435}{8} \lexp w_a w_b\rexp  + \frac{34965}{4} \lexp w_a^2 w_b\rexp \nonumber \\ & & - 
\frac{18549}{2} \lexp w_a^3 w_b\rexp - \frac{6615}{16} \lexp w_b^2\rexp + \frac{16065}{4} \lexp w_a w_b^2\rexp - 
\frac{26649}{4} \lexp w_a^2 w_b^2\rexp + \frac{2835}{4} \lexp w_b^3\rexp - \frac{4725}{2} \lexp w_a w_b^3\rexp - 
\frac{2835}{8} \lexp w_b^4\rexp  \nonumber \\ & & 
 - 7875 \lexp w_a w_c \rexp + 21924 \lexp w_a^2 w_c\rexp - 18549 \lexp w_a^3 w_c\rexp - 
 3780 \lexp w_b w_c\rexp + 20790 \lexp w_a w_b w_c\rexp   - 26649 \lexp w_a^2 w_b w_c\rexp \nonumber \\ & & + 5670 \lexp w_b^2 w_c\rexp - 
 14175 \lexp w_a w_b^2 w_c\rexp - 2835 \lexp w_b^3 w_c\rexp - 6615 \lexp w_c^2 \rexp+ 25515 \lexp w_a w_c^2\rexp    - 
 26649 \lexp w_a^2 w_c^2\rexp  + 14175 \lexp w_b w_c^2 \rexp \nonumber \\ & &- 28350 \lexp w_a w_b w_c^2\rexp - 
 8505 \lexp w_b^2 w_c^2\rexp + 11340 \lexp w_c^3\rexp - 18900 \lexp w_a w_c^3 \rexp   - 11340 \lexp w_b w_c^3 \rexp- 
 5670 \lexp w_c^4 \rexp\,.
\label{E8Evans}
\eeqa 

As mentioned in Sec.~\ref{sec:halo}, for infinite halos $w_b$ drops from these expressions and everything can be written down in terms of $w_a$ due to the $w_a+w_b+w_c=1$ identity. The integrals over the halo of $w_a^n$ can be done analytically, resulting in simple analytic expressions for the expectation value of the cumulants. Defining a real-valued function $f(q)$ for both prolate ($q>1$) and oblate ($q<1$) halos, 
\beq
f(q>1) \equiv \frac{1}{\sqrt{q^2-1}}\  \arccot\Big(\frac{q}{\sqrt{q^2-1}}\Big), \ \ \ \ \ f(q<1) \equiv \frac{1}{\sqrt{1-q^2}}\ \arccoth\Big(\frac{q}{\sqrt{1-q^2}}\Big)\,,
\label{fq}
\eeq
we have for the expectation value of the cumulants
\beq
{\cal E}_2 = 1- {1\over 3} q^2 + {q\over 3}  (2q^2-1)\, f(q), \ \ \ \ \ {\cal E}_4 = {1\over 5} (1-14q^2) + {q\over 5}\, (2q^2-1)(7q^2+6)\, f(q)\,,
\label{E2E4q}
\eeq
\beq
{\cal E}_6 = - {3q^2\over 28}\, (1-234q^2+144q^4) + {3q\over 28} \,(2q^2-1)(108q^4-234q^2+1)\, f(q)\,,
\label{E6q}
\eeq
\beq
{\cal E}_8 =  {q^2\over 24}\, (3183-12308q^2+5292^4-13696q^6) + {q\over 8}\, (2q^2-1)(2568q^6+924q^4+940q^2+1411)\, f(q)\,.
\label{E8q}
\eeq
These are the expressions used to compute the normalized cumulants $\bar{\cal E}_{2n}={\cal E}_{2n}/{\cal E}_2^n$ shown in Fig.~\ref{fig:E4E6E8Evans}.

\begin{figure}[t!]
  \begin{center}
        \includegraphics[width=0.49\columnwidth]{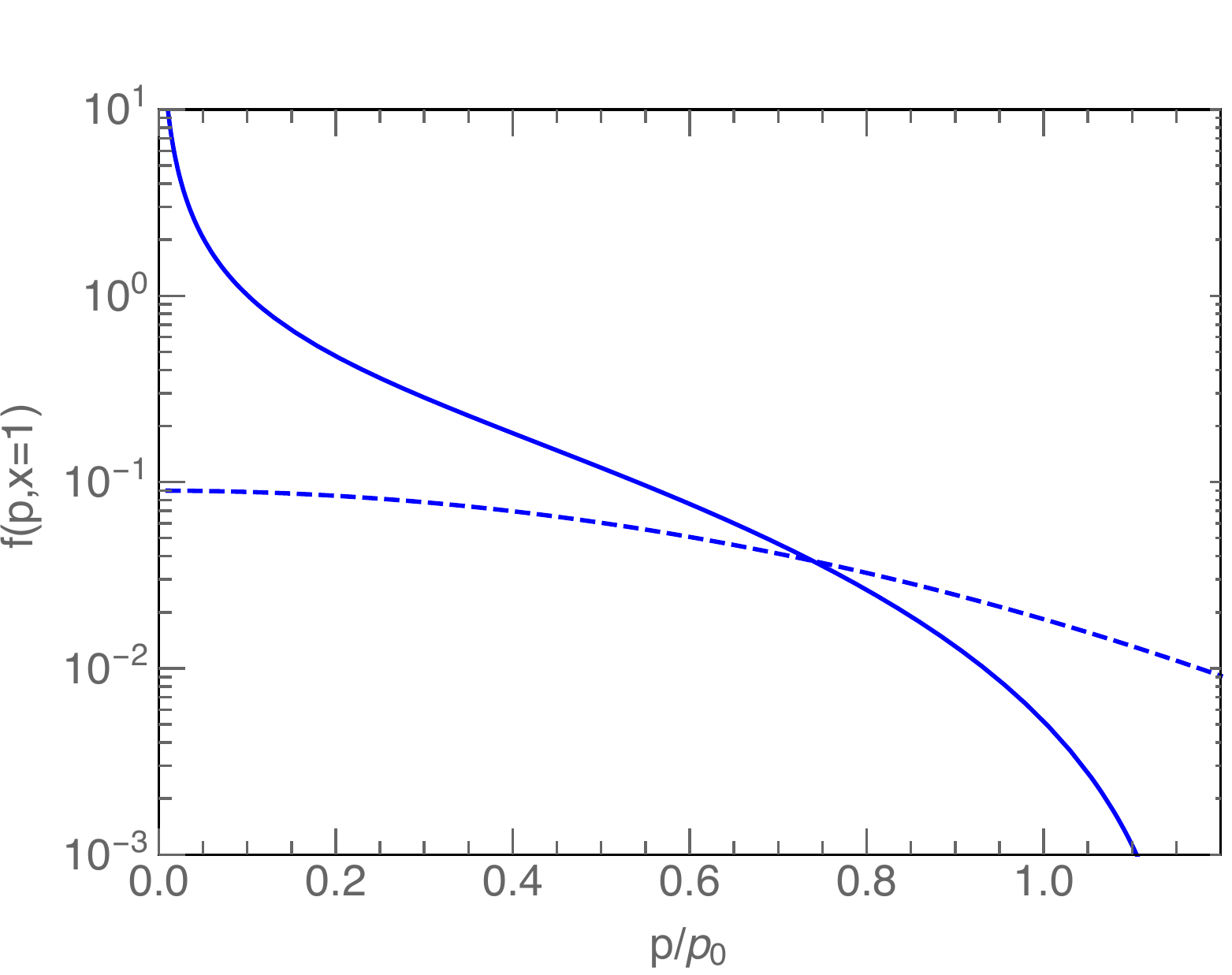}
        \includegraphics[width=0.49\columnwidth]{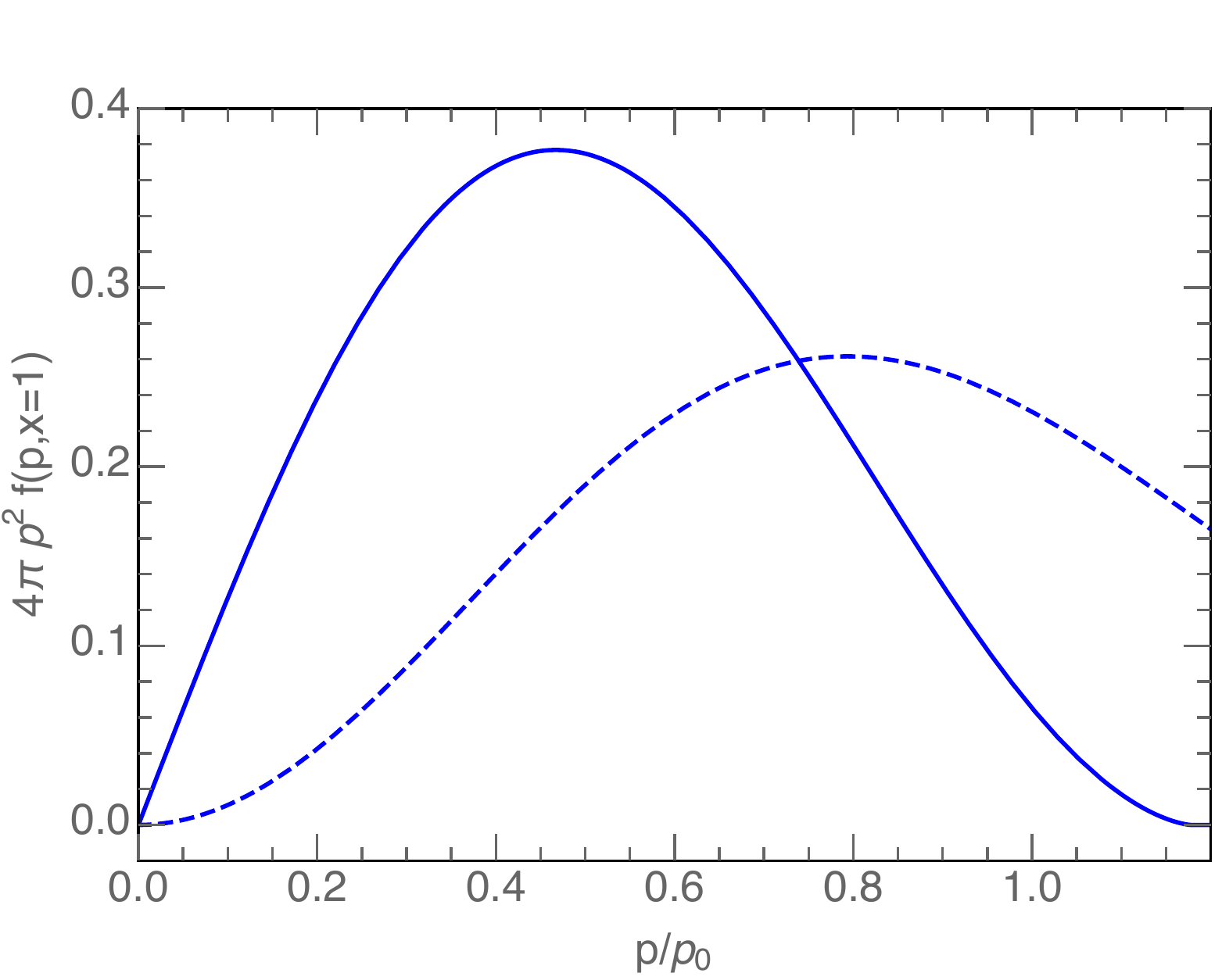}
  \end{center}
  \caption{\label{fig:fNFW} The distribution function monopole for NFW halos with $\beta=1/2$ (solid) compared to a Gaussian distribution with the same second cumulant (dashed), at the scale radius of the halo $x\equiv c r/r_{\rm vir}=1$. The left panel shows $f^{(0)}(p)$ as a function of momentum $p$ in units of $p_0\equiv \sqrt{Gmf(c)c/r_{\rm vir}}$ while the right panel shows $4\pi p^2 f^{(0)}(p)$.}
\end{figure}

\subsection{NFW Halos}

In Sec.~\ref{subsec:NFW} we calculated the cumulants of the phase-space distribution function for a distribution function with constant anisotropy, Eq.~(\ref{fEL}), by integrating the steady Vlasov equation directly, leading to a recursion relation for the moments in terms of the density profile and potential, Eq.~(\ref{recrel2}). Here we take an alternative approach, which is to compute directly the shape of the distribution function. The function $f_E$ in Eq.~(\ref{fEL}) can be computed by a generalization of the so-called Eddington inversion method~\cite{Edd1605,BinTre08} originally developed for the isotropic case ($\beta=0$),
\beq
f_E(\varepsilon) = {2^\beta\over (2\pi)^{3/2} \Gamma(1-\lambda) \Gamma(1-\beta)} \, {d\over d \varepsilon} \int_0^\varepsilon \frac{d\psi}{(\varepsilon-\psi)^\lambda}\,
 {d^n h \over d \psi^n}\,,
\label{FEbeta}
\eeq
where $\varepsilon\equiv-E=\psi-p^2/2\geq0$ is the binding energy per unit mass, $\psi=-\Phi$, $\lambda=3/2-\beta-n$, $n=[3/2-\beta]$ and $h\equiv r^{2\beta} \rho$ is the radially weighted density profile expressed as a function of $\psi$ rather than $r$. When $\beta$ is half-integer ($\beta=1/2,-1/2,\ldots$) this reduces to only derivatives, giving a simple expression for the distribution function~\cite{EvaWil1409}, 
\beq
f(E,L^2)={1\over 2\pi^2}\, {L^{-2\beta} \over (-2\beta)!!}\ {d^{3/2-\beta} h \over d \psi^{3/2-\beta}}\Big|_{\psi=\varepsilon}\,.
\label{fELsemi}
\eeq
In particular, we are interested in $\beta=1/2$ in which case this expression reduces to a first derivative. Using these results, one could compute the cumulants by integration over momentum rather than using the recursion relation Eq.~(\ref{recrel2}), but in practice  the latter is actually easier. However, this approach also gives us explicitly the distribution function. For $\beta=1/2$ we have,
\beq
f(\bm{p},\bm{x})= {1\over 2\pi^2}\, {1 \over p\, r \sqrt{1- (\hat{\bm{p}}\cdot\hat{\bm{x}})^2}}\ {d h \over d \psi}\Big|_{\psi\to\psi-p^2/2}, \ \ \ \ \ \ \ \ \ \ 
f^{(0)}(p,r)=  {1 \over p\, r }\ {d h \over d \psi}\Big|_{\psi\to\psi-p^2/2}\,,
\label{fbeta0p5}
\eeq
where $r\equiv |\bm{x}|$. Taking the monopole of this distribution by integrating over the direction of momentum gives $f^{(0)}$. Figure~\ref{fig:fNFW} shows the result of this calculation for $\beta=1/2$ (solid lines), comparing it to a Gaussian distribution of the same width (dashed). As expected from the form of the distribution function, and the normalized cumulants shown in the left panel of Fig.~\ref{fig:E4E6E8NFW}, the distribution is fairly different from a Maxwellian.

\section{Equations of motion up to the second cumulant}
\label{app:eom}

In this appendix we give the equations of motion up to the second cumulant when neglecting third and higher order cumulants in non-linear terms (see App.\,\ref{app:higher} for those).
The equations for the density contrast $\delta$, the velocity divergence $\theta$ and the log-density field $A$ are given in the main text in Eqs.~(\ref{eq:eom2},~\ref{eq:euler3},~\ref{eq:eomA}), respectively. We use non-bold symbols to denote wavevectors here and below.

\subsection{Equation of motion for the vorticity}

\bea\label{eq:w}
\lefteqn{ w_{k,i}'+\left(\frac32\frac{\Omega_m}{f^2}-1\right)w_{k,i} + k^2\nu_{k,i} }\nn\\
  &=& \int_{pq}  \Bigg\{ 
  -\frac{1}{p^2}\theta_p( k\times( p\times w_q))_i 
  + \frac{1}{p^2}( k\times (( p\times w_p)\times w_q))_i
   - \frac{p\cdot q}{q^2}(p\times q)_iA_p g_q + (p\times q)_iA_p \delta\epsilon_q\nn\\
  &&   {} - \frac{1}{q^2} \Big[(p\cdot q)(k\cdot q\delta_{ij}-k_jq_i)-(p\times q)_i(p\times q)_j\Big] A_p \nu_{q,j}
    - \varepsilon_{ijn} k_j p_m A_p t_{q,nm}\Bigg\}\,. 
\eea

\subsection{Equations of motion for scalar, vector and tensor perturbations of $\epsilon_{ij}$}

\bea
  \lefteqn{ g_k'+2\left(\frac32\frac{\Omega_m}{f^2}-1\right) g_k-2\epsilon \theta_k }\nn\\
  &=&  \int_{pq}  \Bigg\{ \left(3\frac{(k\cdot p)^2}{k^2p^2}-1\right)\theta_p\delta\epsilon_q 
   {} +\frac{p\cdot q}{q^2}\left(\frac{k^2}{p^2}+\frac12-\frac32\frac{(k\cdot p)^2}{k^2p^2}\right)\theta_p g_q + 3\frac{k\cdot p}{p^2k^2}(p\times q)\cdot w_p \delta\epsilon_q \nn\\
  && {} +\frac{1}{p^2}\left(\frac32\frac{(k\cdot q)^2}{k^2q^2}+3\frac{k\cdot q p\cdot q}{k^2q^2}-\frac{p\cdot q}{q^2}-\frac12\right)(p\times q)\cdot w_p g_q 
   {} - \frac{1}{k^2p^2q^2}\left(p\cdot qk^2+3k\cdot p k\cdot q\right)\theta_p (p\times q)\cdot \nu_q \nn\\
  && {} - \frac{p\cdot q}{p^2q^2}(p\cdot q\delta_{ij}-q_ip_j)w_{p,i}\nu_{q,j}
   {} - \frac{6k\cdot q-k^2+3p\cdot q}{k^2p^2q^2}(p\times q)\cdot w_p (p\times q)\cdot \nu_q\nn\\
  && {} - \frac{2k^2-3p\cdot q-6k\cdot p}{2k^2p^2}p_ip_j\theta_p t_{q,ij} +\frac{2k^2(p\times w_p)_i+9(p\times q)\cdot w_p p_i}{2k^2p^2} p_j t_{q,ij} \Bigg\}  \,,
\eea
\bea\label{eq:deleps1}
  \lefteqn{ \delta\epsilon_k'+2\left(\frac32\frac{\Omega_m}{f^2}-1\right)\delta\epsilon_k }\nn\\
  &=&  \int_{pq}  \Bigg\{ \frac{k\cdot p k\cdot q}{k^2p^2}\theta_p\delta\epsilon_q
   {} +\frac{p\cdot q}{2p^2q^2k^2}((k\cdot p)^2-p^2k^2)\theta_p g_q + \frac{k\cdot q}{p^2k^2}(p\times q)\cdot w_p \delta\epsilon_q \nn\\
  && {} + \frac{k^2q^2-(k\cdot q)^2+2p\cdot q k\cdot p}{2p^2q^2k^2}(p\times q)\cdot w_p g_q 
   {} + \frac{1}{q^2}\left(1-\frac{(k\cdot p)^2}{k^2p^2}\right)\theta_p (p\times q)\cdot \nu_q \nn\\
  && {} + \frac{p\cdot q}{p^2q^2}(p\cdot q\delta_{ij}-q_ip_j)w_{p,i}\nu_{q,j}
   {} + \frac{2k\cdot q-k^2+p\cdot q}{k^2p^2q^2}(p\times q)\cdot w_p (p\times q)\cdot \nu_q \nn\\
  && {} + \frac{2k\cdot q-p\cdot q}{2k^2p^2}p_ip_j\theta_p t_{q,ij} -\frac{2k^2(p\times w_p)_i+3(p\times q)\cdot w_p p_i}{2k^2p^2} p_j t_{q,ij} \Bigg\} 
  -  Q\delta^{(3)}(k) \,,
\eea
\bea
  \lefteqn{\nu_{k,i}'+2\left(\frac32\frac{\Omega_m}{f^2}-1\right)\nu_{k,i} -\epsilon w_{k,i} }\nn\\ 
  &=& \int_{pq}  \Bigg\{ -2\frac{k\cdot p (p\times q)_i}{k^2p^2}\theta_p\delta\epsilon_q 
    {} + \frac{p\cdot q k\cdot p (p\times q)_i}{k^2p^2q^2}\theta_p g_q \nn\\
  && {} + \frac{k\cdot p(k\cdot p\delta_{ij}- p_ik_j)-(p\times q)_i(p\times q)_j}{k^2p^2}w_{p,j}\delta\epsilon_q\nn\\
  && {} + \frac{k\cdot q p\cdot q(k\cdot p\delta_{ij}- p_ik_j)+(p\times q)_i(p\times q)_j(k^2-p^2)}{k^2p^2q^2}w_{p,j}g_q\nn\\
  && {} + \frac{k^2 p\cdot q(k\cdot q\delta_{ij}- q_ik_j)-(p\times q)_i(p\times q)_j(p^2-q^2)}{k^2p^2q^2}\theta_p\nu_{q,j} \nn\\
  && {} + \frac{k^2-p^2}{k^2p^2q^2}\left[ (k\times(p\times w_p))_i (p\times q)\cdot\nu_q - (p\times q)\cdot w_p (k\times(q\times\nu_q))_i\right] \nn\\
  && {} -2\frac{(p\times q)_i}{k^2p^2q^2}(p\times q)\cdot w_p  (p\times q)\cdot\nu_q 
    +  \frac{(k^2-q^2)\epsilon_{inm}k_n-(p\times q)_i p_m}{k^2p^2}p_j\theta_p t_{q,mj} \nn\\
  && {} + \frac{2(p\times q)\cdot w_p\epsilon_{inm}k_n-(k\times(p\times w_p))_i p_m}{k^2p^2}p_j t_{q,mj} \Bigg\}\,,
\eea
\bea
  \lefteqn{ t_{k,ij}'+2\left(\frac32\frac{\Omega_m}{f^2}-1\right)t_{k,ij}  }\nn\\
  &=& \int_{pq}  \Bigg\{ \Bigg[ \left(\delta_{ij}-\frac{k_ik_j}{k^2}\right) - 2\frac{(p\times q)_i(p\times q)_j}{(p\times q)^2} \Bigg] 
  \frac{(p\times q)^2}{k^2p^2}\Bigg[\theta_p\delta\epsilon_q - \frac12 \frac{p\cdot q}{q^2}\theta_p g_q +\theta_p\frac{(p\times q)\cdot \nu_q}{q^2}\Bigg]\nn\\
  && {} + \frac{1}{k^2p^2}\Bigg[(q^2k_i-k^2q_i)(p_m\delta_{j\ell}+p_\ell\delta_{jm})+(q^2k_j-k^2q_j)(p_m\delta_{i\ell}+p_\ell\delta_{im})\nn\\
  && {} +p\cdot q\left(\delta_{i\ell}\delta_{jm}+\delta_{im}\delta_{j\ell}-\frac{p_\ell p_m}{k^2}\left(\delta_{ij}-\frac{k_ik_j}{k^2}\right)\right)\nn\\
  && {} -2\frac{p_\ell p_m}{k^2}\left(k_ip_j+k_jp_i-k_ik_j+q^2\delta_{ij}-k\cdot p\frac{k_ik_j}{k^2}\right)\Bigg]\theta_pt_{q,\ell m}\nn\\
  && {} +\frac{1}{k^2p^2}\Bigg[(p\times w_p)_i\left(k_j(k\cdot p)-k^2p_j)\right)+(p\times w_p)_j\left(k_i(k\cdot p)-k^2p_i)\right)\nn\\
  && {} +(p\times q)\cdot w_p\left(k\cdot p\left(\delta_{ij}+\frac{k_ik_j}{k^2}\right)-k_ip_j-k_jp_i\right)\Bigg]\delta\epsilon_q\nn\\
  && {} +\frac{1}{p^2q^2}\Bigg[(p\times w_p)_i(p\cdot q)\left(k\cdot q\frac{k_j}{k^2}-q_j\right)+(p\times w_p)_j(p\cdot q)\left(k\cdot q\frac{k_i}{k^2}-q_i\right)\nn\\
  && {}  +(p\times q)\cdot w_p\Bigg\{\left(\delta_{ij}+\frac{k_ik_j}{k^2}\right)\frac{k\cdot q(k\cdot q+2p\cdot q)}{2k^2}+q_iq_j\nn\\
  && {}  + \frac{1}{2}(k^2-p^2)\left(\delta_{ij}-\frac{k_ik_j}{k^2}+2\frac{k_iq_j+k_jq_i}{k^2}\right)\Bigg\}\Bigg]g_q\nn\\
  && {} + \frac{p^2k_i-k^2p_i}{k^2p^2q^2}\left[(p\times q)\cdot\nu_q(p\times w_p)_j-(p\times q)\cdot w_p(q\times\nu_q)_j\right]\nn\\
  && {}  + \frac{p^2k_j-k^2p_j}{k^2p^2q^2}\left[(p\times q)\cdot\nu_q(p\times w_p)_i-(p\times q)\cdot w_p(q\times\nu_q)_i\right]\nn\\
  && {}  +\frac{p\cdot q}{p^2q^2}\Bigg[(p\times w_p)_i(q\times\nu_q)_j+(p\times w_p)_j(q\times\nu_q)_i-(p\times w_p)\cdot(q\times\nu_q)\left(\delta_{ij}-\frac{k_ik_j}{k^2}\right)\Bigg]\nn\\
  && {}  - \frac{(p\times q)\cdot w_p(p\times q)\cdot\nu_q}{p^2q^2}\Bigg[\left(\delta_{ij}+\frac{k_ik_j}{k^2}\right)\frac{3p\cdot q+2q^2}{k^2}-2\frac{k_iq_j+k_jq_i}{k^2}-\left(\delta_{ij}-\frac{k_ik_j}{k^2}\right)\Bigg]\nn\\
  && {} + \frac{1}{2p^2}\Bigg[2\left(\delta_{ij}-\frac{k_ik_j}{k^2}\right)\left[(p\times w_p)_mp_n+(p\times w_p)_np_m\right]\nn\\
  && {} -2(p\times w_p)_i\left(p_n\delta_{jm}+p_m\delta_{jn}-2\frac{p_mp_nk_j}{k^2}\right) -2(p\times w_p)_j\left(p_n\delta_{im}+p_m\delta_{in}-2\frac{p_mp_nk_i}{k^2}\right)\nn\\
  && {} +(p\times q)\cdot w_p\Bigg(6\frac{p_mp_n}{k^2}\left(\delta_{ij}+\frac{k_ik_j}{k^2}\right)+\delta_{im}\left(\delta_{jn}-4\frac{p_nk_j}{k^2}\right)\nn\\
  && {} +\delta_{jn}\left(\delta_{ik}-4\frac{p_mk_i}{k^2}\right)+\delta_{in}\left(\delta_{jm}-4\frac{p_mk_j}{k^2}\right)+\delta_{jm}\left(\delta_{in}-4\frac{p_nk_i}{k^2}\right)\Bigg)\Bigg]t_{q,mn}\Bigg\}\,.
\eea

\section{Vertices up to the second cumulant}
\label{app:vert}

In this appendix we collect all vertices $\gamma_{abc}(p,q)$ for perturbation modes up to the second cumulant.
We write $\epsilon$ instead of $\delta\epsilon$ in the index for simplicity.
Furthermore $k\equiv p+q$ in the vectorial sense.

\subsection{Vertices involving only scalar perturbations}

\bea
  \gamma_{\delta\theta\delta}(p,q) &=& \frac12\alpha_{pq} = \frac{(p+q)\cdot p}{2p^2}\,,\nn\\
  \gamma_{\theta\theta\theta}(p,q) &=& \beta_{pq} = \frac{(p+q)^2 p\cdot q}{2p^2q^2} \,,\nn\\
  \gamma_{A\theta A}(p,q) &=& \frac12\frac{q\cdot p}{p^2} \,,\nn\\
  \gamma_{\theta A g}(p,q) &=& -\frac12(p+q)\cdot q\ \frac{q\cdot p}{q^2} \,,\nn\\
  \gamma_{\theta A \epsilon}(p,q) &=& -\frac12(p+q)\cdot p  \,,
\eea
\bea
  \gamma_{g \theta g}(p,q) &=& \frac12\frac{p\cdot q}{q^2}\left(\frac{k^2}{p^2}+\frac12-\frac32\frac{(k\cdot p)^2}{k^2p^2}\right) \,,\nn\\
  \gamma_{g \theta \epsilon}(p,q) &=& \frac12\left(3\frac{(k\cdot p)^2}{k^2p^2}-1\right) \,,\nn\\
  \gamma_{\epsilon \theta g}(p,q) &=& \frac12\frac{p\cdot q}{2p^2q^2k^2}((k\cdot p)^2-p^2k^2) \,,\nn\\
  \gamma_{\epsilon \theta \epsilon}(p,q) &=& \frac12 \frac{k\cdot p k\cdot q}{k^2p^2} \,.
\eea

\subsection{Vertices involving at most one vorticity, vector or tensor perturbation}

\bea\label{eq:gamBack}
  \gamma_{\delta w_i \delta}(p,q) &=& \gamma_{A w_i A}(p,q) = \frac12\frac{(p\times q)_i}{p^2} \,,\nn\\  
  \gamma_{\theta w_i \theta}(p,q) &=&  \frac12\left(1+\frac{2p\cdot q}{q^2}\right) \frac{(p\times q)_i}{p^2} \,,\nn\\  
  \gamma_{\theta A \nu_i}(p,q) &=&  \frac12\left(1+\frac{2p\cdot q}{q^2}\right) (p\times q)_i \,,\nn\\  
  \gamma_{\theta A t_{ij}}(p,q) &=&  - \frac12 p_i p_j \,,
\eea
\bea
  \gamma_{g w_i g}(p,q) &=&  \frac12\frac{1}{p^2}\left(\frac32\frac{(k\cdot q)^2}{k^2q^2}+3\frac{k\cdot q p\cdot q}{k^2q^2}-\frac{p\cdot q}{q^2}-\frac12\right)(p\times q)_i \,,\nn\\  
  \gamma_{g w_i \epsilon}(p,q) &=&  \frac12\, 3\frac{k\cdot p}{p^2k^2}(p\times q)_i \,,\nn\\  
  \gamma_{g \theta \nu_i}(p,q) &=&  -\frac12 \frac{1}{k^2p^2q^2}\left(p\cdot qk^2+3k\cdot p k\cdot q\right) (p\times q)_i \,,\nn\\  
  \gamma_{g \theta t_{ij}}(p,q) &=&  -\frac12 \frac{2k^2-3p\cdot q-6k\cdot p}{2k^2p^2}p_ip_j \,,
\eea
\bea
  \gamma_{\epsilon w_i g}(p,q) &=& \frac12\frac{k^2q^2-(k\cdot q)^2+2p\cdot q k\cdot p}{2p^2q^2k^2}(p\times q)_i \,, \nn\\  
  \gamma_{\epsilon w_i \epsilon}(p,q) &=& \frac12 \frac{k\cdot q}{p^2k^2}(p\times q)_i \,,\nn\\  
  \gamma_{\epsilon \theta \nu_i}(p,q) &=& \frac12 \frac{1}{q^2}\left(1-\frac{(k\cdot p)^2}{k^2p^2}\right) (p\times q)_i \,,\nn\\  
  \gamma_{\epsilon \theta t_{ij}}(p,q) &=& \frac12\frac{2k\cdot q-p\cdot q}{2k^2p^2}p_ip_j   \,,
\eea
\bea\label{eq:gamGen}
  \gamma_{w_i A g}(p,q) &=& - \frac12\frac{p\cdot q}{q^2}(p\times q)_i \,,\nn\\  
  \gamma_{w_i A \epsilon}(p,q) &=&\frac12 (p\times q)_i \,,\nn\\  
  \gamma_{\nu_i \theta g}(p,q) &=& \frac12\frac{p\cdot q k\cdot p (p\times q)_i}{k^2p^2q^2} \,,\nn\\  
  \gamma_{\nu_i \theta \epsilon}(p,q) &=& -\frac12\,2\frac{k\cdot p (p\times q)_i}{k^2p^2} \,,
\eea
\bea
  \gamma_{t_{ij} \theta g}(p,q) 
  &=& -\frac12\frac{(p\times q)^2\,p\cdot q}{2k^2p^2q^2}\Bigg\{\left(\delta_{ij}-\frac{k_ik_j}{k^2}\right) -2 \frac{(p\times q)_i(p\times q)_j}{(p\times q)^2}\Bigg\}\,,\nn\\
  \gamma_{t_{ij} \theta \epsilon}(p,q) 
  &=& \frac12\frac{(p\times q)^2}{k^2p^2}\Bigg\{\left(\delta_{ij}-\frac{k_ik_j}{k^2}\right) -2 \frac{(p\times q)_i(p\times q)_j}{(p\times q)^2}\Bigg\}\,.
\eea

\subsection{Vertices involving two vorticity, vector or tensor perturbations}

\bea\label{eq:gamSVV}
 \gamma_{\theta w_iw_j}(p,q) &=&  -\frac{(p \times q)_i(p \times q)_j}{p^2q^2} \,,\nn\\ 
 \gamma_{\epsilon w_i\nu_j}(p,q) &=& \frac12\, \frac{(p\cdot q-p^2+q^2)(p\times q)_i(p\times q)_j}{p^2q^2k^2} \,,\nn\\
 && {} + \frac12\frac{k^2(p\cdot q)\left(\delta_{ij}\,p\cdot q - p_jq_i\right)}{p^2q^2k^2}\,,\nn\\
 \gamma_{g w_i\nu_j}(p,q) &=& \frac12\, \frac{(p^2-5q^2-7p\cdot q)(p\times q)_i(p\times q)_j}{p^2q^2k^2}\,,\nn\\
 && {} -\frac12 \frac{k^2(p\cdot q)\left(\delta_{ij}\,p\cdot q - p_jq_i\right)}{p^2q^2k^2}\,,\nn\\
 \gamma_{g w_it_{jm}}(p,q) &=& \frac12\, \frac{9(p\times q)_ip_j+2k^2\varepsilon_{ijl}p_l}{2k^2p^2}p_m\,,\nn\\
 \gamma_{\epsilon w_it_{jm}}(p,q) &=& -\frac12\, \frac{3(p\times q)_ip_j+2k^2\varepsilon_{ijl}p_l}{2k^2p^2}p_m\,,
\eea
\bea\label{eq:gamVSV}
 \gamma_{w_i\theta w_j}(p,q) &=&  \frac12\,\frac{\delta_{ij}p\cdot k - p_ip_j}{p^2}\,,\nn\\
 \gamma_{w_iA\nu_j}(p,q) &=&  \frac12\,\frac{(p \times q)_i(p \times q)_j + (p\cdot q)\left(p_jq_i - \delta_{ij}(k\cdot q)\right)}{q^2}\,,\nn\\
 \gamma_{w_iAt_{nm}}(p,q) &=& -\frac{1}{2}\varepsilon_{ijn}k_jp_m\,,\nn\\
 \gamma_{\nu_iw_jg}(p,q) &=&  \frac12\, \frac{\left(\delta_{ij} p\cdot k-p_iq_j\right)( q\cdot k)( p\cdot q)}{p^2q^2k^2}\nn\\
 && {} +\frac12 \frac{(p\times q)_i(p\times q)_j(k^2-p^2)}{p^2q^2k^2}\,,\nn\\
 \gamma_{\nu_iw_j\epsilon}(p,q) &=&  \frac12\, \frac{\left(\delta_{ij} p\cdot k-p_iq_j\right)( p\cdot k) - ( p\times q)_i( p\times q)_j}{p^2k^2}\,,\nn\\
 \gamma_{\nu_i\theta \nu_j}(p,q) &=&  \frac12\, \frac{\left(\delta_{ij}q\cdot k-p_jq_i\right)(p\cdot q)k^2 + (p\times q)_i( p\times q)_j(q^2-p^2)}{p^2q^2k^2}\,,\nn\\
 \gamma_{\nu_i\theta t_{mj}}(p,q) &=&  \frac12\, \frac{(k^2-q^2)\varepsilon_{inm}k_n-(p\times q)_ip_m}{k^2p^2}p_j\,,\nn\\
\eea
\bea
\gamma_{t_{ij}w_\ell g}(p,q) &=& \frac12\, \frac{1}{p^2q^2}\Bigg[\varepsilon_{in\ell}\,p_n(p\cdot q)\left(k\cdot q\frac{k_j}{k^2}-q_j\right)+\varepsilon_{jn\ell}\,p_n(p\cdot q)\left(k\cdot q\frac{k_i}{k^2}-q_i\right)\nn\\
  && {}  +\varepsilon_{\ell nm}\,p_nq_m\Bigg\{\left(\delta_{ij}+\frac{k_ik_j}{k^2}\right)\frac{k\cdot q(k\cdot q+2p\cdot q)}{2k^2}+q_iq_j\nn\\
  && {}  + \frac{1}{2}(k^2-p^2)\left(\delta_{ij}-\frac{k_ik_j}{k^2}+2\frac{k_iq_j+k_jq_i}{k^2}\right)\Bigg\}\Bigg]\,,\nn\\
\gamma_{t_{ij}w_\ell\epsilon}(p,q) &=& \frac12\,\frac{p_n}{k^2p^2}\Bigg[\varepsilon_{in\ell}\left(k_j(k\cdot p)-k^2p_j)\right)+\varepsilon_{jn\ell}\left(k_i(k\cdot p)-k^2p_i)\right)\nn\\
  && {} +\varepsilon_{\ell nm}\,q_m\left(k\cdot p\left(\delta_{ij}+\frac{k_ik_j}{k^2}\right)-k_ip_j-k_jp_i\right)\Bigg]\,,\nn\\
\gamma_{t_{ij}\theta\nu_\ell}(p,q) &=& \frac12\, \frac{\varepsilon_{mn\ell}\,p_mq_n(p\times q)^2}{k^2p^2q^2}\Bigg[\left(\delta_{ij}-\frac{k^ik^j}{k^2}\right)-2\frac{(p\times q)_i(p\times q)_j}{(p\times q)^2}\Bigg]\,,\nn\\
\gamma_{t_{ij}\theta t_{\ell m}}(p,q) &=& \frac12\, \frac{1}{k^2p^2}\Bigg[(q^2k_i-k^2q_i)(p_m\delta_{j\ell}+p_\ell\delta_{jm})+(q^2k_j-k^2q_j)(p_m\delta_{i\ell}+p_\ell\delta_{im})\nn\\
  && {} +p\cdot q\left(\delta_{i\ell}\delta_{jm}+\delta_{im}\delta_{j\ell}-\frac{p_\ell p_m}{k^2}\left(\delta_{ij}-\frac{k_ik_j}{k^2}\right)\right)\nn\\
  && {} -2\frac{p_\ell p_m}{k^2}\left(k_ip_j+k_jp_i-k_ik_j+q^2\delta_{ij}-k\cdot p\frac{k_ik_j}{k^2}\right)\Bigg]\,.
\eea

\subsection{Vertices involving three vorticity, vector or tensor perturbations}

\bea\label{eq:gamVVV}
\gamma_{w_iw_jw_\ell}(p,q) &=& \frac{\varepsilon_{imj}\,p_\ell p_m + \delta_{i\ell}(p \times q)_j}{2p^2} + \frac{\varepsilon_{im\ell}\,q_jq_m - \delta_{ij}(p\times q)_\ell}{2q^2}\,,\nn\\
\gamma_{\nu_iw_j\nu_\ell}(p,q) &=& \frac12\, \frac{1}{p^2q^2k^2}\Bigg\{\left[\delta_{i\ell}q\cdot k-p_\ell q_i\right](p\times q)_j(k^2-p^2)\nn\\        && {} - \left[\delta_{ij}p\cdot k-p_iq_j\right](p\times q)_\ell(k^2-p^2)\nn\\
&& {} - 2(p\times q)_i(p\times q)_j(p\times q)_\ell\Bigg\}\,,\nn\\
\gamma_{\nu_iw_\ell t_{mj}}(p,q) &=& \frac12\, \frac{2(p\times q)_\ell\,\varepsilon_{inm}k_n-p_m\left(p_iq_\ell-k\cdot p\delta_{i\ell}\right)}{k^2p^2}p_j\,,
\eea
\bea
\gamma_{t_{ij}w_\ell\nu_m}(p,q) &=& \frac12\,  \frac{p^2k_i-k^2p_i}{k^2p^2q^2}\left[\varepsilon_{mns}\,\varepsilon_{jr\ell}\,p_r-\varepsilon_{\ell ns}\,\varepsilon_{jrm}\,q_r\right]p_nq_s\nn\\
  && {}  + \frac12\frac{p^2k_j-k^2p_j}{k^2p^2q^2}\left[\varepsilon_{mns}\,\varepsilon_{ir\ell}\,p_r-\varepsilon_{\ell ns}\,\varepsilon_{irm}\,q_r\right]p_nq_s\nn\\
  && {}  +\frac12\frac{p\cdot q}{p^2q^2}p_nq_r\Bigg[\varepsilon_{in\ell}\,\varepsilon_{jrm}+\varepsilon_{jn\ell}\,\varepsilon_{irm}-\varepsilon_{sn\ell}\,\varepsilon_{srm}\left(\delta_{ij}-\frac{k_ik_j}{k^2}\right)\Bigg]\nn\\
  && {}  - \frac12\frac{\varepsilon_{\ell ns}\,\varepsilon_{mkr}\,p_np_kq_sq_r}{p^2q^2}\Bigg[\left(\delta_{ij}+\frac{k_ik_j}{k^2}\right)\frac{3p\cdot q+2q^2}{k^2}-2\frac{k_iq_j+k_jq_i}{k^2}-\left(\delta_{ij}-\frac{k_ik_j}{k^2}\right)\Bigg]\,,\nn\\
\gamma_{t_{ij}w_\ell t_{mn}}(p,q) &=& \frac12\, \frac{1}{2p^2}\Bigg[2\left(\delta_{ij}-\frac{k_ik_j}{k^2}\right)\left[\varepsilon_{mr\ell}\,p_n+\varepsilon_{nr\ell}\,p_m\right]p_r\nn\\
  && {} -2\varepsilon_{ir\ell}\,p_r\left(p_n\delta_{jm}+p_m\delta_{jn}-2\frac{p_mp_nk_j}{k^2}\right)-2\varepsilon_{jr\ell}\,p_r\left(p_n\delta_{im}+p_m\delta_{in}-2\frac{p_mp_nk_i}{k^2}\right)\nn\\
  && {} +\varepsilon_{\ell rs}\,p_rq_s\Bigg(6\frac{p_mp_n}{k^2}\left(\delta_{ij}+\frac{k_ik_j}{k^2}\right)+\delta_{im}\left(\delta_{jn}-4\frac{p_nk_j}{k^2}\right)\nn\\
  && {} +\delta_{jn}\left(\delta_{ik}-4\frac{p_mk_i}{k^2}\right)+\delta_{in}\left(\delta_{jm}-4\frac{p_mk_j}{k^2}\right)+\delta_{jm}\left(\delta_{in}-4\frac{p_nk_i}{k^2}\right)\Bigg)\Bigg]\,.
\eea

\section{Third and fourth cumulant}
\label{app:higher}

\subsection{Evolution matrix for scalar perturbations}

The scalar part of the block-diagonal evolution matrix $\Omega_{ab}(k,\eta)$ for the perturbation modes up to the fourth cumulant
\be
  \psi^S=(\delta,\theta,g,\delta\epsilon,A,\pi,\chi,\kappa,\xi,\psi)\,,
\ee
in the approximation $\Omega_m/f^2\to 1$ is given by
\be
  \Omega^S = \left(\begin{array}{cccccccccc}
  & -1 \\
  -3/2 & 1/2 & k^2 & k^2 & k^2\epsilon \\
  & -2\epsilon & 1 & & & 1 & -3/5\\
  & & & 1 & & &  1/5\\
  & -1 \\
  & & -3k^2\epsilon & -5 k^2\epsilon & -k^2\omega & 3/2 & & & -k^2\\
  & &  & -5 k^2\epsilon & -k^2\omega & & 3/2 & -5/2k^2 & 5/2k^2 & 4k^2\\
  & -4\omega & & & & 4\epsilon & & 2 \\
  & -16/5\omega & & & & 4\epsilon & -4/5\epsilon & & 2 \\
  & & & & & & & & & 2
  \end{array}\right)\,.
\ee

\subsection{Vertices involving scalar modes $\pi$ and $\chi$ of the third cumulant}

\bea
 \gamma_{gA\pi}(p,q) &=& \frac14 \frac{p\cdot q}{q^2} - \frac34 \frac{(k\cdot q)^2 p\cdot q}{k^2 q^4}\,,\nn\\
 \gamma_{g A\chi}(p,q) &=& -3\gamma_{\epsilon A\chi}(p,q)\,,\nn\\
 \gamma_{\epsilon A\pi}(p,q) &=& - \frac14 \frac{p\cdot q}{q^2} + \frac14 \frac{(k\cdot q)^2 p\cdot q}{k^2 q^4}\,,\nn\\
 \gamma_{\epsilon A\chi}(p,q) &=& \frac14 \left( \frac{p\cdot q}{5q^2} + \frac25 \frac{k\cdot p k\cdot q}{k^2 q^2} - \frac{(k\cdot q)^2 p\cdot q}{k^2 q^4} \right)\,,
\eea

\bea
  \gamma_{\pi gg}(p,q) &=& \frac{k^2 (p\cdot q)^2}{p^2 q^2} +\frac12 p\cdot q \left(\frac{k\cdot q}{q^2}+\frac{k\cdot p}{p^2}\right)\,,\nn\\
  \gamma_{\pi \epsilon g}(p,q) &=& \frac12 k\cdot q \left(3+\frac{5p\cdot q}{q^2}\right)\,,\nn\\
  \gamma_{\pi \epsilon\epsilon}(p,q) &=& \frac52 k^2\,,\nn\\
  \gamma_{\pi \theta\pi}(p,q) &=& \frac12 \frac{k^2 p\cdot q}{p^2 q^2} +  \frac{k\cdot q (p\cdot q)^2}{p^2q^4}\,, \nn\\
  \gamma_{\pi \theta\chi}(p,q) &=& \frac15 \frac{k\cdot q}{q^2} + \frac25 \frac{k\cdot p p\cdot q}{p^2q^2} - \frac{k\cdot q (p\cdot q)^2}{p^2q^4} \,,
\eea

\bea
  \gamma_{\chi gg}(p,q) &=& -\frac{15}{4} \frac{k\cdot q k\cdot p p\cdot q}{p^2q^2} +\frac52\gamma_{\pi gg}(p,q) \,,\nn\\
  \gamma_{\chi \epsilon g}(p,q) &=& -\frac{15}{4}\left( \frac{k\cdot q p\cdot q }{q^2} + \frac{(k\cdot q)^3}{k^2q^2} \right) +\frac52 \gamma_{\pi \epsilon g}(p,q)\,,\nn\\
  \gamma_{\chi \epsilon\epsilon}(p,q) &=& -\frac{15}{4} k^2 +\frac52\gamma_{\pi \epsilon\epsilon}(p,q)\,,\nn\\
  \gamma_{\chi \theta\pi}(p,q) &=& -\frac54 \frac{(k\cdot q)^2p\cdot q(k^2+2k\cdot p)}{k^2p^2q^4} +\frac52\gamma_{\pi \theta\pi}(p,q)\,,\nn\\
  \gamma_{\chi \theta\chi}(p,q) &=& -\frac34 \frac{k^2 p\cdot q}{p^2q^2}-\frac32\frac{(k\cdot p)^2k\cdot q}{k^2p^2q^2}+\frac54\frac{(k\cdot q)^2p\cdot q(k^2+2k\cdot p)}{k^2p^2q^4} +\frac52\gamma_{\pi \theta\chi}(p,q)\,.
\eea

\section{The local cosmic energy equation}
\label{app:energy}

In this section we review the evolution equation for the total energy (see e.g.~\cite{BinTre08}), given by the sum of kinetic and potential energy,
\bea
  E_\text{kin}&=&\frac12\int d^3p\frac{p^2}{a^2}f(\tau,\vec x,\vec p)=\frac12(1+\delta)(\sigma_{ii}+v_iv_i)\,,\nn\\
  E_\text{pot}&=&\frac12 \Phi\delta\,.
\eea
Using the evolution equation for $\sigma_{ij}$, the continuity, Euler and Poisson equations one obtains
\bea
  \partial_\tau E_\text{kin}+2{\cal H}E_\text{kin}+\nabla_i J_i^\text{kin} &=& -\Phi\partial_\tau\delta\,,\nn\\
  \partial_\tau E_\text{pot}+q{\cal H}E_\text{pot}+\nabla_i J_i^\text{pot} &=& +\Phi\partial_\tau\delta\,,
\eea
where
\bea
  J_i^\text{kin} &=& E_\text{kin} v_i+(1+\delta)\left(\sigma_{ij}v_j+v_i\Phi+\frac12{\cal C}_{ijj}\right)\,,\nn\\
  J_i^\text{pot} &=& \frac{1}{8\pi G\bar\rho_0 a^2}\left(\Phi\partial_\tau\nabla_i\Phi-(\nabla_i\Phi)\partial_\tau\Phi\right)\,,
\eea
with third cumulant ${\cal C}_{ijk}$ and
\be
  q \equiv -\frac{d \ln(a^2\bar\rho_0)}{d\ln a} \to 1\,,
\ee
for the usual scaling $\bar\rho_0\propto a^{-3}$ of the matter rest energy density, that we assume throughout.
The evolution of the total energy is therefore given by
\be
 \partial_\tau E+{\cal H}(2E_\text{kin}+E_\text{pot})+\nabla_i J_i=0\,,
 \label{LCEE}
\ee
with $J_i=J_i^\text{kin}+J_i^\text{pot}$, and recovering the energy conservation law on a static background. Equation~(\ref{LCEE}) is the local version of the cosmic energy equation~\cite{Pee80}. 
Rescaling to $\eta=\ln(D)$ and $\hat X=X/(f{\cal H})^2$ for $X=\Phi,E,E_\text{kin},E_\text{pot}$ and $\hat X_i=X_i/(-f{\cal H})^3$ for $X=J,J^\text{kin},J^\text{pot}$
gives
\bea
  \partial_\eta \hat E_\text{kin}+2\left(\frac32\frac{\Omega_m}{f^2}-1\right)\hat E_\text{kin}-\nabla_i \hat J_i^\text{kin} &=& -\hat\Phi\partial_\eta\delta\,,\nn\\
  \partial_\eta \hat E_\text{pot}+2\left(\frac32\frac{\Omega_m}{f^2}-1-\frac{1}{2f}\right)\hat E_\text{pot}-\nabla_i \hat J_i^\text{pot} &=& +\hat\Phi\partial_\eta\delta\,,
\eea
where
\bea
  \hat J_i^\text{kin} &=& \hat E_\text{kin} u_i+(1+\delta)\left(\epsilon_{ij}u_j+u_i\hat \Phi+\frac12 \pi_{ijj}\right)\,,\nn\\
  \hat J_i^\text{pot} &=& \frac{f^2}{3 \Omega_m}\left(\hat\Phi\partial_\eta\nabla_i\hat\Phi-(\nabla_i\hat\Phi)\partial_\eta\hat\Phi\right)\,,
\eea
with $\pi_{ijk}={\cal C}_{ijk}/(-f{\cal H})^3$ and
\bea
  \hat E_\text{kin}&=&\frac12(1+\delta)(\epsilon_{ii}+u_iu_i)\,,\nn\\
  \hat E_\text{pot}&=&\frac12 \hat\Phi\delta\,,\nn\\
\eea
so that
\be\label{eq:dEhat}
  \partial_\eta \hat E+2\left(\frac32\frac{\Omega_m}{f^2}-1\right)\hat E-\nabla_i \hat J_i = \frac{1}{f}\hat E_\text{pot}\,,
\ee
where $\hat J_i=\hat J_i^\text{kin}+\hat J_i^\text{pot}$.

Let us check that the local cosmic energy equation is indeed satisfied at lowest order in perturbation theory, when including second and higher cumulant contributions
as done in this work. We insert the expansion $\epsilon_{ij}=\epsilon(\eta)\delta_{ij}+\delta\epsilon_{ij}$ as well as the decomposition of $\delta\epsilon_{ij}$ and $\pi_{ijk}$ from Eqs.~(\ref{eq:epsSVT},~\ref{eq:piLamdecomposition}). Keeping terms up to linear order in perturbations (except for contributions involving $\delta\epsilon$, see below)
yields
\bea
  \hat E_\text{kin} &=& \frac32(1+\delta)(\epsilon(\eta)+\delta\epsilon)+\frac12 g \,,\nn\\
  \nabla_i\hat J_i &=& \frac52\epsilon \nabla_iu_i+\frac12\nabla_i\pi_{ijj} = \frac52\epsilon(\eta) \theta -\frac12\pi\,,
\eea
while $\hat E_\text{pot} = 0$ at this order. In the following we use the EdS approximation $\Omega_m=f=1$. One can check that, as expected, Eq.~\eqref{eq:dEhat}
is indeed satisfied up to terms of higher order in perturbation theory when using the linear equations of motion (see Sec.\,\ref{sec:pi})
\be
  \partial_\eta\delta=\theta\,,\quad
  (\partial_\eta+1)\delta\epsilon=-\chi/5-Q\,,\quad
  (\partial_\eta+1)g=2\theta\epsilon(\eta)-\pi-3\chi/5\,,
\ee
as well as the background dispersion evolution equation $(\partial_\eta+1)\epsilon(\eta)=Q$.
Note that here it is important to retain the contribution from $Q$ in the equation for $\delta\epsilon$,
which has to cancel with corresponding contributions from the equation for $\epsilon(\eta)$.
Alternatively, this cancellation can be seen directly by realizing that they appear in the
combination $\epsilon+\delta\epsilon$, and that $Q$ drops out when adding the corresponding equations of motion.

\section{Kernels in linear approximation}
\label{app:lin}

The linear solution for the velocity divergence $\theta_k$ and scalar perturbation mode $\bar g_k=g_k/\epsilon$ of the velocity dispersion tensor in the second-cumulant approximation
and for a power-law dependence $\epsilon=\epsilon_0e^{\alpha\eta}$ is given by Eq.~\eqref{eq:TthetaTg},
with linear kernels
\bea
  F_{1,\theta}(k,\eta) &=& F_{1,\delta}(k,\eta) - \frac{2(4+\alpha)k^2\epsilon}{(2+\alpha)(5+2\alpha)} \,\; {}_1 F_2\left(\frac{4+4\alpha}{3\alpha};2+\frac{2}{\alpha},2+\frac{5}{2\alpha};\frac{-3k^2\epsilon}{\alpha^2}\right)\,, \nn\\
 F_{1,\bar g}(k,\eta) & =& \frac{2}{2+\alpha} \Bigg[ \; {}_1 F_2\left(\frac{4+\alpha}{3\alpha};2+\frac{2}{\alpha},1+\frac{5}{2\alpha};\frac{-3k^2\epsilon}{\alpha^2}\right)\nn\\
  && - \frac{(4+\alpha)k^2\epsilon}{(1+\alpha)(5+2\alpha)} \,\; {}_1 F_2\left(\frac{4+4\alpha}{3\alpha};3+\frac{2}{\alpha},2+\frac{5}{2\alpha};\frac{-3k^2\epsilon}{\alpha^2}\right)\Bigg]\,.
\eea
The coefficients for the expansion Eq.~\eqref{eq:Tlarge} for large $k^2\epsilon$ are given in Table~\ref{tab:Tasymptotic}.

\begin{table*}
  \centering
  \caption{Coefficients in the expansion Eq.~\eqref{eq:Tlarge} of the linear kernels $F_{1,a}(k,\eta)$ for large $k^2\epsilon$.}
  \begin{ruledtabular}
    \begin{tabular}{c|ccccc}
  $a$ & $d_a$ & $e_a$ & $D_a$ & $E_a$ & $\varphi_a$ \\ \hline \\[-2.5ex]
  $\delta$ & $\frac{2}{3\alpha}(4+\alpha)$ & $\frac{19+7\alpha}{6\alpha}$ & 
  $\frac{5\alpha^{4\frac{2-\alpha}{3\alpha}}\Gamma(\frac{2}{\alpha})\Gamma(\frac{5}{2\alpha})}{\Gamma(\frac{4\alpha+7}{6\alpha})\Gamma(\frac{2(1+\alpha)}{3\alpha})}$ & 
  $-\frac{5\alpha^{\frac{19-5\alpha}{6\alpha}}\Gamma(\frac{2}{\alpha})\Gamma(\frac{5}{2\alpha})}{\sqrt{\pi}\Gamma(\frac{4+\alpha}{3\alpha})}$ & 
  $\frac{5\alpha-19}{12\alpha}\pi$
  \\[1.5ex]
  $\theta$ & $\frac{2}{3\alpha}(4+\alpha)$ & $\frac{19+\alpha}{6\alpha}$ & 
  $-\frac{5\alpha^{4\frac{2-\alpha}{3\alpha}}(\alpha+1)\Gamma(\frac{2}{\alpha})\Gamma(\frac{5}{2\alpha})}{3\Gamma(\frac{4\alpha+7}{6\alpha})\Gamma(\frac{2(1+\alpha)}{3\alpha})}$ & 
  $-\frac{5\alpha^{\frac{19-5\alpha}{6\alpha}}\Gamma(\frac{2}{\alpha})\Gamma(\frac{5}{2\alpha})}{\sqrt{\pi}\Gamma(\frac{4+\alpha}{3\alpha})}$ & 
  $\frac{11\alpha-19}{12\alpha}\pi$
  \\[1.5ex]
  $\bar g$ & $\frac{2}{3\alpha}(4+\alpha)$ & $\frac{19+7\alpha}{6\alpha}$ & 
  $-\frac{5\alpha^{4\frac{2-\alpha}{3\alpha}}\Gamma(\frac{2}{\alpha})\Gamma(\frac{5}{2\alpha})}{\Gamma(\frac{4\alpha+7}{6\alpha})\Gamma(\frac{2(1+\alpha)}{3\alpha})}$ & 
  $-\frac{10\alpha^{\frac{19-5\alpha}{6\alpha}}\Gamma(\frac{2}{\alpha})\Gamma(\frac{5}{2\alpha})}{\sqrt{\pi}\Gamma(\frac{4+\alpha}{3\alpha})}$ & 
  $\frac{5\alpha-19}{12\alpha}\pi$\\
    \end{tabular}
  \end{ruledtabular}
  \label{tab:Tasymptotic}
\end{table*}

The full linear solution of the coupled system Eq.~\eqref{eq:lin2nd} for $\delta_k, \theta_k, g_k$, that includes all eigenmodes, can be found by making the ansatz
\be
  \delta_k = e^{\lambda\eta}\sum_{n\geq 0} c_n (\epsilon(\eta)k^2)^n\,,\qquad
  \theta_k = e^{\lambda\eta}\sum_{n\geq 0} d_n (\epsilon(\eta)k^2)^n\,,\qquad
  k^2 g_k = e^{\lambda\eta}\sum_{n\geq 0} e_n (\epsilon(\eta)k^2)^n\,,
\ee
with some exponent $\lambda$ and coefficients $c_n,d_n,e_n$. Inserting this ansatz into Eq.~\eqref{eq:lin2nd} and assuming $\epsilon\propto e^{\alpha\eta}$ yields the recursions
\be
  c_n(\lambda+\alpha n)=d_n\,,\qquad d_n(\lambda+\alpha n+\frac12)=\frac32 c_n-c_{n-1}-e_n\,,\qquad e_n(\lambda+\alpha n+1)=2d_{n-1}\,,
\ee
for $n\geq 1$. They can be combined into
\be
  c_n = -c_{n-1}\frac{3(\lambda+\alpha n)+1-2\alpha}{(\lambda+\alpha n-1)(\lambda+\alpha n+1)(\lambda+\alpha n+\frac32)}\,,
\ee
with explicit expression given by
\be
  c_n = \left(\frac{-3}{\alpha^2}\right)^n c_0 \frac{\Gamma(n+p_1)}{\Gamma(p_1)}\frac{\Gamma(q_1)}{\Gamma(n+q_1)}\frac{\Gamma(q_2)}{\Gamma(n+q_2)}\frac{\Gamma(q_3)}{\Gamma(n+q_3)}\,,
\ee
where 
\be
  p_1\equiv \frac{3\lambda+1+\alpha}{3\alpha}\,,\qquad q_1\equiv 1+\frac{\lambda-1}{\alpha}\,,\qquad q_2\equiv 1+\frac{\lambda+1}{\alpha}\,,\qquad q_3\equiv 1+\frac{\lambda+\frac32}{\alpha}\,.
\ee
The sum over all $n$ yields a solution in terms of generalized hypergeometric functions. To find the allowed values for $\lambda$ it is sufficient to consider Eq.~\eqref{eq:lin2nd} in the limit $\epsilon\to 0$, which yields a linear set of three algebraic equations for $(c_0,d_0,e_0)$,
\be
  \left(\begin{array}{ccc}
  \lambda & -1 & 0 \\ -\frac32 & \lambda+\frac12 & 1 \\ 0 & 0 & \lambda+1
  \end{array}\right) 
  \left(\begin{array}{c}
  c_0\\ d_0\\ e_0
  \end{array}\right) =0\,.
\ee
Non-trivial solutions exist for $\lambda=+1$ (with $c_0=d_0, e_0=0$), $\lambda=-3/2$ (with $c_0=-\frac23 d_0, e_0=0$) and $\lambda=-1$ (with $c_0=-d_0=e_0$).
Each possibility yields one of the three linearly independent solutions. Inserting the values for $\lambda$ in the recursion relation, and building a generic linear combination
of the three solutions, yields the general solution Eq.~\eqref{eq:lincum2gensol}. The solution for all three perturbation modes can be written as
\be\label{eq:cum2lingeneralthreemodes}
  \left(\begin{array}{c}
  \delta_k\\ \theta_k\\ g_k
  \end{array}\right) = {\cal M}_k(\eta)\left(\begin{array}{c}
  A\\ B\\ C
  \end{array}\right)\,,
\ee
with free coefficients $A,B,C$ and $3\times 3$ matrix given by the tensor product
\be 
  {\cal M}_k(\eta)\equiv
  \left(\begin{array}{c}
  1\\ \partial_\eta\\ (-\partial_\eta^2-\frac12\partial_\eta+\frac32-\epsilon k^2)/k^2
  \end{array}\right)\otimes
  \left( e^\eta\, {}_1 F_2\left(p_1;q_2,q_3;x\right),\,
    e^{-\frac32\eta}\, {}_1 F_2\left(p_1;q_1,q_2;x\right),\,
    e^{-\eta}\, {}_1 F_2\left(p_1;q_1,q_3;x\right)
  \right)\,,
\ee
with $x\equiv \frac{-3k^2\epsilon(\eta)}{\alpha^2}$, and $p_i, q_i$ evaluated for $\lambda=1,-3/2,-1$ for the three terms, respectively.
This result can be used to give an analytic expression for the linear propagator
\be
  G_k(\eta,\eta') = {\cal M}_k(\eta)[{\cal M}_k(\eta')]^{-1}\,.
\ee

A full treatment of the scalar modes in second cumulant approximation requires to include also the mode $\delta\epsilon_k$ as well as $A_k$.
Including $\delta\epsilon_k$ leads to an additional decaying mode solution $g_k=-\delta\epsilon_k=D e^{-\eta}$ and $\delta_k=\theta_k=A_k=0$, with free coefficient $D$.
This solution remains valid also for $\epsilon k^2\gg 1$, and can easily be included in the linear propagator by extending Eq.~\eqref{eq:cum2lingeneralthreemodes} by a fourth row.
The solutions given previously remain valid when including $\delta\epsilon_k$.

Finally, taking the $A_k$ mode into account formally yields an additional eigenvalue $\lambda=0$, related to the freedom to choose different initial conditions for $\delta_k$ and $A_k$. 
This is irrelevant for the linear evolution, but the additional linearly independent solution enters in the linear propagator. The corresponding additional solution is given by
\be
  \delta_k = E\, \left[ {}_2 F_3\left(1,p_1;q_1,q_2,q_3;x\right) -1\right]\,,
\ee
with free coefficient $E$ and $p_i, q_i$ evaluated for $\lambda=0$, and $A_k=E+\delta_k, \theta_k=\partial_\eta\delta_k$, $k^2 g_k=(-\partial_\eta^2-\frac12\partial_\eta+\frac32-\epsilon k^2)\delta_k$, and $\delta\epsilon_k=0$. The most general solution for all five scalar modes is therefore given by
\be
  \left(\begin{array}{c}
  \delta_k\\ \theta_k\\ g_k \\ \delta\epsilon_k \\ A_k
  \end{array}\right) = {\cal M}_k^{(5\times 5)}(\eta)\left(\begin{array}{c}
  A\\ B\\ C\\ D\\ E
  \end{array}\right)\,,
\ee
with ${\cal M}_k^{(5\times 5)}(\eta)$ given by the $5\times 5$ matrix
\be 
  \left(\begin{array}{ccc|cc}
  &&&0& {}_2 F_3\left(1,p_1;q_1,q_2,q_3;x\right) -1 \\
  &{\cal M}_k(\eta)&&0& \partial_\eta\,[ {}_2 F_3\left(1,p_1;q_1,q_2,q_3;x\right) -1 ] \\
  &&&-e^{-\eta}& D_\eta [{}_2 F_3\left(1,p_1;q_1,q_2,q_3;x\right) -1] \\ \hline
  0&0&0 & e^{-\eta} &0\\
  e^\eta\, {}_1 F_2\left(p_1;q_2,q_3;x\right) & e^{-\frac32\eta}\, {}_1 F_2\left(p_1;q_1,q_2;x\right) & e^{-\eta}\, {}_1 F_2\left(p_1;q_1,q_3;x\right) & 0 & {}_2 F_3\left(1,p_1;q_1,q_2,q_3;x\right) 
  \end{array}\right)\,,
\ee
where $D_\eta\equiv (-\partial_\eta^2-\frac12\partial_\eta+\frac32-\epsilon k^2)/k^2$. The corresponding linear propagator reads
\be
  G_k^{(5\times 5)}(\eta,\eta') = {\cal M}_k^{(5\times 5)}(\eta)[{\cal M}_k^{(5\times 5)}(\eta')]^{-1}\,.
\ee

\section{Evolution equations for the full hierarchy of cumulants}
\label{app:hierarchy}

The linear evolution equation obtained from expanding Eq.~\eqref{eq:eomCell} in powers of $L$ is given by
\bea
  \left[\partial_\eta+1+\left(\frac32\frac{\Omega_m}{f^2}-1\right)\left(\ell+2n\right)\right] {\cal C}_{\ell, 2n}  
  &=&  \frac{\{1,k^2\}}{2\ell+1}{\cal R}_{\ell, 2n}  + \frac{1}{2}\delta_{\ell 1}\delta_{n0}\frac{\Omega_m}{f^2}{\cal C}_{0, 0} \,,
\eea
where $\{A,B\}=A$ for even $\ell$, and $B$ for odd $\ell$, and
\bea
  {\cal R}_{\ell, 2n}  
  &\equiv&  \sum_{m=0}^n\frac{(2n)!}{(2m+1)!(2n-2m)!} {\cal E}_{2m+2}\left((\ell+1)(2n-2m)(2n-2m-1){\cal C}_{\ell+1,2(n-m-1)}-\ell{\cal C}_{\ell-1,2(n-m)}\right)\nn\\
  && + (\ell+1)\left(2\ell+3+2n\right){\cal C}_{\ell+1, 2n}
  - \frac{\ell}{2n+1} {\cal C}_{\ell-1,2(n+1)} \,.
\eea
We solve this system of equations with growing mode initial conditions
\bea
 {\cal C}_{0,0}&\to& 1\,,\nn\\
 {\cal C}_{1,0}&\to& \frac{1}{3}\,,\nn\\
 {\cal C}_{\ell, 2n}&\to& 0,\qquad \ell+2n\geq 2\,,
\eea
for $\eta\to-\infty$, as appropriate for cold dark matter and adiabatic initial conditions.

The evolution equation for the expectation values of even cumulants obtained from Eq.~\eqref{eq:E} and Taylor expanding Eq.~\eqref{eq:Qe} in $L$ is given by
\be\label{eq:eomE2n}
  \left[\partial_\eta + \left(\frac32\frac{\Omega_m}{f^2}-1\right)2n\right]{\cal E}_{2n}=Q_{{\cal E}_{2n}}\,,
\ee
with
\bea
  Q_{{\cal E}_{2n}}&=&4\pi\int_0^\infty dk\,k^2\,e^{2\eta}\,P_0(k)\sum_{\ell=0}^n (\ell+1) \sum_{m_1,m_2=0}^{n-\ell}\delta_{m_1+m_2,n-\ell}\frac{(2n)!}{(2m_1)!(2m_2)!}\nn\\
  &&(2m_2-2m_1-2\ell-3){\cal C}_{\ell+1,2m_1}{\cal C}_{\ell,2m_2}\,.
\eea
The evolution equations for the rescaled variables $T_{\ell,2n}={\cal C}_{\ell,2n}/\epsilon^{n+[\ell/2]}$ and $\bar{\cal E}_{2n}={\cal E}_{2n}/\epsilon^n$ read
\bea\label{eq:eomTl2n}
  \left[\partial_\eta+1+\left(\frac32\frac{\Omega_m}{f^2}-1\right)\left(\ell+2n\right)+(n+\{\ell/2,(\ell-1)/2\})(\partial_\eta\ln\epsilon)\right] T_{\ell, 2n}  
  &=&  \frac{\{1,\epsilon k^2\}}{2\ell+1}\bar {\cal R}_{\ell, 2n}  + \frac{1}{2}\delta_{\ell 1}\delta_{n0}\frac{\Omega_m}{f^2}T_{0, 0} \,,
\eea
\be
  \left[\partial_\eta + \left(\frac32\frac{\Omega_m}{f^2}-1\right)2n+ n(\partial_\eta\ln\epsilon) \right]\bar{\cal E}_{2n}=\bar Q_{{\cal E}_{2n}}\,,
\ee
where $\bar {\cal R}_{\ell, 2n}={\cal R}_{\ell, 2n}|_{{\cal C}\to T,{\cal E}\to\bar{\cal E}}$ and $\bar Q_{{\cal E}_{2n}}=Q_{{\cal E}_{2n}}|_{{\cal C}\to T}$.

\begin{table*}
  \centering
  \caption{Factor $q_{c_\text{max}}(\lambda)$ contributing to the characteristic polynomial related to the asymptotic behavior of the linear kernels for large $\epsilon k^2$ when taking scalar perturbation modes of cumulants up to order $c_\text{max}$ into account. Stability requires that all roots of $q_{c_\text{max}}(\lambda)$ lie on the negative real axis or are zero. 
  We set $X_{10}\equiv 945 + 1575 \bar{\cal E}_4^2 + 630 \bar{\cal E}_6 - 210 \bar{\cal E}_4 (15 + \bar{\cal E}_6) - 45 \bar{\cal E}_8 +
    \bar{\cal E}_{10}$ and $X_{12}\equiv 10395 + 51975 \bar{\cal E}_4^2 - 5775 \bar{\cal E}_4^3 + 13860 \bar{\cal E}_6 + 462 \bar{\cal E}_6^2 - 
 495 \bar{\cal E}_4 (105 + 28 \bar{\cal E}_6 - \bar{\cal E}_8) - 1485 \bar{\cal E}_8 + 66 \bar{\cal E}_{10} - \bar{\cal E}_{12}$.}
  \begin{ruledtabular}
    \begin{tabular}{c@{\ }|l}\\[-1.5ex] 
  $c_\text{max}$ & $q_{c_\text{max}}(\lambda)$ \\[1.5ex] \hline \\[-1.5ex]
  $3$ & $\lambda^2+2\lambda+\frac19(3-\bar{\cal E}_4)$
  \\[1.5ex]
  $4$ & $\lambda^2+\frac{10}{3}\lambda+\frac59(3-\bar{\cal E}_4)$
  \\[1.5ex]
  $5$ & $\lambda^3+5\lambda^2+\frac53\lambda(3-\bar{\cal E}_4) +\frac{1}{27}(15-15\bar{\cal E}_4+\bar{\cal E}_6)$
    \\[1.5ex]
  $6$ & $\lambda^3+7\lambda^2+\frac{35}{9}\lambda(3-\bar{\cal E}_4) +\frac{7}{27}(15-15\bar{\cal E}_4+\bar{\cal E}_6)$
    \\[1.5ex]
  $7$ & $\lambda^4+\frac{28}{3}\lambda^3+\frac{70}{9}\lambda^2(3-\bar{\cal E}_4) +\frac{28}{27}\lambda(15-15\bar{\cal E}_4+\bar{\cal E}_6)+\frac{1}{81} (105 - 210 \bar{\cal E}_4 + 35 \bar{\cal E}_4^2 + 28 \bar{\cal E}_6 - \bar{\cal E}_8)$
    \\[1.5ex]
  $8$ & $\lambda^4+12\lambda^3+14\lambda^2(3-\bar{\cal E}_4) +\frac{28}{9}\lambda(15-15\bar{\cal E}_4+\bar{\cal E}_6)+\frac{1}{9} (105 - 210 \bar{\cal E}_4 + 35 \bar{\cal E}_4^2 + 28 \bar{\cal E}_6 - \bar{\cal E}_8)$  
    \\[1.5ex]
  $9$ & $\lambda^5+15\lambda^4+\frac{70}{3}\lambda^3(3-\bar{\cal E}_4) +\frac{70}{9}\lambda^2(15-15\bar{\cal E}_4+\bar{\cal E}_6)+\frac{5}{9}\lambda (105 - 210 \bar{\cal E}_4 + 35 \bar{\cal E}_4^2 + 28 \bar{\cal E}_6 - \bar{\cal E}_8) + \frac{1}{243} X_{10}$
    \\[1.5ex]
  $10$ & $\lambda^5+\frac{55}{3}\lambda^4+\frac{110}{3}\lambda^3(3-\bar{\cal E}_4) +\frac{154}{9}\lambda^2(15-15\bar{\cal E}_4+\bar{\cal E}_6)+\frac{55}{27}\lambda (105 - 210 \bar{\cal E}_4 + 35 \bar{\cal E}_4^2 + 28 \bar{\cal E}_6 - \bar{\cal E}_8) + \frac{11}{243} X_{10}$
    \\[1.5ex]
  $11$ & $\lambda^6+22\lambda^5+55\lambda^4(3-\bar{\cal E}_4) +\frac{308}{9}\lambda^3(15-15\bar{\cal E}_4+\bar{\cal E}_6)+\frac{55}{9}\lambda^2 (105 - 210 \bar{\cal E}_4 + 35 \bar{\cal E}_4^2 + 28 \bar{\cal E}_6 - \bar{\cal E}_8) + \frac{22}{81} \lambda X_{10} + \frac{1}{729} X_{12}$
    \\[1.5ex]
  $12$ & $\lambda^6+26\lambda^5+\frac{715}{9}\lambda^4(3-\bar{\cal E}_4) +\frac{572}{9}\lambda^3(15-15\bar{\cal E}_4+\bar{\cal E}_6)+\frac{143}{9}\lambda^2 (105 - 210 \bar{\cal E}_4 + 35 \bar{\cal E}_4^2 + 28 \bar{\cal E}_6 - \bar{\cal E}_8) + \frac{286}{243} \lambda X_{10} + \frac{13}{729} X_{12}$
    \end{tabular}
  \end{ruledtabular}
  \label{tab:Tasymptotichierarchy}
\end{table*}

The stability conditions for $c_\text{max}=7$ are given by those for $c_\text{max}=5$, and in addition
\bea
  0 &\leq& 6+5\bar{\cal E}_4\,,\nn\\
  0 &\leq& 105 - 210 \bar{\cal E}_4 + 35 \bar{\cal E}_4^2 + 28 \bar{\cal E}_6 - \bar{\cal E}_8\,,\nn\\
  0 &\leq& 630 - 525 \bar{\cal E}_4 - 35 \bar{\cal E}_4^2 + 21 \bar{\cal E}_6 + \bar{\cal E}_8\,,\nn\\  
  0 &\leq& 70 (216 + 108 \bar{\cal E}_4 - 378 \bar{\cal E}_4^2 + 215 \bar{\cal E}_4^3)
  -14 (-108 + 180 \bar{\cal E}_4 + 95 \bar{\cal E}_4^2)\bar{\cal E}_6 +63\bar{\cal E}_6^2 + (66 + 95 \bar{\cal E}_4 + 3 \bar{\cal E}_6) \bar{\cal E}_8\,,\nn\\
  0 &\leq& 70 (72 + 132 \bar{\cal E}_4 + 2 \bar{\cal E}_4^2 + 85 \bar{\cal E}_4^3) -1680 \bar{\cal E}_4 \bar{\cal E}_6 - 63 \bar{\cal E}_6^2+(6 + 5 \bar{\cal E}_4) \bar{\cal E}_8\,,\nn\\
  0 &\leq& 34300 (5184 + 5184 \bar{\cal E}_4 - 8208 \bar{\cal E}_4^2 + 7008 \bar{\cal E}_4^3 - 9204 \bar{\cal E}_4^4 + 16340 \bar{\cal E}_4^5 + 1445 \bar{\cal E}_4^6) \nn\\
   && -54880 (432 + 1296 \bar{\cal E}_4 - 2016 \bar{\cal E}_4^2 + 4100 \bar{\cal E}_4^3 + 1805 \bar{\cal E}_4^4) \bar{\cal E}_6
   + 1372 (-4104 + 13500 \bar{\cal E}_4 + 15390 \bar{\cal E}_4^2 + 2275 \bar{\cal E}_4^3) \bar{\cal E}_6^2 \nn\\
   && -592704 \bar{\cal E}_6^3 - 64827 \bar{\cal E}_6^4
   + \big[ -7840 (-108 - 324 \bar{\cal E}_4 - 441 \bar{\cal E}_4^2 - 290 \bar{\cal E}_4^3 + 170 \bar{\cal E}_4^4) + 4704 (-72 - 150 \bar{\cal E}_4 + 95 \bar{\cal E}_4^2) \bar{\cal E}_6 \nn\\
   && + 882 (-54 + 35 \bar{\cal E}_4) \bar{\cal E}_6^2 \big] \bar{\cal E}_8
   -7 (396 + 1140 \bar{\cal E}_4 + 335 \bar{\cal E}_4^2 + 72 \bar{\cal E}_6)\bar{\cal E}_8^2 - \bar{\cal E}_8^3\,.
\eea
The stability conditions for $c_\text{max}=8$ are given by those for $c_\text{max}=6$, and in addition
\bea
  0 &\leq& 6+7\bar{\cal E}_4\,,\nn\\
  0 &\leq& 105 - 210 \bar{\cal E}_4 + 35 \bar{\cal E}_4^2 + 28 \bar{\cal E}_6 - \bar{\cal E}_8\,,\nn\\
  0 &\leq& 210 - 105 \bar{\cal E}_4 - 35 \bar{\cal E}_4^2 - 7 \bar{\cal E}_6 + \bar{\cal E}_8\,,\nn\\
  0 &\leq& 210 (24 + 36 \bar{\cal E}_4 - 34 \bar{\cal E}_4^2 + 49 \bar{\cal E}_4^3) -14 (-12 + 252 \bar{\cal E}_4 + 91 \bar{\cal E}_4^2) \bar{\cal E}_6  -49 \bar{\cal E}_6 ^2 +(78 + 147 \bar{\cal E}_4 + 7 \bar{\cal E}_6 ) \bar{\cal E}_8\,,\nn\\
  0 &\leq& 14 (72 + 204 \bar{\cal E}_4 + 90 \bar{\cal E}_4^2 + 203 \bar{\cal E}_4^3) -784 \bar{\cal E}_4 \bar{\cal E}_6  - 49 \bar{\cal E}_6 ^2 + (6 + 7 \bar{\cal E}_4) \bar{\cal E}_8 \,,\nn\\  
  0 &\leq& 980 (1728 + 5184 \bar{\cal E}_4 + 3024 \bar{\cal E}_4^2 + 7776 \bar{\cal E}_4^3 + 2628 \bar{\cal E}_4^4 + 22932 \bar{\cal E}_4^5 + 5887 \bar{\cal E}_4^6) \nn\\
   && -4704 (144 + 720 \bar{\cal E}_4 + 384 \bar{\cal E}_4^2 + 2436 \bar{\cal E}_4^3 + 1183 \bar{\cal E}_4^4)\bar{\cal E}_6
   +588 (-120 + 1932 \bar{\cal E}_4 + 1274 \bar{\cal E}_4^2 + 147 \bar{\cal E}_4^3)\bar{\cal E}_6^2 \nn\\
   && + 21952 \bar{\cal E}_6^3 -7203 \bar{\cal E}_6^4
   + \big[ -672 (-36 - 180 \bar{\cal E}_4 - 411 \bar{\cal E}_4^2 - 294 \bar{\cal E}_4^3 + 203 \bar{\cal E}_4^4) + 672 (-24 - 42 \bar{\cal E}_4 + 91 \bar{\cal E}_4^2) \bar{\cal E}_6 \nn\\
   && + 2058 (-2 + 3 \bar{\cal E}_4) \bar{\cal E}_6^2 \big] \bar{\cal E}_8
   -3 (156 + 588 \bar{\cal E}_4 + 259 \bar{\cal E}_4^2 + 56 \bar{\cal E}_6) \bar{\cal E}_8^2 - \bar{\cal E}_8^3\,.
\eea

\end{widetext}


\begin{thebibliography}{82}%
\makeatletter
\providecommand \@ifxundefined [1]{%
 \@ifx{#1\undefined}
}%
\providecommand \@ifnum [1]{%
 \ifnum #1\expandafter \@firstoftwo
 \else \expandafter \@secondoftwo
 \fi
}%
\providecommand \@ifx [1]{%
 \ifx #1\expandafter \@firstoftwo
 \else \expandafter \@secondoftwo
 \fi
}%
\providecommand \natexlab [1]{#1}%
\providecommand \enquote  [1]{``#1''}%
\providecommand \bibnamefont  [1]{#1}%
\providecommand \bibfnamefont [1]{#1}%
\providecommand \citenamefont [1]{#1}%
\providecommand \href@noop [0]{\@secondoftwo}%
\providecommand \href [0]{\begingroup \@sanitize@url \@href}%
\providecommand \@href[1]{\@@startlink{#1}\@@href}%
\providecommand \@@href[1]{\endgroup#1\@@endlink}%
\providecommand \@sanitize@url [0]{\catcode `\\12\catcode `\$12\catcode
  `\&12\catcode `\#12\catcode `\^12\catcode `\_12\catcode `\%12\relax}%
\providecommand \@@startlink[1]{}%
\providecommand \@@endlink[0]{}%
\providecommand \url  [0]{\begingroup\@sanitize@url \@url }%
\providecommand \@url [1]{\endgroup\@href {#1}{\urlprefix }}%
\providecommand \urlprefix  [0]{URL }%
\providecommand \Eprint [0]{\href }%
\providecommand \doibase [0]{http://dx.doi.org/}%
\providecommand \selectlanguage [0]{\@gobble}%
\providecommand \bibinfo  [0]{\@secondoftwo}%
\providecommand \bibfield  [0]{\@secondoftwo}%
\providecommand \translation [1]{[#1]}%
\providecommand \BibitemOpen [0]{}%
\providecommand \bibitemStop [0]{}%
\providecommand \bibitemNoStop [0]{.\EOS\space}%
\providecommand \EOS [0]{\spacefactor3000\relax}%
\providecommand \BibitemShut  [1]{\csname bibitem#1\endcsname}%
\let\auto@bib@innerbib\@empty
\bibitem [{\citenamefont {{Bernardeau}}\ \emph
  {et~al.}(2002{\natexlab{a}})\citenamefont {{Bernardeau}}, \citenamefont
  {{Colombi}}, \citenamefont {{Gaztanaga}},\ and\ \citenamefont
  {{Scoccimarro}}}]{BerColGaz02}%
  \BibitemOpen
  \bibfield  {author} {\bibinfo {author} {\bibfnamefont {F.}~\bibnamefont
  {{Bernardeau}}}, \bibinfo {author} {\bibfnamefont {S.}~\bibnamefont
  {{Colombi}}}, \bibinfo {author} {\bibfnamefont {E.}~\bibnamefont
  {{Gaztanaga}}}, \ and\ \bibinfo {author} {\bibfnamefont {R.}~\bibnamefont
  {{Scoccimarro}}},\ }\href@noop {} {\bibfield  {journal} {\bibinfo  {journal}
  {\physrep}\ }\textbf {\bibinfo {volume} {367}},\ \bibinfo {pages} {1}
  (\bibinfo {year} {2002}{\natexlab{a}})}\BibitemShut {NoStop}%
\bibitem [{\citenamefont {{Colombi}}\ \emph {et~al.}(1996)\citenamefont
  {{Colombi}}, \citenamefont {{Bouchet}},\ and\ \citenamefont
  {{Hernquist}}}]{ColBouHer96}%
  \BibitemOpen
  \bibfield  {author} {\bibinfo {author} {\bibfnamefont {S.}~\bibnamefont
  {{Colombi}}}, \bibinfo {author} {\bibfnamefont {F.~R.}\ \bibnamefont
  {{Bouchet}}}, \ and\ \bibinfo {author} {\bibfnamefont {L.}~\bibnamefont
  {{Hernquist}}},\ }\href {\doibase 10.1086/177398} {\bibfield  {journal}
  {\bibinfo  {journal} {\apj}\ }\textbf {\bibinfo {volume} {465}},\ \bibinfo
  {pages} {14} (\bibinfo {year} {1996})}\BibitemShut {NoStop}%
\bibitem [{\citenamefont {{Makino}}\ \emph {et~al.}(1992)\citenamefont
  {{Makino}}, \citenamefont {{Sasaki}},\ and\ \citenamefont
  {{Suto}}}]{MakSasSut92}%
  \BibitemOpen
  \bibfield  {author} {\bibinfo {author} {\bibfnamefont {N.}~\bibnamefont
  {{Makino}}}, \bibinfo {author} {\bibfnamefont {M.}~\bibnamefont {{Sasaki}}},
  \ and\ \bibinfo {author} {\bibfnamefont {Y.}~\bibnamefont {{Suto}}},\
  }\href@noop {} {\bibfield  {journal} {\bibinfo  {journal} {\prd}\ }\textbf
  {\bibinfo {volume} {46}},\ \bibinfo {pages} {585} (\bibinfo {year}
  {1992})}\BibitemShut {NoStop}%
\bibitem [{\citenamefont {{Scoccimarro}}\ and\ \citenamefont
  {{Frieman}}(1996)}]{ScoFri9612}%
  \BibitemOpen
  \bibfield  {author} {\bibinfo {author} {\bibfnamefont {R.}~\bibnamefont
  {{Scoccimarro}}}\ and\ \bibinfo {author} {\bibfnamefont {J.~A.}\ \bibnamefont
  {{Frieman}}},\ }\href {\doibase 10.1086/178177} {\bibfield  {journal}
  {\bibinfo  {journal} {\apj}\ }\textbf {\bibinfo {volume} {473}},\ \bibinfo
  {pages} {620} (\bibinfo {year} {1996})},\ \Eprint
  {http://arxiv.org/abs/arXiv:astro-ph/9602070} {arXiv:astro-ph/9602070}
  \BibitemShut {NoStop}%
\bibitem [{\citenamefont {{Cooray}}\ and\ \citenamefont
  {{Sheth}}(2002)}]{CooShe02}%
  \BibitemOpen
  \bibfield  {author} {\bibinfo {author} {\bibfnamefont {A.}~\bibnamefont
  {{Cooray}}}\ and\ \bibinfo {author} {\bibfnamefont {R.}~\bibnamefont
  {{Sheth}}},\ }\href@noop {} {\bibfield  {journal} {\bibinfo  {journal}
  {\physrep}\ }\textbf {\bibinfo {volume} {372}},\ \bibinfo {pages} {1}
  (\bibinfo {year} {2002})}\BibitemShut {NoStop}%
\bibitem [{\citenamefont {{Nishimichi}}\ \emph {et~al.}(2016)\citenamefont
  {{Nishimichi}}, \citenamefont {{Bernardeau}},\ and\ \citenamefont
  {{Taruya}}}]{NisBerTar1611}%
  \BibitemOpen
  \bibfield  {author} {\bibinfo {author} {\bibfnamefont {T.}~\bibnamefont
  {{Nishimichi}}}, \bibinfo {author} {\bibfnamefont {F.}~\bibnamefont
  {{Bernardeau}}}, \ and\ \bibinfo {author} {\bibfnamefont {A.}~\bibnamefont
  {{Taruya}}},\ }\href {\doibase 10.1016/j.physletb.2016.09.035} {\bibfield
  {journal} {\bibinfo  {journal} {Physics Letters B}\ }\textbf {\bibinfo
  {volume} {762}},\ \bibinfo {pages} {247} (\bibinfo {year} {2016})},\ \Eprint
  {http://arxiv.org/abs/1411.2970} {arXiv:1411.2970} \BibitemShut {NoStop}%
\bibitem [{\citenamefont {{Nishimichi}}\ \emph {et~al.}(2017)\citenamefont
  {{Nishimichi}}, \citenamefont {{Bernardeau}},\ and\ \citenamefont
  {{Taruya}}}]{NisBerTar1712}%
  \BibitemOpen
  \bibfield  {author} {\bibinfo {author} {\bibfnamefont {T.}~\bibnamefont
  {{Nishimichi}}}, \bibinfo {author} {\bibfnamefont {F.}~\bibnamefont
  {{Bernardeau}}}, \ and\ \bibinfo {author} {\bibfnamefont {A.}~\bibnamefont
  {{Taruya}}},\ }\href {\doibase 10.1103/PhysRevD.96.123515} {\bibfield
  {journal} {\bibinfo  {journal} {\prd}\ }\textbf {\bibinfo {volume} {96}},\
  \bibinfo {eid} {123515} (\bibinfo {year} {2017})},\ \Eprint
  {http://arxiv.org/abs/1708.08946} {arXiv:1708.08946 [astro-ph.CO]}
  \BibitemShut {NoStop}%
\bibitem [{\citenamefont {{Carrasco}}\ \emph {et~al.}(2012)\citenamefont
  {{Carrasco}}, \citenamefont {{Hertzberg}},\ and\ \citenamefont
  {{Senatore}}}]{CarHerSen1206}%
  \BibitemOpen
  \bibfield  {author} {\bibinfo {author} {\bibfnamefont {J.~J.~M.}\
  \bibnamefont {{Carrasco}}}, \bibinfo {author} {\bibfnamefont {M.~P.}\
  \bibnamefont {{Hertzberg}}}, \ and\ \bibinfo {author} {\bibfnamefont
  {L.}~\bibnamefont {{Senatore}}},\ }\href@noop {} {\bibfield  {journal}
  {\bibinfo  {journal} {ArXiv e-prints}\ } (\bibinfo {year} {2012})},\ \Eprint
  {http://arxiv.org/abs/1206.2926} {arXiv:1206.2926 [astro-ph.CO]} \BibitemShut
  {NoStop}%
\bibitem [{\citenamefont {{Baumann}}\ \emph {et~al.}(2012)\citenamefont
  {{Baumann}}, \citenamefont {{Nicolis}}, \citenamefont {{Senatore}},\ and\
  \citenamefont {{Zaldarriaga}}}]{BauNicSen1207}%
  \BibitemOpen
  \bibfield  {author} {\bibinfo {author} {\bibfnamefont {D.}~\bibnamefont
  {{Baumann}}}, \bibinfo {author} {\bibfnamefont {A.}~\bibnamefont
  {{Nicolis}}}, \bibinfo {author} {\bibfnamefont {L.}~\bibnamefont
  {{Senatore}}}, \ and\ \bibinfo {author} {\bibfnamefont {M.}~\bibnamefont
  {{Zaldarriaga}}},\ }\href {\doibase 10.1088/1475-7516/2012/07/051} {\bibfield
   {journal} {\bibinfo  {journal} {\jcap}\ }\textbf {\bibinfo {volume} {7}},\
  \bibinfo {eid} {051} (\bibinfo {year} {2012})},\ \Eprint
  {http://arxiv.org/abs/1004.2488} {arXiv:1004.2488 [astro-ph.CO]} \BibitemShut
  {NoStop}%
\bibitem [{\citenamefont {{Pueblas}}\ and\ \citenamefont
  {{Scoccimarro}}(2009)}]{PueSco0908}%
  \BibitemOpen
  \bibfield  {author} {\bibinfo {author} {\bibfnamefont {S.}~\bibnamefont
  {{Pueblas}}}\ and\ \bibinfo {author} {\bibfnamefont {R.}~\bibnamefont
  {{Scoccimarro}}},\ }\href {\doibase 10.1103/PhysRevD.80.043504} {\bibfield
  {journal} {\bibinfo  {journal} {\prd}\ }\textbf {\bibinfo {volume} {80}},\
  \bibinfo {pages} {043504} (\bibinfo {year} {2009})},\ \Eprint
  {http://arxiv.org/abs/0809.4606} {arXiv:0809.4606} \BibitemShut {NoStop}%
\bibitem [{\citenamefont {{Garny}}\ \emph {et~al.}(2022)\citenamefont
  {{Garny}}, \citenamefont {{Laxhuber}},\ and\ \citenamefont
  {{Scoccimarro}}}]{cumPT2}%
  \BibitemOpen
  \bibfield  {author} {\bibinfo {author} {\bibfnamefont {M.}~\bibnamefont
  {{Garny}}}, \bibinfo {author} {\bibfnamefont {D.}~\bibnamefont {{Laxhuber}}},
  \ and\ \bibinfo {author} {\bibfnamefont {R.}~\bibnamefont {{Scoccimarro}}},\
  }\href@noop {} {\bibfield  {journal} {\bibinfo  {journal} {to appear}\ }
  (\bibinfo {year} {2022})}\BibitemShut {NoStop}%
\bibitem [{\citenamefont {{McDonald}}(2011)}]{McD1104}%
  \BibitemOpen
  \bibfield  {author} {\bibinfo {author} {\bibfnamefont {P.}~\bibnamefont
  {{McDonald}}},\ }\href {\doibase 10.1088/1475-7516/2011/04/032} {\bibfield
  {journal} {\bibinfo  {journal} {\jcap}\ }\textbf {\bibinfo {volume} {2011}},\
  \bibinfo {eid} {032} (\bibinfo {year} {2011})},\ \Eprint
  {http://arxiv.org/abs/0910.1002} {arXiv:0910.1002 [astro-ph.CO]} \BibitemShut
  {NoStop}%
\bibitem [{\citenamefont {{Erschfeld}}\ and\ \citenamefont
  {{Floerchinger}}(2019)}]{ErsFlo1906}%
  \BibitemOpen
  \bibfield  {author} {\bibinfo {author} {\bibfnamefont {A.}~\bibnamefont
  {{Erschfeld}}}\ and\ \bibinfo {author} {\bibfnamefont {S.}~\bibnamefont
  {{Floerchinger}}},\ }\href {\doibase 10.1088/1475-7516/2019/06/039}
  {\bibfield  {journal} {\bibinfo  {journal} {\jcap}\ }\textbf {\bibinfo
  {volume} {2019}},\ \bibinfo {eid} {039} (\bibinfo {year} {2019})},\ \Eprint
  {http://arxiv.org/abs/1812.06891} {arXiv:1812.06891 [astro-ph.CO]}
  \BibitemShut {NoStop}%
\bibitem [{\citenamefont {Erschfeld}(2021)}]{Erschfeld:2021kem}%
  \BibitemOpen
  \bibfield  {author} {\bibinfo {author} {\bibfnamefont {A.~A.}\ \bibnamefont
  {Erschfeld}},\ }\emph {\bibinfo {title} {{Functional methods for cosmic
  structure formation.}}},\ \href {\doibase 10.11588/heidok.00030982} {Ph.D.
  thesis},\ \bibinfo  {school} {U. Heidelberg (main)} (\bibinfo {year}
  {2021})\BibitemShut {NoStop}%
\bibitem [{\citenamefont {{Buchert}}\ and\ \citenamefont
  {{Dom{\'\i}nguez}}(1998)}]{BucDom9807}%
  \BibitemOpen
  \bibfield  {author} {\bibinfo {author} {\bibfnamefont {T.}~\bibnamefont
  {{Buchert}}}\ and\ \bibinfo {author} {\bibfnamefont {A.}~\bibnamefont
  {{Dom{\'\i}nguez}}},\ }\href@noop {} {\bibfield  {journal} {\bibinfo
  {journal} {\aap}\ }\textbf {\bibinfo {volume} {335}},\ \bibinfo {pages} {395}
  (\bibinfo {year} {1998})},\ \Eprint {http://arxiv.org/abs/astro-ph/9702139}
  {arXiv:astro-ph/9702139 [astro-ph]} \BibitemShut {NoStop}%
\bibitem [{\citenamefont {{Adler}}\ and\ \citenamefont
  {{Buchert}}(1999)}]{AdlBuc9903}%
  \BibitemOpen
  \bibfield  {author} {\bibinfo {author} {\bibfnamefont {S.}~\bibnamefont
  {{Adler}}}\ and\ \bibinfo {author} {\bibfnamefont {T.}~\bibnamefont
  {{Buchert}}},\ }\href@noop {} {\bibfield  {journal} {\bibinfo  {journal}
  {\aap}\ }\textbf {\bibinfo {volume} {343}},\ \bibinfo {pages} {317} (\bibinfo
  {year} {1999})},\ \Eprint {http://arxiv.org/abs/astro-ph/9806320}
  {arXiv:astro-ph/9806320 [astro-ph]} \BibitemShut {NoStop}%
\bibitem [{\citenamefont {{Tatekawa}}\ \emph {et~al.}(2002)\citenamefont
  {{Tatekawa}}, \citenamefont {{Suda}}, \citenamefont {{Maeda}}, \citenamefont
  {{Morita}},\ and\ \citenamefont {{Anzai}}}]{TatSudMae0209}%
  \BibitemOpen
  \bibfield  {author} {\bibinfo {author} {\bibfnamefont {T.}~\bibnamefont
  {{Tatekawa}}}, \bibinfo {author} {\bibfnamefont {M.}~\bibnamefont {{Suda}}},
  \bibinfo {author} {\bibfnamefont {K.-I.}\ \bibnamefont {{Maeda}}}, \bibinfo
  {author} {\bibfnamefont {M.}~\bibnamefont {{Morita}}}, \ and\ \bibinfo
  {author} {\bibfnamefont {H.}~\bibnamefont {{Anzai}}},\ }\href {\doibase
  10.1103/PhysRevD.66.064014} {\bibfield  {journal} {\bibinfo  {journal}
  {\prd}\ }\textbf {\bibinfo {volume} {66}},\ \bibinfo {eid} {064014} (\bibinfo
  {year} {2002})},\ \Eprint {http://arxiv.org/abs/astro-ph/0205017}
  {arXiv:astro-ph/0205017 [astro-ph]} \BibitemShut {NoStop}%
\bibitem [{\citenamefont {{Morita}}\ and\ \citenamefont
  {{Tatekawa}}(2001)}]{MorTat0112}%
  \BibitemOpen
  \bibfield  {author} {\bibinfo {author} {\bibfnamefont {M.}~\bibnamefont
  {{Morita}}}\ and\ \bibinfo {author} {\bibfnamefont {T.}~\bibnamefont
  {{Tatekawa}}},\ }\href {\doibase 10.1046/j.1365-8711.2001.04904.x} {\bibfield
   {journal} {\bibinfo  {journal} {\mnras}\ }\textbf {\bibinfo {volume}
  {328}},\ \bibinfo {pages} {815} (\bibinfo {year} {2001})},\ \Eprint
  {http://arxiv.org/abs/astro-ph/0108289} {arXiv:astro-ph/0108289 [astro-ph]}
  \BibitemShut {NoStop}%
\bibitem [{\citenamefont {{Colombi}}(2015)}]{Col1501}%
  \BibitemOpen
  \bibfield  {author} {\bibinfo {author} {\bibfnamefont {S.}~\bibnamefont
  {{Colombi}}},\ }\href {\doibase 10.1093/mnras/stu2308} {\bibfield  {journal}
  {\bibinfo  {journal} {\mnras}\ }\textbf {\bibinfo {volume} {446}},\ \bibinfo
  {pages} {2902} (\bibinfo {year} {2015})},\ \Eprint
  {http://arxiv.org/abs/1411.4165} {arXiv:1411.4165} \BibitemShut {NoStop}%
\bibitem [{\citenamefont {{Aviles}}(2016)}]{Avi1603}%
  \BibitemOpen
  \bibfield  {author} {\bibinfo {author} {\bibfnamefont {A.}~\bibnamefont
  {{Aviles}}},\ }\href {\doibase 10.1103/PhysRevD.93.063517} {\bibfield
  {journal} {\bibinfo  {journal} {\prd}\ }\textbf {\bibinfo {volume} {93}},\
  \bibinfo {eid} {063517} (\bibinfo {year} {2016})},\ \Eprint
  {http://arxiv.org/abs/1512.07198} {arXiv:1512.07198 [astro-ph.CO]}
  \BibitemShut {NoStop}%
\bibitem [{\citenamefont {{Cusin}}\ \emph {et~al.}(2017)\citenamefont
  {{Cusin}}, \citenamefont {{Tansella}},\ and\ \citenamefont
  {{Durrer}}}]{CusTanDur1703}%
  \BibitemOpen
  \bibfield  {author} {\bibinfo {author} {\bibfnamefont {G.}~\bibnamefont
  {{Cusin}}}, \bibinfo {author} {\bibfnamefont {V.}~\bibnamefont {{Tansella}}},
  \ and\ \bibinfo {author} {\bibfnamefont {R.}~\bibnamefont {{Durrer}}},\
  }\href {\doibase 10.1103/PhysRevD.95.063527} {\bibfield  {journal} {\bibinfo
  {journal} {\prd}\ }\textbf {\bibinfo {volume} {95}},\ \bibinfo {eid} {063527}
  (\bibinfo {year} {2017})},\ \Eprint {http://arxiv.org/abs/1612.00783}
  {arXiv:1612.00783 [astro-ph.CO]} \BibitemShut {NoStop}%
\bibitem [{\citenamefont {{Taruya}}\ and\ \citenamefont
  {{Colombi}}(2017)}]{TarCol1710}%
  \BibitemOpen
  \bibfield  {author} {\bibinfo {author} {\bibfnamefont {A.}~\bibnamefont
  {{Taruya}}}\ and\ \bibinfo {author} {\bibfnamefont {S.}~\bibnamefont
  {{Colombi}}},\ }\href {\doibase 10.1093/mnras/stx1501} {\bibfield  {journal}
  {\bibinfo  {journal} {\mnras}\ }\textbf {\bibinfo {volume} {470}},\ \bibinfo
  {pages} {4858} (\bibinfo {year} {2017})},\ \Eprint
  {http://arxiv.org/abs/1701.09088} {arXiv:1701.09088 [astro-ph.CO]}
  \BibitemShut {NoStop}%
\bibitem [{\citenamefont {{Rampf}}\ and\ \citenamefont
  {{Frisch}}(2017)}]{RamFri1710}%
  \BibitemOpen
  \bibfield  {author} {\bibinfo {author} {\bibfnamefont {C.}~\bibnamefont
  {{Rampf}}}\ and\ \bibinfo {author} {\bibfnamefont {U.}~\bibnamefont
  {{Frisch}}},\ }\href {\doibase 10.1093/mnras/stx1613} {\bibfield  {journal}
  {\bibinfo  {journal} {\mnras}\ }\textbf {\bibinfo {volume} {471}},\ \bibinfo
  {pages} {671} (\bibinfo {year} {2017})},\ \Eprint
  {http://arxiv.org/abs/1705.08456} {arXiv:1705.08456 [astro-ph.CO]}
  \BibitemShut {NoStop}%
\bibitem [{\citenamefont {{McDonald}}\ and\ \citenamefont
  {{Vlah}}(2018)}]{McDVla1801}%
  \BibitemOpen
  \bibfield  {author} {\bibinfo {author} {\bibfnamefont {P.}~\bibnamefont
  {{McDonald}}}\ and\ \bibinfo {author} {\bibfnamefont {Z.}~\bibnamefont
  {{Vlah}}},\ }\href {\doibase 10.1103/PhysRevD.97.023508} {\bibfield
  {journal} {\bibinfo  {journal} {\prd}\ }\textbf {\bibinfo {volume} {97}},\
  \bibinfo {eid} {023508} (\bibinfo {year} {2018})},\ \Eprint
  {http://arxiv.org/abs/1709.02834} {arXiv:1709.02834 [astro-ph.CO]}
  \BibitemShut {NoStop}%
\bibitem [{\citenamefont {{Saga}}\ \emph {et~al.}(2018)\citenamefont {{Saga}},
  \citenamefont {{Taruya}},\ and\ \citenamefont {{Colombi}}}]{SagTarCol1812}%
  \BibitemOpen
  \bibfield  {author} {\bibinfo {author} {\bibfnamefont {S.}~\bibnamefont
  {{Saga}}}, \bibinfo {author} {\bibfnamefont {A.}~\bibnamefont {{Taruya}}}, \
  and\ \bibinfo {author} {\bibfnamefont {S.}~\bibnamefont {{Colombi}}},\ }\href
  {\doibase 10.1103/PhysRevLett.121.241302} {\bibfield  {journal} {\bibinfo
  {journal} {\prl}\ }\textbf {\bibinfo {volume} {121}},\ \bibinfo {eid}
  {241302} (\bibinfo {year} {2018})},\ \Eprint
  {http://arxiv.org/abs/1805.08787} {arXiv:1805.08787 [astro-ph.CO]}
  \BibitemShut {NoStop}%
\bibitem [{\citenamefont {{Halle}}\ \emph {et~al.}(2020)\citenamefont
  {{Halle}}, \citenamefont {{Nishimichi}}, \citenamefont {{Taruya}},
  \citenamefont {{Colombi}},\ and\ \citenamefont
  {{Bernardeau}}}]{HalNisTar2012}%
  \BibitemOpen
  \bibfield  {author} {\bibinfo {author} {\bibfnamefont {A.}~\bibnamefont
  {{Halle}}}, \bibinfo {author} {\bibfnamefont {T.}~\bibnamefont
  {{Nishimichi}}}, \bibinfo {author} {\bibfnamefont {A.}~\bibnamefont
  {{Taruya}}}, \bibinfo {author} {\bibfnamefont {S.}~\bibnamefont {{Colombi}}},
  \ and\ \bibinfo {author} {\bibfnamefont {F.}~\bibnamefont {{Bernardeau}}},\
  }\href {\doibase 10.1093/mnras/staa2878} {\bibfield  {journal} {\bibinfo
  {journal} {\mnras}\ }\textbf {\bibinfo {volume} {499}},\ \bibinfo {pages}
  {1769} (\bibinfo {year} {2020})},\ \Eprint {http://arxiv.org/abs/2001.10417}
  {arXiv:2001.10417 [astro-ph.CO]} \BibitemShut {NoStop}%
\bibitem [{\citenamefont {{Rampf}}\ and\ \citenamefont
  {{Hahn}}(2021)}]{RamHah2102}%
  \BibitemOpen
  \bibfield  {author} {\bibinfo {author} {\bibfnamefont {C.}~\bibnamefont
  {{Rampf}}}\ and\ \bibinfo {author} {\bibfnamefont {O.}~\bibnamefont
  {{Hahn}}},\ }\href {\doibase 10.1093/mnrasl/slaa198} {\bibfield  {journal}
  {\bibinfo  {journal} {\mnras}\ }\textbf {\bibinfo {volume} {501}},\ \bibinfo
  {pages} {L71} (\bibinfo {year} {2021})},\ \Eprint
  {http://arxiv.org/abs/2010.12584} {arXiv:2010.12584 [astro-ph.CO]}
  \BibitemShut {NoStop}%
\bibitem [{\citenamefont {{Valageas}}(2010)}]{Val1009}%
  \BibitemOpen
  \bibfield  {author} {\bibinfo {author} {\bibfnamefont {P.}~\bibnamefont
  {{Valageas}}},\ }\href@noop {} {\bibfield  {journal} {\bibinfo  {journal}
  {ArXiv e-prints}\ } (\bibinfo {year} {2010})},\ \Eprint
  {http://arxiv.org/abs/1009.0106} {arXiv:1009.0106 [astro-ph.CO]} \BibitemShut
  {NoStop}%
\bibitem [{\citenamefont {{Valageas}}\ and\ \citenamefont
  {{Nishimichi}}(2011{\natexlab{a}})}]{ValNis1103}%
  \BibitemOpen
  \bibfield  {author} {\bibinfo {author} {\bibfnamefont {P.}~\bibnamefont
  {{Valageas}}}\ and\ \bibinfo {author} {\bibfnamefont {T.}~\bibnamefont
  {{Nishimichi}}},\ }\href {\doibase 10.1051/0004-6361/201015685} {\bibfield
  {journal} {\bibinfo  {journal} {\aap}\ }\textbf {\bibinfo {volume} {527}},\
  \bibinfo {eid} {A87} (\bibinfo {year} {2011}{\natexlab{a}})},\ \Eprint
  {http://arxiv.org/abs/1009.0597} {arXiv:1009.0597 [astro-ph.CO]} \BibitemShut
  {NoStop}%
\bibitem [{\citenamefont {{Valageas}}\ and\ \citenamefont
  {{Nishimichi}}(2011{\natexlab{b}})}]{ValNis1108}%
  \BibitemOpen
  \bibfield  {author} {\bibinfo {author} {\bibfnamefont {P.}~\bibnamefont
  {{Valageas}}}\ and\ \bibinfo {author} {\bibfnamefont {T.}~\bibnamefont
  {{Nishimichi}}},\ }\href {\doibase 10.1051/0004-6361/201116638} {\bibfield
  {journal} {\bibinfo  {journal} {\aap}\ }\textbf {\bibinfo {volume} {532}},\
  \bibinfo {eid} {A4} (\bibinfo {year} {2011}{\natexlab{b}})},\ \Eprint
  {http://arxiv.org/abs/1102.0641} {arXiv:1102.0641 [astro-ph.CO]} \BibitemShut
  {NoStop}%
\bibitem [{\citenamefont {{Valageas}}\ \emph {et~al.}(2013)\citenamefont
  {{Valageas}}, \citenamefont {{Nishimichi}},\ and\ \citenamefont
  {{Taruya}}}]{ValNisTar1304}%
  \BibitemOpen
  \bibfield  {author} {\bibinfo {author} {\bibfnamefont {P.}~\bibnamefont
  {{Valageas}}}, \bibinfo {author} {\bibfnamefont {T.}~\bibnamefont
  {{Nishimichi}}}, \ and\ \bibinfo {author} {\bibfnamefont {A.}~\bibnamefont
  {{Taruya}}},\ }\href {\doibase 10.1103/PhysRevD.87.083522} {\bibfield
  {journal} {\bibinfo  {journal} {\prd}\ }\textbf {\bibinfo {volume} {87}},\
  \bibinfo {eid} {083522} (\bibinfo {year} {2013})},\ \Eprint
  {http://arxiv.org/abs/1302.4533} {arXiv:1302.4533 [astro-ph.CO]} \BibitemShut
  {NoStop}%
\bibitem [{\citenamefont {{Seljak}}\ and\ \citenamefont
  {{Vlah}}(2015)}]{SelVla1506}%
  \BibitemOpen
  \bibfield  {author} {\bibinfo {author} {\bibfnamefont {U.}~\bibnamefont
  {{Seljak}}}\ and\ \bibinfo {author} {\bibfnamefont {Z.}~\bibnamefont
  {{Vlah}}},\ }\href {\doibase 10.1103/PhysRevD.91.123516} {\bibfield
  {journal} {\bibinfo  {journal} {\prd}\ }\textbf {\bibinfo {volume} {91}},\
  \bibinfo {eid} {123516} (\bibinfo {year} {2015})},\ \Eprint
  {http://arxiv.org/abs/1501.07512} {arXiv:1501.07512 [astro-ph.CO]}
  \BibitemShut {NoStop}%
\bibitem [{\citenamefont {{Abel}}\ \emph {et~al.}(2012)\citenamefont {{Abel}},
  \citenamefont {{Hahn}},\ and\ \citenamefont {{Kaehler}}}]{AbeHahKae1211}%
  \BibitemOpen
  \bibfield  {author} {\bibinfo {author} {\bibfnamefont {T.}~\bibnamefont
  {{Abel}}}, \bibinfo {author} {\bibfnamefont {O.}~\bibnamefont {{Hahn}}}, \
  and\ \bibinfo {author} {\bibfnamefont {R.}~\bibnamefont {{Kaehler}}},\ }\href
  {\doibase 10.1111/j.1365-2966.2012.21754.x} {\bibfield  {journal} {\bibinfo
  {journal} {\mnras}\ }\textbf {\bibinfo {volume} {427}},\ \bibinfo {pages}
  {61} (\bibinfo {year} {2012})},\ \Eprint {http://arxiv.org/abs/1111.3944}
  {arXiv:1111.3944 [astro-ph.CO]} \BibitemShut {NoStop}%
\bibitem [{\citenamefont {{Hahn}}\ \emph {et~al.}(2013)\citenamefont {{Hahn}},
  \citenamefont {{Abel}},\ and\ \citenamefont {{Kaehler}}}]{HahAbeKae1309}%
  \BibitemOpen
  \bibfield  {author} {\bibinfo {author} {\bibfnamefont {O.}~\bibnamefont
  {{Hahn}}}, \bibinfo {author} {\bibfnamefont {T.}~\bibnamefont {{Abel}}}, \
  and\ \bibinfo {author} {\bibfnamefont {R.}~\bibnamefont {{Kaehler}}},\ }\href
  {\doibase 10.1093/mnras/stt1061} {\bibfield  {journal} {\bibinfo  {journal}
  {\mnras}\ }\textbf {\bibinfo {volume} {434}},\ \bibinfo {pages} {1171}
  (\bibinfo {year} {2013})},\ \Eprint {http://arxiv.org/abs/1210.6652}
  {arXiv:1210.6652 [astro-ph.CO]} \BibitemShut {NoStop}%
\bibitem [{\citenamefont {{Colombi}}\ and\ \citenamefont
  {{Touma}}(2014)}]{ColTou1407}%
  \BibitemOpen
  \bibfield  {author} {\bibinfo {author} {\bibfnamefont {S.}~\bibnamefont
  {{Colombi}}}\ and\ \bibinfo {author} {\bibfnamefont {J.}~\bibnamefont
  {{Touma}}},\ }\href {\doibase 10.1093/mnras/stu739} {\bibfield  {journal}
  {\bibinfo  {journal} {\mnras}\ }\textbf {\bibinfo {volume} {441}},\ \bibinfo
  {pages} {2414} (\bibinfo {year} {2014})}\BibitemShut {NoStop}%
\bibitem [{\citenamefont {{Hahn}}\ and\ \citenamefont
  {{Angulo}}(2016)}]{HahAng1601}%
  \BibitemOpen
  \bibfield  {author} {\bibinfo {author} {\bibfnamefont {O.}~\bibnamefont
  {{Hahn}}}\ and\ \bibinfo {author} {\bibfnamefont {R.~E.}\ \bibnamefont
  {{Angulo}}},\ }\href {\doibase 10.1093/mnras/stv2304} {\bibfield  {journal}
  {\bibinfo  {journal} {\mnras}\ }\textbf {\bibinfo {volume} {455}},\ \bibinfo
  {pages} {1115} (\bibinfo {year} {2016})},\ \Eprint
  {http://arxiv.org/abs/1501.01959} {arXiv:1501.01959 [astro-ph.CO]}
  \BibitemShut {NoStop}%
\bibitem [{\citenamefont {{Sousbie}}\ and\ \citenamefont
  {{Colombi}}(2016)}]{SouCol1609}%
  \BibitemOpen
  \bibfield  {author} {\bibinfo {author} {\bibfnamefont {T.}~\bibnamefont
  {{Sousbie}}}\ and\ \bibinfo {author} {\bibfnamefont {S.}~\bibnamefont
  {{Colombi}}},\ }\href {\doibase 10.1016/j.jcp.2016.05.048} {\bibfield
  {journal} {\bibinfo  {journal} {Journal of Computational Physics}\ }\textbf
  {\bibinfo {volume} {321}},\ \bibinfo {pages} {644} (\bibinfo {year}
  {2016})},\ \Eprint {http://arxiv.org/abs/1509.07720} {arXiv:1509.07720
  [physics.comp-ph]} \BibitemShut {NoStop}%
\bibitem [{\citenamefont {Stucker}\ \emph {et~al.}(2020)\citenamefont
  {Stucker}, \citenamefont {Hahn}, \citenamefont {Angulo},\ and\ \citenamefont
  {White}}]{Stucker:2019txm}%
  \BibitemOpen
  \bibfield  {author} {\bibinfo {author} {\bibfnamefont {J.}~\bibnamefont
  {Stucker}}, \bibinfo {author} {\bibfnamefont {O.}~\bibnamefont {Hahn}},
  \bibinfo {author} {\bibfnamefont {R.~E.}\ \bibnamefont {Angulo}}, \ and\
  \bibinfo {author} {\bibfnamefont {S.~D.~M.}\ \bibnamefont {White}},\ }\href
  {\doibase 10.1093/mnras/staa1468} {\bibfield  {journal} {\bibinfo  {journal}
  {Mon. Not. Roy. Astron. Soc.}\ }\textbf {\bibinfo {volume} {495}},\ \bibinfo
  {pages} {4943} (\bibinfo {year} {2020})},\ \Eprint
  {http://arxiv.org/abs/1909.00008} {arXiv:1909.00008 [astro-ph.CO]}
  \BibitemShut {NoStop}%
\bibitem [{\citenamefont {{Colombi}}(2021)}]{Col2103}%
  \BibitemOpen
  \bibfield  {author} {\bibinfo {author} {\bibfnamefont {S.}~\bibnamefont
  {{Colombi}}},\ }\href {\doibase 10.1051/0004-6361/202039719} {\bibfield
  {journal} {\bibinfo  {journal} {\aap}\ }\textbf {\bibinfo {volume} {647}},\
  \bibinfo {eid} {A66} (\bibinfo {year} {2021})},\ \Eprint
  {http://arxiv.org/abs/2012.04409} {arXiv:2012.04409 [astro-ph.CO]}
  \BibitemShut {NoStop}%
\bibitem [{\citenamefont {{Saga}}\ \emph {et~al.}(2022)\citenamefont {{Saga}},
  \citenamefont {{Taruya}},\ and\ \citenamefont {{Colombi}}}]{SagTarCol2208}%
  \BibitemOpen
  \bibfield  {author} {\bibinfo {author} {\bibfnamefont {S.}~\bibnamefont
  {{Saga}}}, \bibinfo {author} {\bibfnamefont {A.}~\bibnamefont {{Taruya}}}, \
  and\ \bibinfo {author} {\bibfnamefont {S.}~\bibnamefont {{Colombi}}},\ }\href
  {\doibase 10.1051/0004-6361/202142756} {\bibfield  {journal} {\bibinfo
  {journal} {\aap}\ }\textbf {\bibinfo {volume} {664}},\ \bibinfo {eid} {A3}
  (\bibinfo {year} {2022})},\ \Eprint {http://arxiv.org/abs/2111.08836}
  {arXiv:2111.08836 [astro-ph.CO]} \BibitemShut {NoStop}%
\bibitem [{\citenamefont {{Angulo}}\ and\ \citenamefont
  {{Hahn}}(2022)}]{AngHah2212}%
  \BibitemOpen
  \bibfield  {author} {\bibinfo {author} {\bibfnamefont {R.~E.}\ \bibnamefont
  {{Angulo}}}\ and\ \bibinfo {author} {\bibfnamefont {O.}~\bibnamefont
  {{Hahn}}},\ }\href {\doibase 10.1007/s41115-021-00013-z} {\bibfield
  {journal} {\bibinfo  {journal} {Living Reviews in Computational
  Astrophysics}\ }\textbf {\bibinfo {volume} {8}},\ \bibinfo {eid} {1}
  (\bibinfo {year} {2022})},\ \Eprint {http://arxiv.org/abs/2112.05165}
  {arXiv:2112.05165 [astro-ph.CO]} \BibitemShut {NoStop}%
\bibitem [{\citenamefont {{Widrow}}\ and\ \citenamefont
  {{Kaiser}}(1993)}]{WidKai9310}%
  \BibitemOpen
  \bibfield  {author} {\bibinfo {author} {\bibfnamefont {L.~M.}\ \bibnamefont
  {{Widrow}}}\ and\ \bibinfo {author} {\bibfnamefont {N.}~\bibnamefont
  {{Kaiser}}},\ }\href {\doibase 10.1086/187073} {\bibfield  {journal}
  {\bibinfo  {journal} {\apjl}\ }\textbf {\bibinfo {volume} {416}},\ \bibinfo
  {pages} {L71+} (\bibinfo {year} {1993})}\BibitemShut {NoStop}%
\bibitem [{\citenamefont {Schive}\ \emph {et~al.}(2014)\citenamefont {Schive},
  \citenamefont {Chiueh},\ and\ \citenamefont {Broadhurst}}]{Schive:2014dra}%
  \BibitemOpen
  \bibfield  {author} {\bibinfo {author} {\bibfnamefont {H.-Y.}\ \bibnamefont
  {Schive}}, \bibinfo {author} {\bibfnamefont {T.}~\bibnamefont {Chiueh}}, \
  and\ \bibinfo {author} {\bibfnamefont {T.}~\bibnamefont {Broadhurst}},\
  }\href {\doibase 10.1038/nphys2996} {\bibfield  {journal} {\bibinfo
  {journal} {Nature Phys.}\ }\textbf {\bibinfo {volume} {10}},\ \bibinfo
  {pages} {496} (\bibinfo {year} {2014})},\ \Eprint
  {http://arxiv.org/abs/1406.6586} {arXiv:1406.6586 [astro-ph.GA]} \BibitemShut
  {NoStop}%
\bibitem [{\citenamefont {Kopp}\ \emph {et~al.}(2017)\citenamefont {Kopp},
  \citenamefont {Vattis},\ and\ \citenamefont {Skordis}}]{Kopp:2017hbb}%
  \BibitemOpen
  \bibfield  {author} {\bibinfo {author} {\bibfnamefont {M.}~\bibnamefont
  {Kopp}}, \bibinfo {author} {\bibfnamefont {K.}~\bibnamefont {Vattis}}, \ and\
  \bibinfo {author} {\bibfnamefont {C.}~\bibnamefont {Skordis}},\ }\href
  {\doibase 10.1103/PhysRevD.96.123532} {\bibfield  {journal} {\bibinfo
  {journal} {Phys. Rev. D}\ }\textbf {\bibinfo {volume} {96}},\ \bibinfo
  {pages} {123532} (\bibinfo {year} {2017})},\ \Eprint
  {http://arxiv.org/abs/1711.00140} {arXiv:1711.00140 [astro-ph.CO]}
  \BibitemShut {NoStop}%
\bibitem [{\citenamefont {{Garny}}\ and\ \citenamefont
  {{Konstandin}}(2018)}]{GarKon1801}%
  \BibitemOpen
  \bibfield  {author} {\bibinfo {author} {\bibfnamefont {M.}~\bibnamefont
  {{Garny}}}\ and\ \bibinfo {author} {\bibfnamefont {T.}~\bibnamefont
  {{Konstandin}}},\ }\href {\doibase 10.1088/1475-7516/2018/01/009} {\bibfield
  {journal} {\bibinfo  {journal} {\jcap}\ }\textbf {\bibinfo {volume} {2018}},\
  \bibinfo {eid} {009} (\bibinfo {year} {2018})},\ \Eprint
  {http://arxiv.org/abs/1710.04846} {arXiv:1710.04846 [astro-ph.CO]}
  \BibitemShut {NoStop}%
\bibitem [{\citenamefont {{Uhlemann}}(2018)}]{Uhl1810}%
  \BibitemOpen
  \bibfield  {author} {\bibinfo {author} {\bibfnamefont {C.}~\bibnamefont
  {{Uhlemann}}},\ }\href {\doibase 10.1088/1475-7516/2018/10/030} {\bibfield
  {journal} {\bibinfo  {journal} {\jcap}\ }\textbf {\bibinfo {volume} {2018}},\
  \bibinfo {eid} {030} (\bibinfo {year} {2018})},\ \Eprint
  {http://arxiv.org/abs/1807.07274} {arXiv:1807.07274 [astro-ph.CO]}
  \BibitemShut {NoStop}%
\bibitem [{\citenamefont {Mocz}\ \emph {et~al.}(2018)\citenamefont {Mocz},
  \citenamefont {Lancaster}, \citenamefont {Fialkov}, \citenamefont {Becerra},\
  and\ \citenamefont {Chavanis}}]{Mocz:2018ium}%
  \BibitemOpen
  \bibfield  {author} {\bibinfo {author} {\bibfnamefont {P.}~\bibnamefont
  {Mocz}}, \bibinfo {author} {\bibfnamefont {L.}~\bibnamefont {Lancaster}},
  \bibinfo {author} {\bibfnamefont {A.}~\bibnamefont {Fialkov}}, \bibinfo
  {author} {\bibfnamefont {F.}~\bibnamefont {Becerra}}, \ and\ \bibinfo
  {author} {\bibfnamefont {P.-H.}\ \bibnamefont {Chavanis}},\ }\href {\doibase
  10.1103/PhysRevD.97.083519} {\bibfield  {journal} {\bibinfo  {journal} {Phys.
  Rev. D}\ }\textbf {\bibinfo {volume} {97}},\ \bibinfo {pages} {083519}
  (\bibinfo {year} {2018})},\ \Eprint {http://arxiv.org/abs/1801.03507}
  {arXiv:1801.03507 [astro-ph.CO]} \BibitemShut {NoStop}%
\bibitem [{\citenamefont {Li}\ \emph {et~al.}(2019)\citenamefont {Li},
  \citenamefont {Hui},\ and\ \citenamefont {Bryan}}]{Li:2018kyk}%
  \BibitemOpen
  \bibfield  {author} {\bibinfo {author} {\bibfnamefont {X.}~\bibnamefont
  {Li}}, \bibinfo {author} {\bibfnamefont {L.}~\bibnamefont {Hui}}, \ and\
  \bibinfo {author} {\bibfnamefont {G.~L.}\ \bibnamefont {Bryan}},\ }\href
  {\doibase 10.1103/PhysRevD.99.063509} {\bibfield  {journal} {\bibinfo
  {journal} {Phys. Rev. D}\ }\textbf {\bibinfo {volume} {99}},\ \bibinfo
  {pages} {063509} (\bibinfo {year} {2019})},\ \Eprint
  {http://arxiv.org/abs/1810.01915} {arXiv:1810.01915 [astro-ph.CO]}
  \BibitemShut {NoStop}%
\bibitem [{\citenamefont {{Uhlemann}}\ \emph {et~al.}(2019)\citenamefont
  {{Uhlemann}}, \citenamefont {{Rampf}}, \citenamefont {{Gosenca}},\ and\
  \citenamefont {{Hahn}}}]{UhlRamGos1904}%
  \BibitemOpen
  \bibfield  {author} {\bibinfo {author} {\bibfnamefont {C.}~\bibnamefont
  {{Uhlemann}}}, \bibinfo {author} {\bibfnamefont {C.}~\bibnamefont {{Rampf}}},
  \bibinfo {author} {\bibfnamefont {M.}~\bibnamefont {{Gosenca}}}, \ and\
  \bibinfo {author} {\bibfnamefont {O.}~\bibnamefont {{Hahn}}},\ }\href
  {\doibase 10.1103/PhysRevD.99.083524} {\bibfield  {journal} {\bibinfo
  {journal} {\prd}\ }\textbf {\bibinfo {volume} {99}},\ \bibinfo {eid} {083524}
  (\bibinfo {year} {2019})},\ \Eprint {http://arxiv.org/abs/1812.05633}
  {arXiv:1812.05633 [astro-ph.CO]} \BibitemShut {NoStop}%
\bibitem [{\citenamefont {Garny}\ \emph {et~al.}(2020)\citenamefont {Garny},
  \citenamefont {Konstandin},\ and\ \citenamefont {Rubira}}]{Garny:2019noq}%
  \BibitemOpen
  \bibfield  {author} {\bibinfo {author} {\bibfnamefont {M.}~\bibnamefont
  {Garny}}, \bibinfo {author} {\bibfnamefont {T.}~\bibnamefont {Konstandin}}, \
  and\ \bibinfo {author} {\bibfnamefont {H.}~\bibnamefont {Rubira}},\ }\href
  {\doibase 10.1088/1475-7516/2020/04/003} {\bibfield  {journal} {\bibinfo
  {journal} {JCAP}\ }\textbf {\bibinfo {volume} {04}},\ \bibinfo {pages} {003}
  (\bibinfo {year} {2020})},\ \Eprint {http://arxiv.org/abs/1911.04505}
  {arXiv:1911.04505 [astro-ph.CO]} \BibitemShut {NoStop}%
\bibitem [{\citenamefont {May}\ and\ \citenamefont
  {Springel}(2021)}]{May:2021wwp}%
  \BibitemOpen
  \bibfield  {author} {\bibinfo {author} {\bibfnamefont {S.}~\bibnamefont
  {May}}\ and\ \bibinfo {author} {\bibfnamefont {V.}~\bibnamefont {Springel}},\
  }\href {\doibase 10.1093/mnras/stab1764} {\bibfield  {journal} {\bibinfo
  {journal} {Mon. Not. Roy. Astron. Soc.}\ }\textbf {\bibinfo {volume} {506}},\
  \bibinfo {pages} {2603} (\bibinfo {year} {2021})},\ \Eprint
  {http://arxiv.org/abs/2101.01828} {arXiv:2101.01828 [astro-ph.CO]}
  \BibitemShut {NoStop}%
\bibitem [{\citenamefont {{Evans}}(1993)}]{Eva9301}%
  \BibitemOpen
  \bibfield  {author} {\bibinfo {author} {\bibfnamefont {N.~W.}\ \bibnamefont
  {{Evans}}},\ }\href {\doibase 10.1093/mnras/260.1.191} {\bibfield  {journal}
  {\bibinfo  {journal} {\mnras}\ }\textbf {\bibinfo {volume} {260}},\ \bibinfo
  {pages} {191} (\bibinfo {year} {1993})}\BibitemShut {NoStop}%
\bibitem [{\citenamefont {{Evans}}(1994)}]{Eva9403}%
  \BibitemOpen
  \bibfield  {author} {\bibinfo {author} {\bibfnamefont {N.~W.}\ \bibnamefont
  {{Evans}}},\ }\href {\doibase 10.1093/mnras/267.2.333} {\bibfield  {journal}
  {\bibinfo  {journal} {\mnras}\ }\textbf {\bibinfo {volume} {267}},\ \bibinfo
  {pages} {333} (\bibinfo {year} {1994})}\BibitemShut {NoStop}%
\bibitem [{\citenamefont {{Frieman}}\ and\ \citenamefont
  {{Scoccimarro}}(1994)}]{FriSco9408}%
  \BibitemOpen
  \bibfield  {author} {\bibinfo {author} {\bibfnamefont {J.}~\bibnamefont
  {{Frieman}}}\ and\ \bibinfo {author} {\bibfnamefont {R.}~\bibnamefont
  {{Scoccimarro}}},\ }\href {\doibase 10.1086/187463} {\bibfield  {journal}
  {\bibinfo  {journal} {\apjl}\ }\textbf {\bibinfo {volume} {431}},\ \bibinfo
  {pages} {L23} (\bibinfo {year} {1994})},\ \Eprint
  {http://arxiv.org/abs/arXiv:astro-ph/9312043} {arXiv:astro-ph/9312043}
  \BibitemShut {NoStop}%
\bibitem [{\citenamefont {{Alcock}}\ \emph {et~al.}(1995)\citenamefont
  {{Alcock}}, \citenamefont {{Allsman}}, \citenamefont {{Axelrod}},
  \citenamefont {{Bennett}}, \citenamefont {{Cook}}, \citenamefont {{Evans}},
  \citenamefont {{Freeman}}, \citenamefont {{Griest}}, \citenamefont
  {{Jijina}}, \citenamefont {{Lehner}}, \citenamefont {{Marshall}},
  \citenamefont {{Perlmutter}}, \citenamefont {{Peterson}}, \citenamefont
  {{Pratt}}, \citenamefont {{Quinn}}, \citenamefont {{Rodgers}}, \citenamefont
  {{Stubbs}}, \citenamefont {{Sutherland}},\ and\ \citenamefont {{MACHO
  Collaboration}}}]{AlcAllAxe9508}%
  \BibitemOpen
  \bibfield  {author} {\bibinfo {author} {\bibfnamefont {C.}~\bibnamefont
  {{Alcock}}}, \bibinfo {author} {\bibfnamefont {R.~A.}\ \bibnamefont
  {{Allsman}}}, \bibinfo {author} {\bibfnamefont {T.~S.}\ \bibnamefont
  {{Axelrod}}}, \bibinfo {author} {\bibfnamefont {D.~P.}\ \bibnamefont
  {{Bennett}}}, \bibinfo {author} {\bibfnamefont {K.~H.}\ \bibnamefont
  {{Cook}}}, \bibinfo {author} {\bibfnamefont {N.~W.}\ \bibnamefont {{Evans}}},
  \bibinfo {author} {\bibfnamefont {K.~C.}\ \bibnamefont {{Freeman}}}, \bibinfo
  {author} {\bibfnamefont {K.}~\bibnamefont {{Griest}}}, \bibinfo {author}
  {\bibfnamefont {J.}~\bibnamefont {{Jijina}}}, \bibinfo {author}
  {\bibfnamefont {M.}~\bibnamefont {{Lehner}}}, \bibinfo {author}
  {\bibfnamefont {S.~L.}\ \bibnamefont {{Marshall}}}, \bibinfo {author}
  {\bibfnamefont {S.}~\bibnamefont {{Perlmutter}}}, \bibinfo {author}
  {\bibfnamefont {B.~A.}\ \bibnamefont {{Peterson}}}, \bibinfo {author}
  {\bibfnamefont {M.~R.}\ \bibnamefont {{Pratt}}}, \bibinfo {author}
  {\bibfnamefont {P.~J.}\ \bibnamefont {{Quinn}}}, \bibinfo {author}
  {\bibfnamefont {A.~W.}\ \bibnamefont {{Rodgers}}}, \bibinfo {author}
  {\bibfnamefont {C.~W.}\ \bibnamefont {{Stubbs}}}, \bibinfo {author}
  {\bibfnamefont {W.}~\bibnamefont {{Sutherland}}}, \ and\ \bibinfo {author}
  {\bibnamefont {{MACHO Collaboration}}},\ }\href {\doibase 10.1086/176028}
  {\bibfield  {journal} {\bibinfo  {journal} {\apj}\ }\textbf {\bibinfo
  {volume} {449}},\ \bibinfo {pages} {28} (\bibinfo {year} {1995})},\ \Eprint
  {http://arxiv.org/abs/astro-ph/9411019} {arXiv:astro-ph/9411019 [astro-ph]}
  \BibitemShut {NoStop}%
\bibitem [{\citenamefont {{Calcino}}\ \emph {et~al.}(2018)\citenamefont
  {{Calcino}}, \citenamefont {{Garc{\'\i}a-Bellido}},\ and\ \citenamefont
  {{Davis}}}]{CalGarDav1809}%
  \BibitemOpen
  \bibfield  {author} {\bibinfo {author} {\bibfnamefont {J.}~\bibnamefont
  {{Calcino}}}, \bibinfo {author} {\bibfnamefont {J.}~\bibnamefont
  {{Garc{\'\i}a-Bellido}}}, \ and\ \bibinfo {author} {\bibfnamefont {T.~M.}\
  \bibnamefont {{Davis}}},\ }\href {\doibase 10.1093/mnras/sty1368} {\bibfield
  {journal} {\bibinfo  {journal} {\mnras}\ }\textbf {\bibinfo {volume} {479}},\
  \bibinfo {pages} {2889} (\bibinfo {year} {2018})},\ \Eprint
  {http://arxiv.org/abs/1803.09205} {arXiv:1803.09205 [astro-ph.CO]}
  \BibitemShut {NoStop}%
\bibitem [{\citenamefont {{Navarro}}\ \emph {et~al.}(1997)\citenamefont
  {{Navarro}}, \citenamefont {{Frenk}},\ and\ \citenamefont
  {{White}}}]{NavFreWhi97}%
  \BibitemOpen
  \bibfield  {author} {\bibinfo {author} {\bibfnamefont {J.}~\bibnamefont
  {{Navarro}}}, \bibinfo {author} {\bibfnamefont {C.}~\bibnamefont {{Frenk}}},
  \ and\ \bibinfo {author} {\bibfnamefont {S.}~\bibnamefont {{White}}},\
  }\href@noop {} {\bibfield  {journal} {\bibinfo  {journal} {\apj}\ }\textbf
  {\bibinfo {volume} {490}},\ \bibinfo {pages} {493} (\bibinfo {year}
  {1997})}\BibitemShut {NoStop}%
\bibitem [{\citenamefont {{Chua}}\ \emph {et~al.}(2019)\citenamefont {{Chua}},
  \citenamefont {{Pillepich}}, \citenamefont {{Vogelsberger}},\ and\
  \citenamefont {{Hernquist}}}]{ChuPilVog1903}%
  \BibitemOpen
  \bibfield  {author} {\bibinfo {author} {\bibfnamefont {K.~T.~E.}\
  \bibnamefont {{Chua}}}, \bibinfo {author} {\bibfnamefont {A.}~\bibnamefont
  {{Pillepich}}}, \bibinfo {author} {\bibfnamefont {M.}~\bibnamefont
  {{Vogelsberger}}}, \ and\ \bibinfo {author} {\bibfnamefont {L.}~\bibnamefont
  {{Hernquist}}},\ }\href {\doibase 10.1093/mnras/sty3531} {\bibfield
  {journal} {\bibinfo  {journal} {\mnras}\ }\textbf {\bibinfo {volume} {484}},\
  \bibinfo {pages} {476} (\bibinfo {year} {2019})},\ \Eprint
  {http://arxiv.org/abs/1809.07255} {arXiv:1809.07255 [astro-ph.GA]}
  \BibitemShut {NoStop}%
\bibitem [{\citenamefont {{Prada}}\ \emph {et~al.}(2019)\citenamefont
  {{Prada}}, \citenamefont {{Forero-Romero}}, \citenamefont {{Grand}},
  \citenamefont {{Pakmor}},\ and\ \citenamefont {{Springel}}}]{PraForGra1912}%
  \BibitemOpen
  \bibfield  {author} {\bibinfo {author} {\bibfnamefont {J.}~\bibnamefont
  {{Prada}}}, \bibinfo {author} {\bibfnamefont {J.~E.}\ \bibnamefont
  {{Forero-Romero}}}, \bibinfo {author} {\bibfnamefont {R.~J.~J.}\ \bibnamefont
  {{Grand}}}, \bibinfo {author} {\bibfnamefont {R.}~\bibnamefont {{Pakmor}}}, \
  and\ \bibinfo {author} {\bibfnamefont {V.}~\bibnamefont {{Springel}}},\
  }\href {\doibase 10.1093/mnras/stz2873} {\bibfield  {journal} {\bibinfo
  {journal} {\mnras}\ }\textbf {\bibinfo {volume} {490}},\ \bibinfo {pages}
  {4877} (\bibinfo {year} {2019})},\ \Eprint {http://arxiv.org/abs/1910.04045}
  {arXiv:1910.04045 [astro-ph.GA]} \BibitemShut {NoStop}%
\bibitem [{\citenamefont {{Binney}}\ and\ \citenamefont
  {{Tremaine}}(2008)}]{BinTre08}%
  \BibitemOpen
  \bibfield  {author} {\bibinfo {author} {\bibfnamefont {J.}~\bibnamefont
  {{Binney}}}\ and\ \bibinfo {author} {\bibfnamefont {S.}~\bibnamefont
  {{Tremaine}}},\ }\href@noop {} {\emph {\bibinfo {title} {{Galactic Dynamics:
  Second Edition}}}}\ (\bibinfo {year} {2008})\BibitemShut {NoStop}%
\bibitem [{\citenamefont {{Hansen}}\ and\ \citenamefont
  {{Moore}}(2006)}]{HanMoo0603}%
  \BibitemOpen
  \bibfield  {author} {\bibinfo {author} {\bibfnamefont {S.~H.}\ \bibnamefont
  {{Hansen}}}\ and\ \bibinfo {author} {\bibfnamefont {B.}~\bibnamefont
  {{Moore}}},\ }\href {\doibase 10.1016/j.newast.2005.09.001} {\bibfield
  {journal} {\bibinfo  {journal} {\na}\ }\textbf {\bibinfo {volume} {11}},\
  \bibinfo {pages} {333} (\bibinfo {year} {2006})},\ \Eprint
  {http://arxiv.org/abs/astro-ph/0411473} {arXiv:astro-ph/0411473 [astro-ph]}
  \BibitemShut {NoStop}%
\bibitem [{\citenamefont {{Wojtak}}\ \emph {et~al.}(2008)\citenamefont
  {{Wojtak}}, \citenamefont {{{\L}okas}}, \citenamefont {{Mamon}},
  \citenamefont {{Gottl{\"o}ber}}, \citenamefont {{Klypin}},\ and\
  \citenamefont {{Hoffman}}}]{WojlokMam0808}%
  \BibitemOpen
  \bibfield  {author} {\bibinfo {author} {\bibfnamefont {R.}~\bibnamefont
  {{Wojtak}}}, \bibinfo {author} {\bibfnamefont {E.~L.}\ \bibnamefont
  {{{\L}okas}}}, \bibinfo {author} {\bibfnamefont {G.~A.}\ \bibnamefont
  {{Mamon}}}, \bibinfo {author} {\bibfnamefont {S.}~\bibnamefont
  {{Gottl{\"o}ber}}}, \bibinfo {author} {\bibfnamefont {A.}~\bibnamefont
  {{Klypin}}}, \ and\ \bibinfo {author} {\bibfnamefont {Y.}~\bibnamefont
  {{Hoffman}}},\ }\href {\doibase 10.1111/j.1365-2966.2008.13441.x} {\bibfield
  {journal} {\bibinfo  {journal} {\mnras}\ }\textbf {\bibinfo {volume} {388}},\
  \bibinfo {pages} {815} (\bibinfo {year} {2008})},\ \Eprint
  {http://arxiv.org/abs/0802.0429} {arXiv:0802.0429 [astro-ph]} \BibitemShut
  {NoStop}%
\bibitem [{\citenamefont {{Sparre}}\ and\ \citenamefont
  {{Hansen}}(2012)}]{SpaHan1210}%
  \BibitemOpen
  \bibfield  {author} {\bibinfo {author} {\bibfnamefont {M.}~\bibnamefont
  {{Sparre}}}\ and\ \bibinfo {author} {\bibfnamefont {S.~H.}\ \bibnamefont
  {{Hansen}}},\ }\href {\doibase 10.1088/1475-7516/2012/10/049} {\bibfield
  {journal} {\bibinfo  {journal} {\jcap}\ }\textbf {\bibinfo {volume} {2012}},\
  \bibinfo {eid} {049} (\bibinfo {year} {2012})},\ \Eprint
  {http://arxiv.org/abs/1210.2392} {arXiv:1210.2392 [astro-ph.CO]} \BibitemShut
  {NoStop}%
\bibitem [{\citenamefont {{Osipkov}}(1979)}]{Osi7901}%
  \BibitemOpen
  \bibfield  {author} {\bibinfo {author} {\bibfnamefont {L.~P.}\ \bibnamefont
  {{Osipkov}}},\ }\href@noop {} {\bibfield  {journal} {\bibinfo  {journal}
  {Soviet Astronomy Letters}\ }\textbf {\bibinfo {volume} {5}},\ \bibinfo
  {pages} {42} (\bibinfo {year} {1979})}\BibitemShut {NoStop}%
\bibitem [{\citenamefont {{Merritt}}(1985)}]{Mer8506}%
  \BibitemOpen
  \bibfield  {author} {\bibinfo {author} {\bibfnamefont {D.}~\bibnamefont
  {{Merritt}}},\ }\href {\doibase 10.1086/113810} {\bibfield  {journal}
  {\bibinfo  {journal} {\aj}\ }\textbf {\bibinfo {volume} {90}},\ \bibinfo
  {pages} {1027} (\bibinfo {year} {1985})}\BibitemShut {NoStop}%
\bibitem [{\citenamefont {{Cole}}\ and\ \citenamefont
  {{Lacey}}(1996)}]{ColLac9607}%
  \BibitemOpen
  \bibfield  {author} {\bibinfo {author} {\bibfnamefont {S.}~\bibnamefont
  {{Cole}}}\ and\ \bibinfo {author} {\bibfnamefont {C.}~\bibnamefont
  {{Lacey}}},\ }\href {\doibase 10.1093/mnras/281.2.716} {\bibfield  {journal}
  {\bibinfo  {journal} {\mnras}\ }\textbf {\bibinfo {volume} {281}},\ \bibinfo
  {pages} {716} (\bibinfo {year} {1996})},\ \Eprint
  {http://arxiv.org/abs/astro-ph/9510147} {arXiv:astro-ph/9510147 [astro-ph]}
  \BibitemShut {NoStop}%
\bibitem [{\citenamefont {{Sheth}}\ \emph {et~al.}(2001)\citenamefont
  {{Sheth}}, \citenamefont {{Hui}}, \citenamefont {{Diaferio}},\ and\
  \citenamefont {{Scoccimarro}}}]{SheHuiDia01}%
  \BibitemOpen
  \bibfield  {author} {\bibinfo {author} {\bibfnamefont {R.}~\bibnamefont
  {{Sheth}}}, \bibinfo {author} {\bibfnamefont {L.}~\bibnamefont {{Hui}}},
  \bibinfo {author} {\bibfnamefont {A.}~\bibnamefont {{Diaferio}}}, \ and\
  \bibinfo {author} {\bibfnamefont {R.}~\bibnamefont {{Scoccimarro}}},\
  }\href@noop {} {\bibfield  {journal} {\bibinfo  {journal} {\mnras}\ }\textbf
  {\bibinfo {volume} {325}},\ \bibinfo {pages} {1288} (\bibinfo {year}
  {2001})}\BibitemShut {NoStop}%
\bibitem [{\citenamefont {{Pichon}}\ and\ \citenamefont
  {{Bernardeau}}(1999)}]{PicBer9903}%
  \BibitemOpen
  \bibfield  {author} {\bibinfo {author} {\bibfnamefont {C.}~\bibnamefont
  {{Pichon}}}\ and\ \bibinfo {author} {\bibfnamefont {F.}~\bibnamefont
  {{Bernardeau}}},\ }\href@noop {} {\bibfield  {journal} {\bibinfo  {journal}
  {\aap}\ }\textbf {\bibinfo {volume} {343}},\ \bibinfo {pages} {663} (\bibinfo
  {year} {1999})},\ \Eprint {http://arxiv.org/abs/astro-ph/9902142}
  {arXiv:astro-ph/9902142 [astro-ph]} \BibitemShut {NoStop}%
\bibitem [{\citenamefont {{Szapudi}}\ and\ \citenamefont
  {{Kaiser}}(2003)}]{SzaKai0301}%
  \BibitemOpen
  \bibfield  {author} {\bibinfo {author} {\bibfnamefont {I.}~\bibnamefont
  {{Szapudi}}}\ and\ \bibinfo {author} {\bibfnamefont {N.}~\bibnamefont
  {{Kaiser}}},\ }\href {\doibase 10.1086/368013} {\bibfield  {journal}
  {\bibinfo  {journal} {\apjl}\ }\textbf {\bibinfo {volume} {583}},\ \bibinfo
  {pages} {L1} (\bibinfo {year} {2003})},\ \Eprint
  {http://arxiv.org/abs/arXiv:astro-ph/0211065} {arXiv:astro-ph/0211065}
  \BibitemShut {NoStop}%
\bibitem [{\citenamefont {{Neyrinck}}\ \emph {et~al.}(2009)\citenamefont
  {{Neyrinck}}, \citenamefont {{Szapudi}},\ and\ \citenamefont
  {{Szalay}}}]{NeySzaSza0906}%
  \BibitemOpen
  \bibfield  {author} {\bibinfo {author} {\bibfnamefont {M.~C.}\ \bibnamefont
  {{Neyrinck}}}, \bibinfo {author} {\bibfnamefont {I.}~\bibnamefont
  {{Szapudi}}}, \ and\ \bibinfo {author} {\bibfnamefont {A.~S.}\ \bibnamefont
  {{Szalay}}},\ }\href {\doibase 10.1088/0004-637X/698/2/L90} {\bibfield
  {journal} {\bibinfo  {journal} {\apjl}\ }\textbf {\bibinfo {volume} {698}},\
  \bibinfo {pages} {L90} (\bibinfo {year} {2009})},\ \Eprint
  {http://arxiv.org/abs/0903.4693} {arXiv:0903.4693 [astro-ph.CO]} \BibitemShut
  {NoStop}%
\bibitem [{\citenamefont {Wang}\ \emph {et~al.}(2011)\citenamefont {Wang},
  \citenamefont {Neyrinck}, \citenamefont {Szapudi}, \citenamefont {Szalay},
  \citenamefont {Chen}, \citenamefont {Lesgourgues}, \citenamefont {Riotto},\
  and\ \citenamefont {Sloth}}]{Wang:2011fj}%
  \BibitemOpen
  \bibfield  {author} {\bibinfo {author} {\bibfnamefont {X.}~\bibnamefont
  {Wang}}, \bibinfo {author} {\bibfnamefont {M.}~\bibnamefont {Neyrinck}},
  \bibinfo {author} {\bibfnamefont {I.}~\bibnamefont {Szapudi}}, \bibinfo
  {author} {\bibfnamefont {A.}~\bibnamefont {Szalay}}, \bibinfo {author}
  {\bibfnamefont {X.}~\bibnamefont {Chen}}, \bibinfo {author} {\bibfnamefont
  {J.}~\bibnamefont {Lesgourgues}}, \bibinfo {author} {\bibfnamefont
  {A.}~\bibnamefont {Riotto}}, \ and\ \bibinfo {author} {\bibfnamefont
  {M.}~\bibnamefont {Sloth}},\ }\href {\doibase 10.1088/0004-637X/735/1/32}
  {\bibfield  {journal} {\bibinfo  {journal} {Astrophys. J.}\ }\textbf
  {\bibinfo {volume} {735}},\ \bibinfo {pages} {32} (\bibinfo {year} {2011})},\
  \Eprint {http://arxiv.org/abs/1103.2166} {arXiv:1103.2166 [astro-ph.CO]}
  \BibitemShut {NoStop}%
\bibitem [{\citenamefont {{Carron}}\ and\ \citenamefont
  {{Szapudi}}(2013)}]{CarSza1310}%
  \BibitemOpen
  \bibfield  {author} {\bibinfo {author} {\bibfnamefont {J.}~\bibnamefont
  {{Carron}}}\ and\ \bibinfo {author} {\bibfnamefont {I.}~\bibnamefont
  {{Szapudi}}},\ }\href {\doibase 10.1093/mnras/stt1215} {\bibfield  {journal}
  {\bibinfo  {journal} {\mnras}\ }\textbf {\bibinfo {volume} {434}},\ \bibinfo
  {pages} {2961} (\bibinfo {year} {2013})},\ \Eprint
  {http://arxiv.org/abs/1306.1230} {arXiv:1306.1230 [astro-ph.CO]} \BibitemShut
  {NoStop}%
\bibitem [{\citenamefont {Rubira}\ and\ \citenamefont
  {Voivodic}(2021)}]{Rubira:2020inb}%
  \BibitemOpen
  \bibfield  {author} {\bibinfo {author} {\bibfnamefont {H.}~\bibnamefont
  {Rubira}}\ and\ \bibinfo {author} {\bibfnamefont {R.}~\bibnamefont
  {Voivodic}},\ }\href {\doibase 10.1088/1475-7516/2021/03/070} {\bibfield
  {journal} {\bibinfo  {journal} {JCAP}\ }\textbf {\bibinfo {volume} {03}},\
  \bibinfo {pages} {070} (\bibinfo {year} {2021})},\ \Eprint
  {http://arxiv.org/abs/2011.12280} {arXiv:2011.12280 [astro-ph.CO]}
  \BibitemShut {NoStop}%
\bibitem [{\citenamefont {{Scoccimarro}}(2001)}]{Sco01}%
  \BibitemOpen
  \bibfield  {author} {\bibinfo {author} {\bibfnamefont {R.}~\bibnamefont
  {{Scoccimarro}}},\ }\href@noop {} {\bibfield  {journal} {\bibinfo  {journal}
  {ArXiv:astro-ph/0008277, Annals New York Academy Sciences}\ }\textbf
  {\bibinfo {volume} {927}},\ \bibinfo {pages} {13} (\bibinfo {year}
  {2001})}\BibitemShut {NoStop}%
\bibitem [{\citenamefont {{Goroff}}\ \emph {et~al.}(1986)\citenamefont
  {{Goroff}}, \citenamefont {{Grinstein}}, \citenamefont {{Rey}},\ and\
  \citenamefont {{Wise}}}]{GorGriRey86}%
  \BibitemOpen
  \bibfield  {author} {\bibinfo {author} {\bibfnamefont {M.}~\bibnamefont
  {{Goroff}}}, \bibinfo {author} {\bibfnamefont {B.}~\bibnamefont
  {{Grinstein}}}, \bibinfo {author} {\bibfnamefont {S.-J.}\ \bibnamefont
  {{Rey}}}, \ and\ \bibinfo {author} {\bibfnamefont {M.}~\bibnamefont
  {{Wise}}},\ }\href@noop {} {\bibfield  {journal} {\bibinfo  {journal} {\apj}\
  }\textbf {\bibinfo {volume} {311}},\ \bibinfo {pages} {6} (\bibinfo {year}
  {1986})}\BibitemShut {NoStop}%
\bibitem [{\citenamefont {{Bernardeau}}\ \emph
  {et~al.}(2002{\natexlab{b}})\citenamefont {{Bernardeau}}, \citenamefont
  {{Colombi}}, \citenamefont {{Gazta{\~n}aga}},\ and\ \citenamefont
  {{Scoccimarro}}}]{Bernardeau:2002}%
  \BibitemOpen
  \bibfield  {author} {\bibinfo {author} {\bibfnamefont {F.}~\bibnamefont
  {{Bernardeau}}}, \bibinfo {author} {\bibfnamefont {S.}~\bibnamefont
  {{Colombi}}}, \bibinfo {author} {\bibfnamefont {E.}~\bibnamefont
  {{Gazta{\~n}aga}}}, \ and\ \bibinfo {author} {\bibfnamefont {R.}~\bibnamefont
  {{Scoccimarro}}},\ }\href {\doibase 10.1016/S0370-1573(02)00135-7} {\bibfield
   {journal} {\bibinfo  {journal} {\physrep}\ }\textbf {\bibinfo {volume}
  {367}},\ \bibinfo {pages} {1} (\bibinfo {year} {2002}{\natexlab{b}})},\
  \Eprint {http://arxiv.org/abs/astro-ph/0112551} {astro-ph/0112551}
  \BibitemShut {NoStop}%
\bibitem [{\citenamefont {Blas}\ \emph {et~al.}(2015)\citenamefont {Blas},
  \citenamefont {Floerchinger}, \citenamefont {Garny}, \citenamefont
  {Tetradis},\ and\ \citenamefont {Wiedemann}}]{Blas:2015tla}%
  \BibitemOpen
  \bibfield  {author} {\bibinfo {author} {\bibfnamefont {D.}~\bibnamefont
  {Blas}}, \bibinfo {author} {\bibfnamefont {S.}~\bibnamefont {Floerchinger}},
  \bibinfo {author} {\bibfnamefont {M.}~\bibnamefont {Garny}}, \bibinfo
  {author} {\bibfnamefont {N.}~\bibnamefont {Tetradis}}, \ and\ \bibinfo
  {author} {\bibfnamefont {U.~A.}\ \bibnamefont {Wiedemann}},\ }\href {\doibase
  10.1088/1475-7516/2015/11/049} {\bibfield  {journal} {\bibinfo  {journal}
  {JCAP}\ }\textbf {\bibinfo {volume} {11}},\ \bibinfo {pages} {049} (\bibinfo
  {year} {2015})},\ \Eprint {http://arxiv.org/abs/1507.06665} {arXiv:1507.06665
  [astro-ph.CO]} \BibitemShut {NoStop}%
\bibitem [{\citenamefont {Garny}\ and\ \citenamefont
  {Taule}(2021)}]{Garny:2020ilv}%
  \BibitemOpen
  \bibfield  {author} {\bibinfo {author} {\bibfnamefont {M.}~\bibnamefont
  {Garny}}\ and\ \bibinfo {author} {\bibfnamefont {P.}~\bibnamefont {Taule}},\
  }\href {\doibase 10.1088/1475-7516/2021/01/020} {\bibfield  {journal}
  {\bibinfo  {journal} {JCAP}\ }\textbf {\bibinfo {volume} {01}},\ \bibinfo
  {pages} {020} (\bibinfo {year} {2021})},\ \Eprint
  {http://arxiv.org/abs/2008.00013} {arXiv:2008.00013 [astro-ph.CO]}
  \BibitemShut {NoStop}%
\bibitem [{\citenamefont {Abolhasani}\ \emph {et~al.}(2016)\citenamefont
  {Abolhasani}, \citenamefont {Mirbabayi},\ and\ \citenamefont
  {Pajer}}]{Abolhasani:2015mra}%
  \BibitemOpen
  \bibfield  {author} {\bibinfo {author} {\bibfnamefont {A.~A.}\ \bibnamefont
  {Abolhasani}}, \bibinfo {author} {\bibfnamefont {M.}~\bibnamefont
  {Mirbabayi}}, \ and\ \bibinfo {author} {\bibfnamefont {E.}~\bibnamefont
  {Pajer}},\ }\href {\doibase 10.1088/1475-7516/2016/05/063} {\bibfield
  {journal} {\bibinfo  {journal} {JCAP}\ }\textbf {\bibinfo {volume} {05}},\
  \bibinfo {pages} {063} (\bibinfo {year} {2016})},\ \Eprint
  {http://arxiv.org/abs/1509.07886} {arXiv:1509.07886 [hep-th]} \BibitemShut
  {NoStop}%
\bibitem [{\citenamefont {{Peebles}}(1980)}]{Pee80}%
  \BibitemOpen
  \bibfield  {author} {\bibinfo {author} {\bibfnamefont {P.}~\bibnamefont
  {{Peebles}}},\ }\href@noop {} {\emph {\bibinfo {title} {{The large-scale
  structure of the universe}}}}\ (\bibinfo  {publisher} {Princeton University
  Press},\ \bibinfo {year} {1980})\BibitemShut {NoStop}%
\bibitem [{\citenamefont {{Eddington}}(1916)}]{Edd1605}%
  \BibitemOpen
  \bibfield  {author} {\bibinfo {author} {\bibfnamefont {A.~S.}\ \bibnamefont
  {{Eddington}}},\ }\href {\doibase 10.1093/mnras/76.7.572} {\bibfield
  {journal} {\bibinfo  {journal} {\mnras}\ }\textbf {\bibinfo {volume} {76}},\
  \bibinfo {pages} {572} (\bibinfo {year} {1916})}\BibitemShut {NoStop}%
\bibitem [{\citenamefont {{Evans}}\ and\ \citenamefont
  {{Williams}}(2014)}]{EvaWil1409}%
  \BibitemOpen
  \bibfield  {author} {\bibinfo {author} {\bibfnamefont {N.~W.}\ \bibnamefont
  {{Evans}}}\ and\ \bibinfo {author} {\bibfnamefont {A.~A.}\ \bibnamefont
  {{Williams}}},\ }\href {\doibase 10.1093/mnras/stu1172} {\bibfield  {journal}
  {\bibinfo  {journal} {\mnras}\ }\textbf {\bibinfo {volume} {443}},\ \bibinfo
  {pages} {791} (\bibinfo {year} {2014})},\ \Eprint
  {http://arxiv.org/abs/1406.3730} {arXiv:1406.3730 [astro-ph.GA]} \BibitemShut
  {NoStop}%
\end{thebibliography}
%

\end{document}